\documentclass[preprint,12pt]{elsarticle}

\usepackage{amssymb}
\usepackage{amsmath}
\usepackage{soul}

\usepackage{subcaption}
\usepackage[export]{adjustbox}
\usepackage{graphicx, wrapfig}
\usepackage{graphics}

\usepackage{multirow}
\usepackage{booktabs}

\usepackage{dsfont}

\usepackage{xcolor}         % colors

\usepackage{hyperref}

\usepackage{adjustbox}

\usepackage{amsthm}
\theoremstyle{plain}

\theoremstyle{remark}

\usepackage[normalem]{ulem}
\usepackage{xcolor}
\journal{Journal of Computational Physics}

\begin{document}
% \linenumbers %% LIne numers visible

\begin{frontmatter}



\title{Learning mechanical systems from real-world data using discrete forced Lagrangian dynamics}


\author[sintef,ntnu]{Martine ~Dyring Hansen} %% Author name
\author[ntnu]{Elena ~Celledoni} %% Author name
\author[sintef]{Benjamin Kwanen ~Tapley} %% Author name

\affiliation[sintef]{organization={Department of Mathematics and Cybernetics, SINTEF Digital},%Department and Organization
            % city={},
            postcode={0373 Oslo}, 
            %state={},
            country={Norway}}
\affiliation[ntnu]{organization={Department of Mathematical Sciences, Norwegian University of Science and Technology},%Department and Organization
            % city={},
            postcode={7043 Trondheim}, 
            country={Norway}}
%% Abstract
\begin{abstract}
We introduce a data-driven method for learning the equations of motion of mechanical systems directly from position measurements.
This is particularly relevant in system identification tasks where only positional information is available, such as motion capture, pixel data or low-resolution tracking. Our approach takes advantage of the discrete Lagrange-d'Alembert principle and the forced discrete Euler-Lagrange equations to construct a physically grounded model of the system’s dynamics. We decompose the dynamics into conservative and non-conservative components, which are learned separately using feed-forward neural networks. In the absence of external forces, our method reduces to a variational discretization of the action principle naturally preserving the symplectic structure of the underlying Hamiltonian system. We validate our approach on a variety of synthetic and real-world datasets, demonstrating its effectiveness compared to baseline methods. In particular, we apply our model to (1) measured human motion data and (2)  latent embeddings obtained via an autoencoder trained on image sequences. We demonstrate that we can faithfully reconstruct and separate both the conservative and forced dynamics, yielding interpretable and physically consistent predictions.
\end{abstract}


%     \includegraphics[width=0.8\linewidth]{graphic_sketch.png}


% \begin{highlights}
% \item Learns forced Lagrangian dynamics without velocity measurements
% \item Discrete variational model preserving the Lagrange-d'Alembert structure
% \item Interpretable decomposition of energy and dissipative forces
% \item Latent dynamics aligned with underlying physical structure
% \item Robust multi-step prediction on synthetic and real motion data
% \end{highlights}

% %% Keywords
% \begin{keyword}
% Discrete variational mechanics \sep Lagrangian neural networks \sep Structure-preserving learning \sep Dissipative dynamical systems \sep Physics-informed machine learning \sep Latent dynamics modeling

% \end{keyword}

\end{frontmatter}

%% INTRODUCTION

\begin{figure}[htbp]
    \centering
    \includegraphics[width=0.8\linewidth]{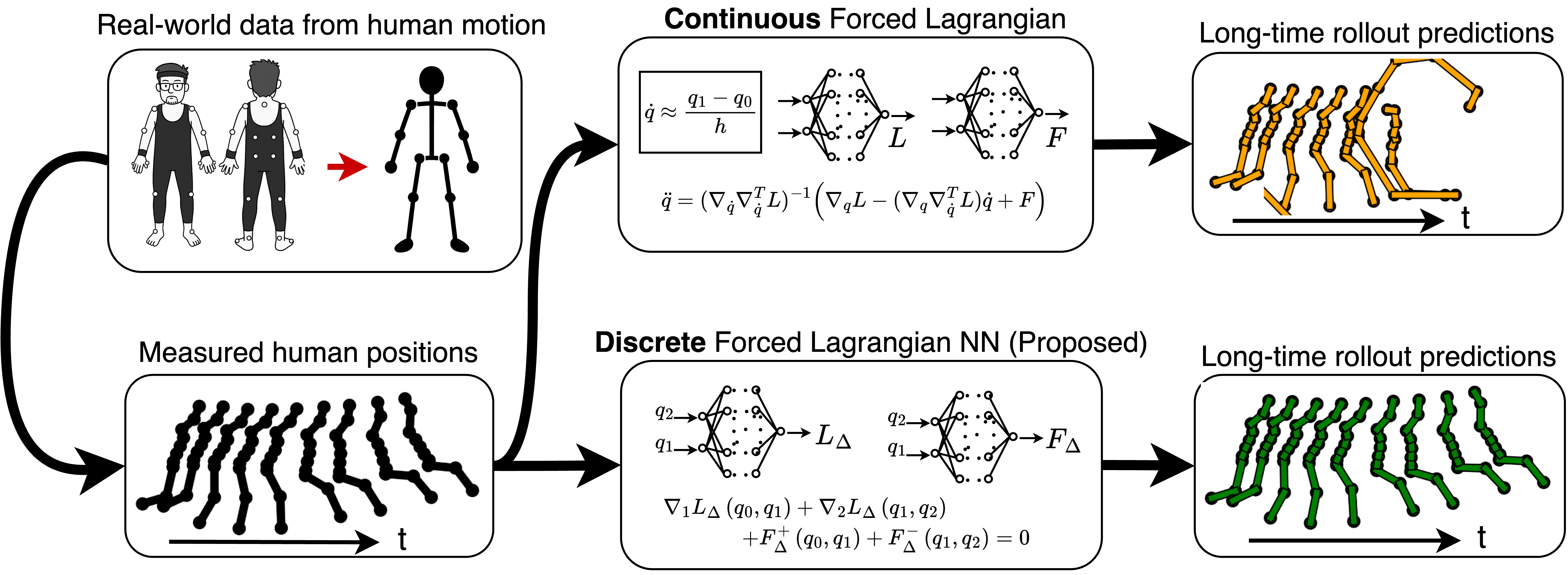}
    \caption{
    In this paper, we propose a structure-preserving approach for learning non-conservative system that directly learns from position data only (i.e., does not require velocities or momenta).}
    \label{fig:front_figure}
\end{figure}

\section{Introduction}
    Incorporating physics principles into machine learning has become a popular strategy for system identification and for accurately predicting dynamics. Seminal examples are the Hamiltonian and Lagrangian neural networks, which are designed to preserve structure when applied to canonical systems \cite{GreydanusHamiltonian2019, CranmerLagrangian2020}. These architectures have naturally been extended to include dissipation, enabling accurate modeling of realistic mechanical systems with physically motivated inductive biases \cite{xiao_generalized_2024, zhong_dissipative_2020}. Such models offer advantages over traditional ``black-box" modeling strategies that face challenges such as noise sensitivity, data sparsity, poor generalization, and limited interpretability \cite{Arridge_Maass_Öktem_Schönlieb_2019, HabibData2021, 
   YinData2014, GuptaNoise2019, WangGeneralization2023}. Incorporating physics into the training processes is becoming an increasingly popular way to increase model generalization while giving physically interpretable predictions. 
    
      Almost all known methods for learning dynamics rely on being able to observe momentum or velocity data, in addition to position data. However, velocity and momentum are not always observed in practice and estimates are usually made from sequential position measurements using finite differences, which lead to inaccuracies due to noise and truncation errors \cite{chartrand2011numerical}. This creates a significant challenge for the two cases explored in this paper: learning from pixel data and motion tracking data, where instantaneous velocities are indeed unavailable. 
    
    Building on ideas of geometric mechanics and variational integrators, \cite{Marsden_West_2001}, we propose a structure-preserving approach that directly learns from \textit{position data only} (i.e., does not require velocities or momenta) and naturally incorporates dissipation and other non-conservative forces such as control terms. The method, which we refer to as a Discrete Forced Lagrangian Neural Network (DFLNN), is based on the discrete Lagrange-d'Alembert principle, which naturally reduces to a symplectic (variational) integrator on the configuration manifold in absence of non-conservative forces. Table~\ref{tab:model_comparison} situates the DFLNN among existing structure-preserving learning methods. A detailed account of what is new relative to these methods is given in Section~\ref{our-work}.
    
    In a number of synthetic problems, the DFLNN is demonstrated to yield significant advantages over a non-structure-preserving neural ODE baseline model and, notably, a structure-preserving continuous analogue of the proposed method based on the (continuous) forced Euler-Lagrange equations where velocities are approximated using finite differences. 
    
    Additionally, we emphasize the flexibility of combining the proposed model with an autoencoder trained on images of damped pendulum dynamics. We show that we can separate the dissipative forces in the latent space to accurately recover and simulate pixel sequences of a \textit{conservative} pendulum, despite the model only learning from the dissipative pixel images. Finally, we apply the model to real-world 3D human motion skeleton tracking data. The model successfully captures the underlying physics and shows excellent prediction performance over long times. Human motion data is extensively employed in simulations, animations, and biomechanical research \cite{MoeslundSurvey2006}. Examining these movements has promising applications across various domains, including healthcare \cite{NguyenSkeletion2016, SardariSkeletion2023} and sports \cite{GuoAttention2021}. Ensuring that models generalize effectively to unseen scenarios is crucial, particularly in medical applications.

\begin{table}[ht]
    \centering
    \caption{Comparison to other models. Models marked $\approx$ learn a continuous-time model that is integrated with a generic scheme, so structure is preserved only approximately.
    } 
    \small 
    \setlength{\tabcolsep}{6pt} % default is 6pt; shrinking this often helps more than font size
    \begin{tabular}{lcccccccc}
        \toprule
        & \textbf{HNN} 
        & \textbf{LNN} 
        & \textbf{LSI}
        & \textbf{D-HNN} 
        & \textbf{GLNN} 
        %& \textbf{\dmA{LieFVIN}}
        & \textbf{DFLNN} \\
        & \cite{GreydanusHamiltonian2019} 
        & \cite{CranmerLagrangian2020} 
        & \cite{OberBlobaumVariational2023} 
        & \cite{desai_port-hamiltonian_2021, sosanya_dissipative_2022} 
        & \cite{xiao_generalized_2024} 
        % & \cite{DuruisseauxLieFVIN2023}
        & (Ours)\\
        \midrule
        Position only   &&& \checkmark &&& %\dmA{\checkmark}  &
        \checkmark \\
        Struct.-pres.   & $\approx$&$\approx$ & \checkmark & $\approx$ & $\approx$ & %\dmA{\checkmark} &
        \checkmark  \\
        Dissipation    & & & & \checkmark & \checkmark &  %&
        \checkmark  \\
        \bottomrule
    \end{tabular}
    \label{tab:model_comparison}
\end{table}

\subsection{Related work}
Understanding and modeling mechanical systems from data is vital in fields such as robotics, biomechanics, and structural health monitoring. While conventional system identification often assumes access to fully known physics, recent efforts focus on physics-informed learning, leveraging partial physical knowledge to improve generalization and interpretability \cite{WuExpertSystems2024, ReichsteinDeep2019, ItenDiscovering2020, KarpatneTheoryGuided2017, AbbasiPhysicsInformed2024, OberBlobaumVariational2023, OffenSymplectic2022, MatsubaraDeep2020, ChenNeural2021, sholokhov2023physics, celledoni24sha, TapleySymplectic2024}. Early approaches like Neural ODEs \cite{ChenNeural2018} and sparse identification of nonlinear dynamics \cite{brunton2016discovering} rely on black-box models or predefined basis functions. These have been paired with auto-encoders to identify low-dimensional latent ODE models from pixel data \cite{champion2019data}, but often neglect structural constraints such as conservation laws.

To address this, recent work embeds physical priors directly into the learning process. Hamiltonian and Lagrangian-based learning methods learn energy-based formulations that define the equations of motion, yielding models with physically consistent inductive biases \cite{GreydanusHamiltonian2019, finzi2020simplifying, celledoni2023learning, ortega24asp, ortega24eas, grigoryeva2023learning, CranmerLagrangian2020, GuptaNoise2019, aoshima2021deep, qin2020machine, lutter2019deep, offen_learning_2024, offen2024machine, OberBlobaumVariational2023}. Such structured models have also been learned directly from image
observations \cite{GreydanusHamiltonian2019, toth2019hamiltonian, ZhongLeonard2020}. When paired with symplectic (variational) integrators or with other geometric integrators, these networks can preserve geometric properties such as the symplectic form, a modified energy over long times and other invariants \cite{MatsubaraDeep2020, TapleySymplectic2024, ChenNeural2021, OffenSymplectic2022, celledoni24sha}. More recently, extensions to non-conservative systems have emerged, combining structure-preserving principles with dissipative or driven dynamics \cite{ortega24eas, ortega2024learnability, xiao_generalized_2024, eidnes2024pseudo, TraskMetriplectic2025, DuruisseauxLieFVIN2023}. These approaches mark a shift toward hybrid models that respect both physical structure and the realities of real-world data.

\subsection{Our contributions.}
\label{our-work}
This paper introduces the Discrete Forced Lagrangian Neural Network (DFLNN), a framework for learning non-conservative, dissipative dynamics from position data alone. 

We build on a range of established concepts, e.g. the discrete Lagrange d'Alembert principle for forced Lagrangian discrete mechanics, and latent dynamics learning. The novelty of our work lies in their combination. 
As Table~\ref{tab:model_comparison} summarizes, existing structure-preserving methods offer either position-only learning or non-conservative forcing, but not both.  We show that this is important when dealing with real world data, such as human motion dynamics.
Our primary contributions are as follows:
   \begin{itemize}
        \item \textit{Velocity-Free Variational Learning}: DFLNN avoids the need for instantaneous velocity or momentum observations. Unlike other approaches where velocity measurements are required \cite{CranmerLagrangian2020, xiao_generalized_2024}, DFLNN naturally incorporates finite-differences approximations from equally spaced position data into a discrete-time structure that preserves the underlying variational mechanics.
        \item  \textit{Structure Preservation}: We demonstrate that at zero external force, the DFLNN reduces to a variational discretization of the action principle and is therefore a symplectic map, providing superior long term stability. More in general,   the scheme is derived from a discrete Lagrange-d'Alemberts principle for forced Hamiltonian systems, \cite{Marsden_West_2001, cortes2001non}. 
        \item \textit{Physical Interpretability}: By leveraging the Lagrange-d'Alembert principle, the DFLNN offers a transparent decomposition of dynamics. This allows for the explicit separation of conservative energy-based terms from non-conservative forces, such as friction or control inputs.
        \item \textit{Latent Space Decomposition}: We show that DFLNN can be integrated with autoencoder embeddings to learn dynamics directly from pixels. Notably, we demonstrate that coupling the DFLNN model together with an autoencoder shapes the learned latent space to follow an underlying physical structure. This results in the ability to identify dissipative forces within the latent representation and to simulate conservative counterparts of observed dissipative systems.
        \item \textit{Empirical Robustness}: We validate the method across a spectrum of complexity, from synthetic benchmarks to high-dimensional, real-world 3D human motion tracking. Our results show that DFLNN significantly outperforms non-structure-preserving baselines in multi-step prediction and generalization. 
    \end{itemize}

%% BACKGROUND
\section{Background}\label{sc:Background}    
    \paragraph{Continuous d'Alembert Principle and Forced Euler-Lagrange Equations}
    In the presence of non-conservative forces, it is useful to consider the generalized formulation of Hamilton's principle, commonly referred to as the Lagrange-d'Alemberts principle. Let $Q$ denote a configuration manifold with $TQ$ its tangent bundle. The Lagrange-d'Alemberts principle states that a curve $q:[t_0,t_N]\rightarrow \mathbb{R}^d$ is an admissible evolution of the mechanical system if it satisfies:
    \begin{equation}\label{eq:ContinuousDAalembertsPrinciple}
        \delta \left( \int_{t_0}^{t_N} L\left( q(t), \dot{q}(t) \right) dt \right) 
        + \int_{t_0}^{t_N} F \left( q(t), \dot{q}(t), u(t)\right) \cdot  \delta q(t)\, dt = 0,
    \end{equation}
    where $\delta$ denotes variations that vanish at the endpoints $q(t_0)=q_0$ and $q(t_N)=q_N$, $L:TQ\rightarrow \mathbb{R}$ is the Lagrangian function and $F:TQ\rightarrow TQ^*$ is the Lagrangian force, which is a fiber preserving map $F: (q,\dot{q})\mapsto (q,F(q,\dot{q}))$, \cite[p. 421]{Marsden_West_2001}. 
    This principle is equivalent to the \textit{forced Euler-Lagrange equations},
    \begin{equation}\label{eq:ForcedEulerLagrange}
        0 = \mathcal{E}(L, F)(q, t) := \frac{\partial L}{\partial q}\left( q(t), \dot{q}(t) \right) - \frac{d}{dt} \left(\frac{\partial L}{\partial \dot{q}} \left( q(t), \dot{q}(t) \right) \right) + F\left( q(t), \dot{q}(t)\right).
    \end{equation}
    In absence of non-conservative forces, \eqref{eq:ContinuousDAalembertsPrinciple} expresses the extremization of the action functional, realizing Hamilton's principle, and \eqref{eq:ForcedEulerLagrange} are the Euler-Lagrange equations. 
    In previous work \cite{xiao_generalized_2024} Xiao et al. formulated their learning problem starting from equations \eqref{eq:ForcedEulerLagrange}, and proposed the Generalised Lagrangian Neural Networks (GLNNs), see also \cite{CranmerLagrangian2020}. In their framework, $L$ and $F$ are functions approximated by neural networks using position and velocity data $(q,\dot{q})$ observed at discrete times; see \cite{xiao_generalized_2024}, and~\ref{apx:GLNN} for details on this method and our implementation.
    
    Since we here assume that only position data are observed,
    following \cite{Marsden_West_2001, de2018variational}, we propose instead to discretize the Lagrange d'Alembert principle, and derive discrete equations directly from the discrete principle. 
    \paragraph{Discrete d'Alembert Principle and Forced Euler-Lagrange Equations}  
    A discretization of \eqref{eq:ContinuousDAalembertsPrinciple} leads to the Discrete Lagrange-d'Alembert principle \cite{oberblobaum2011dmoc, kane2000variational}.
    Consider a partition of $[t_0,t_N]$ in $N$ subintervals of equal size, $h$.  
    Consider a choice of numerical approximations of $F$ and $L$ on each subinterval $[t_n,t_{n+1}]$, denoted by
    \begin{equation}\label{eq:discrete to continous relation}
    \begin{aligned}
        L_\Delta(q_n, q_{n+1}) &\approx 
        \int_{t_n}^{t_{n}+h} L\big(q(t), \dot{q}(t)\big) \, dt, \\
        F_\Delta^+(q_n, q_{n+1}, h) &\approx \int_{t_n}^{t_{n}+h} F\big(q(t), \dot{q}(t)\big) \cdot \frac{\partial q(t)}{\partial q_{n+1}} \, dt, \\ 
        F_\Delta^-(q_n, q_{n+1}, h) &\approx \int_{t_n}^{t_{n}+h} F\big(q(t), \dot{q}(t)\big) \cdot \frac{\partial q(t)}{\partial q_{n}} \, dt,
    \end{aligned}
    \end{equation}
    where $q(t_n)=q_n$ and $q(t_n + h) = q_{n+1}$.
     Here, $L_\Delta: Q \times Q \rightarrow \mathbb{R}$ and $F_\Delta^\pm: Q \times Q \rightarrow T^*Q$  represent the discrete Lagrangian and the discrete force, respectively.
    The discrete Lagrange-d'Alembert principle  (dLDA) seeks a discrete trajectory, $\{q_n\}_{n=1}^N$, that satisfies the following
    condition:
    \begin{equation}\label{eq:DiscreteLagrangeDalemberts}
        \delta \sum_{n=0}^{N-1} L_\Delta(q_n, q_{n+1}) + \sum_{n=0}^{N-1} \left[ F_\Delta^-(q_n, q_{n+1}) \cdot \delta q_n + F_\Delta^+(q_n, q_{n+1}) \cdot \delta q_{n+1} \right] = 0,
    \end{equation} 
    for all variations $\{\delta q_n\}_{n=0}^N$ vanishing at the endpoints $\delta q_0=\delta q_N=0$.
    Equivalently, \eqref{eq:DiscreteLagrangeDalemberts} can be expressed in terms of the discrete forced Euler-Lagrange equations:
    \begin{equation}\label{eq:DEL_force}
    \begin{aligned}
        0 = \mathcal{E}_{\Delta}(L_{\Delta}, F_{\Delta})(q_{n-1}, q_{n}, q_{n+1}) := \nabla_2 L_\Delta(q_{n-1}, q_n) + \nabla_1 L_\Delta(q_n, q_{n+1}) &\\
        + F_\Delta^+(q_{n-1}, q_n) + F_\Delta^-(q_n, q_{n+1})& \\
        \text{ for } n=1, \dots, N-1,&
    \end{aligned}
    \end{equation}
 where $\nabla_1$ and $\nabla_2$ denote differentiation with respect to the first and second variable.

%% METHODOLOGY
\section{Discrete Forced Lagrangian Neural Networks}\label{sc:Methodology}
    The proposed model learns the dynamics of an observed system through neural network approximations of the Lagrangian and external forces,
    $$L_\theta\approx L\quad L_\theta:TQ\rightarrow \mathbb{R},\qquad F_\theta\approx F,\quad F_\theta:TQ\rightarrow TQ^*,$$
    where $L_\theta$ and $F_\theta$ are parameterized with learnable parameters $\theta$. 
    
    The loss function minimized during training combines a physics term and a regularization term:
    \begin{equation}\label{eq:overall_loss_definition}
    \begin{aligned}
        \mathcal{L} =\frac{\omega_\text{physics}}{N_\mathcal{T}(N+1)}\sum_{\mathcal{T}}\sum_{n=1}^{N-1} \mathcal{L}_{\text{physics}}(L_\theta, F_\theta)(q_{n-1}, q_{n}, q_{n+1}) &\\
        + \frac{\omega_\text{reg}}{R} \sum_{r=1}^{R} \mathcal{L}_{\text{reg}}(L_\theta)(q_{r}, q_{r+1}).&
    \end{aligned}
    \end{equation}

    Here, $\mathcal{L}_{\text{physics}}$, defined in~\eqref{eq:DataLoss}, is deduced from the discrete forced Euler-Lagrange equations ~\eqref{eq:DEL_force} where the neural network approximations $L_\theta$ and $F_\theta$ are used to approximate the discrete Lagrangian $L_\Delta$ and discrete force $F_\Delta$, while $\mathcal{L}_{\text{reg}}$ promotes the regularity of $L_\theta$~\cite[p. 379]{Marsden_West_2001}, each weighted with the hyperparameters $\omega_\text{physics}$ and $\omega_\text{reg}$. 
    The dataset $\mathcal{T}$ contains $N_\mathcal{T}$ trajectories of $N$ steps each, and $R$ denotes the number of point pairs $(q_n,q_{n+1})$ used for regularization.
    The regularization may be applied over all such pairs in the training dataset. However, in the experiments, we demonstrate that it is sufficient to regularize only over a moderate number of pairs of data points.
    
    As~\eqref{eq:overall_loss_definition} only evaluates $\mathcal{L}_{\text{physics}}$ on local triplets $(q_{n-1}, q_n, q_{n+1})$, training is performed independently on such segments, without leveraging longer-range temporal dependencies present in the full trajectories. 
    
    \subsection{Learning Physics from the Observed Data ($\mathcal{L}_{\text{physics}}$)} \label{Learning-Physics}

    \begin{figure}[htb!]
        \centering
        \includegraphics[width=0.87\linewidth]{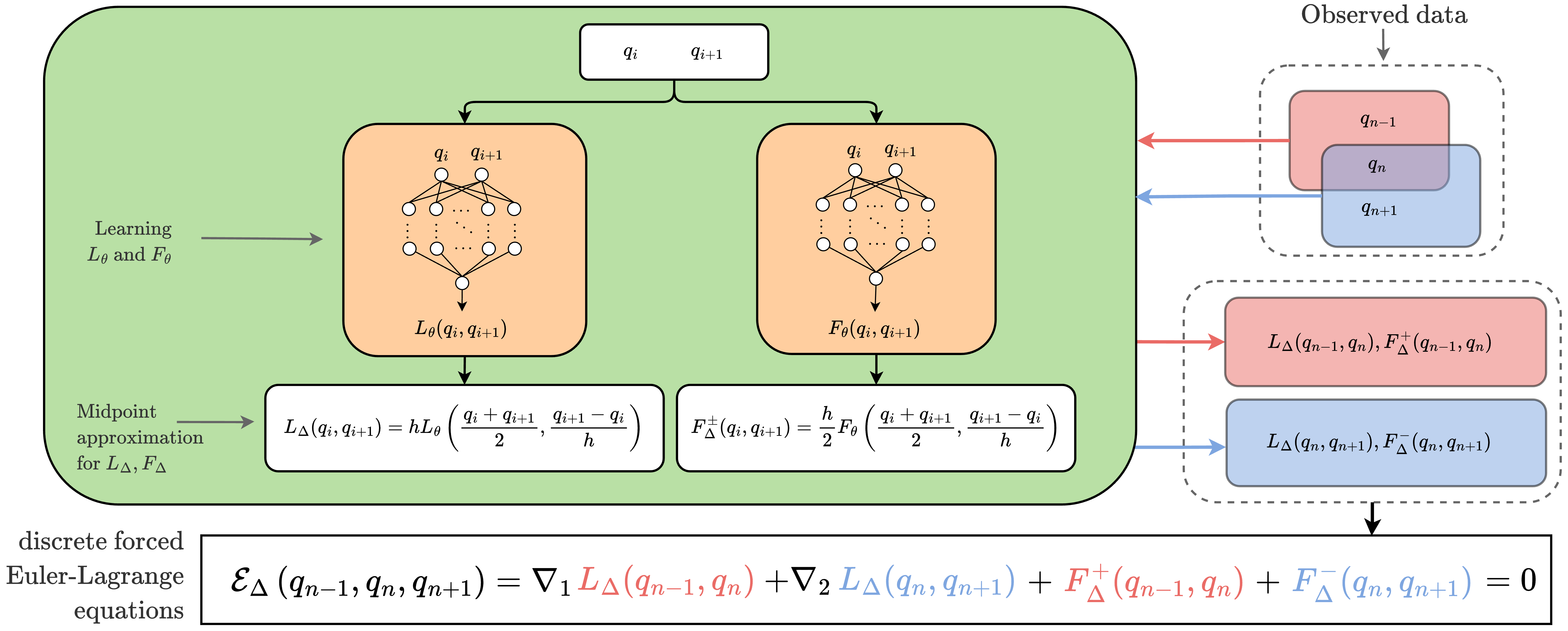}
        \caption{A schematic illustration of the data flow through the proposed model. We use segments of observed $(q_{n-1}, q_{n}, q_{n+1})$ to evaluate the discrete forced Euler-Lagrange equations.}
        \label{fig:dLDA dataFlow}
    \end{figure}

    Usually, the numerical discretization of a forced Lagrangian system would suggest for the exact $L$ and $F$ to be known, while the discrete trajectories $\{q_n\}_{n=0}^N$ are the unknowns of the problem. 
    In the inverse setting considered here, the roles are reversed: the trajectories $\{q_n\}_{n=0}^N$ are given, and the goal is to recover $(L_\theta, F_\theta)$ as approximations to the underlying Lagrangian and forces. 
    
    A discretization scheme can be applied to obtain the discrete Lagrangian and discrete force ($L_\Delta, F_\Delta$) as a function of  $(L_\theta, F_\theta)$, \eqref{eq:discrete to continous relation}. A concrete example using a mid-point approximation which makes the discrete Lagrangian a central difference about the midpoint,
    \begin{align}
        L_\Delta(q_n, q_{n+1}) &:=  h \, L_{\theta}\left(\frac{q_{n}+q_{n+1}}{2}, \frac{q_{n+1}-q_{n}}{h}\right) 
        = h \, L_{\theta}\left(\bar{q}_{n+\frac{1}{2}}, \bar{\dot{q}}_{n+\frac{1}{2}}\right)\label{eq:discretization-L} \\ 
        F_\Delta^{\pm}(q_n, q_{n+1}, h) &:=
        \frac{h}{2} \, F_{\theta}\left(\frac{q_{n}+q_{n+1}}{2}, \frac{q_{n+1}-q_{n}}{h}\right) 
        = \frac{h}{2} \, F_{\theta}\left(\bar{q}_{n+\frac{1}{2}}, \bar{\dot{q}}_{n+\frac{1}{2}}\right)\label{eq:discretization-F}
    \end{align}
    where we have introduced the notation
    $\bar{q}_{n+\frac{1}{2}} := \frac{1}{2}(q_{n}+q_{n + 1})$, $\bar{\dot{q}}_{n+\frac{1}{2}} := \frac{1}{h}(q_{n + 1} - q_{n})$
    for brevity\footnote{We observe that discretizing the Lagrangian with the midpoint discretization \eqref{eq:discretization-L} gives a method equivalent to the implicit midpoint applied to the corresponding Hamiltonian equations, as shown in \cite[p. 208, example 6.3]{HairerLubichWanner2006}. }. We will use this discretization in our numerical experiments.

    The key distinctive feature of our approach is the preservation of the variational structure on $Q\times Q$, without introducing  approximations that break symplecticity, as is the case when finite-differences are applied to a scheme that is symplectic on $TQ$. As demonstrated in the numerical experiments in section \ref{sc:Experiments}, this is advantageous and less prone to instabilities.
    
    By applying this procedure, the finite-difference approximation of the velocity is naturally incorporated into the model’s architecture without violating its variational structure. As a result, when $F$ is set to zero, we obtain a symplectic method \cite{Marsden_West_2001}. Discretizing the second order Euler-Lagrange equations, e.g. introducing arbitrary finite differences to approximate the velocities or applying other arbitrary discretizations will not in general preserve symplecticity in the $F=0$ setting. As demonstrated in the numerical experiments in section \ref{sc:Experiments}, this is advantageous and less prone to instabilities.

    Assuming that the observed position data satisfy the dLDA principle \eqref{eq:DiscreteLagrangeDalemberts} and, equivalently, the dLDA equations \eqref{eq:DEL_force}, a schematic illustration considering the flow of an observed trajectory of data points through the neural networks $L_\theta$ and $F_\theta$-used to construct the dLDA equations in \eqref{eq:DEL_force}-is provided in Figure~\ref{fig:dLDA dataFlow}. Summing over all observed trajectories, we define the following physics term for the loss function:
    \begin{equation}\begin{aligned}\label{eq:DataLoss}
        \mathcal{L}_{\text{physics}}(L_\theta, F_\theta)\left(q_{n-1}, q_{n}, q_{n+1} \right) = \, 
        \frac{h}{2} \Bigg\| 
        F_\theta \left(\bar{q}_{n-\frac{1}{2}}, \bar{\dot{q}}_{n-\frac{1}{2}}\right) 
        + F_\theta \left(\bar{q}_{n+\frac{1}{2}}, \bar{\dot{q}}_{n+\frac{1}{2}}\right)& \\
        + \nabla_1 L_\theta \left(\bar{q}_{n-\frac{1}{2}}, \bar{\dot{q}}_{n-\frac{1}{2}}\right) + \frac{2}{h} \, \nabla_2 L_\theta \left(\bar{q}_{n-\frac{1}{2}}, \bar{\dot{q}}_{n-\frac{1}{2}}\right)& \\
        + \nabla_1 L_\theta \left(\bar{q}_{n+\frac{1}{2}}, \bar{\dot{q}}_{n+\frac{1}{2}}\right)
        - \frac{2}{h}  \, \nabla_2 L_\theta \left(\bar{q}_{n+\frac{1}{2}}, \bar{\dot{q}}_{n+\frac{1}{2}}\right)&
        \Bigg\|_2  .
    \end{aligned}\end{equation}

These residual equations are later solved for $q_{n+1}$ to produce predictions (Section~\ref{sc:Experiments}). For a fixed set of parameters $\theta$, the resulting prediction map is the discrete Lagrange--d'Alembert flow of $(L_\theta, F_\theta)$, and the learned dynamics describe in general a forced Lagrangian system. However, when $F_\theta = 0$, the prediction map is symplectic. All non-conservative effects in the model therefore enter through $F_\theta$. With the Rayleigh form, which we return to in
Section~\ref{sc:TheForceTerm}, this non-conservative force is dissipative, so it can only dissipate energy, never inject it. We refer to \cite[\S~3.2]{Marsden_West_2001} and \cite{kane2000variational, cortes2001non} for a full account of discrete forced mechanics.

\subsection{The Learned Lagrangian ($L_\theta$)}\label{sc:TheLagrangianTerm}
    The Lagrangian function $L_\theta$ can be learned either as a generic function or with embedded physical structure. In the most general case $L_\theta$ can be a feed-forward neural network. Alternatively, structural priors can be imposed. Assuming the Lagrangian takes a mechanical form, we can let
    \begin{equation}\label{eq:structuredNetworkEquations}
        L_\theta (\bar{q}_k, \bar{\dot{q}}_k) = \bar{\dot{q}}_k^T M_{\theta}(\bar{q}_k) \bar{\dot{q}}_k - U_{\theta}(\cdot), \qquad k=n\pm \frac{1}{2}
    \end{equation}
    with $M_{\theta}(\bar{q}_k) = \epsilon
    \mathbb{I} + {\Lambda_{\theta}^T}(\bar{q}_k)\Lambda_{\theta}(\bar{q}_k)$ ensuring $M_{\theta}$ is symmetric and positive definite. Here, $\Lambda_{\theta}:\mathbb{R}^{d} \rightarrow \mathbb{R}^{d \times d}$ is a lower triangular matrix parameterized by a feed-forward neural network, 
    and $\epsilon > 0$ is a small constant ensuring  positive-definiteness of $M_\theta$.
    $U_{\theta}(\cdot)$ is the potential energy function, also modeled as a feed-forward neural network, and may depend on $\bar{q}_k$ and/or $\bar{\dot{q}}_k$.
    
    \paragraph{Maximizing the regularity of the Lagrangian}
    Learning $L_\theta$ by minimizing \eqref{eq:DataLoss} can lead to trivial solutions, 
    preventing the discovery of meaningful physical relationships (e.g., the network might learn $L_\theta$ as a constant, causing the derivative terms in \eqref{eq:DataLoss} to vanish without properly fitting the data). 
    This is a well-known challenge in data-driven Lagrangian-based models, and there are different strategies proposed to address this problem \cite{offen2024machine, OberBlobaumVariational2023, lishkova2023discrete}. 
    A Lagrangian is regular, or non-degenerate, if and only if its Hessian with respect to the second argument is invertible, \cite[379]{Marsden_West_2001}. We will require the invertibility to hold point-wise: specifically on a selected number of pairs of points $(\bar{q}_{r+\frac{1}{2}}, \bar{\dot{q}}_{r+\frac{1}{2}})$, we require that
    \begin{equation}
        S(\bar{q}_{r+\frac{1}{2}}, \bar{\dot{q}}_{r+\frac{1}{2}}):= \left( \frac{\partial^2 L(\bar{q}_{r+\frac{1}{2}}, \bar{\dot{q}}_{r+\frac{1}{2}})}{\partial \bar{\dot{q}}_{r+\frac{1}{2}}^{\,\, 2}} \right),
    \end{equation}
    is invertible. Here $\bar{q}_{r+\frac{1}{2}}$ and $ \bar{\dot{q}}_{r+\frac{1}{2}}$ are as defined in Section~\ref{Learning-Physics} and depend on the data $(q_r,q_{r+1})$. We then include in the loss function a regularization term $\mathcal{L}_\text{reg}$ as a logarithmic barrier that maximizes the regularity of $S$ by penalizing the absolute value of its determinant \cite{offen2024machine}, 
    \begin{equation}
    \begin{aligned}
        \mathcal{L}_\text{reg}(q_{r}, q_{r+1}) &= 
         \mathrm{ReLU} \left(c - \lvert \det(S(\bar{q}_{r+\frac{1}{2}}, \bar{\dot{q}}_{r+\frac{1}{2}}) \rvert \right),
    \end{aligned}
    \end{equation}
    where $c$ is an arbitrary constant. In all experiments included, we have used $c=10$.
    
\subsection{The Learned External Force ($F_\theta$)}\label{sc:TheForceTerm}

    To learn the dynamics of systems with completely unknown forces, we parameterize $F_\theta = F_{\theta}^{\text{Free}}$ by a neural network. When there is strong prior belief about a particular structure of the external force, this bias can be incorporated into the learned quantity. For instance, if we suspect that the external force has a dissipative character, we can design the learned function $F_\theta$ so that it satisfies the properties of a Rayleigh dissipation function \cite{minguzzi2015rayleigh},
    \begin{equation}\label{eq:structureExternalForceNetwork}
        \begin{aligned}
            F_\theta(\bar{q}_k, \bar{\dot{q}}_k) = - K_\theta(\bar{q}_k) \, \bar{\dot{q}}_k, \qquad k=n\pm \frac{1}{2}
        \end{aligned}
    \end{equation}
    with $K_\theta (\bar{q}_k) = A_{\theta}^T(\bar{q}_k)A_{\theta}(\bar{q}_k)$ parameterized by a Cholesky factorization with the lower triangular matrix $A_{\theta}$ to ensure positive-semi-definiteness. For linear dissipation, $F_\theta$ is independent of $\bar{q}_k$.

\paragraph{Identifiability of the decomposition}
The forced Euler--Lagrange equations~\eqref{eq:ForcedEulerLagrange} constrain only the sum $\mathcal{E}(L) + F = 0$, so the split into $L_\theta$ and $F_\theta$ is not unique in general. With a free force, conservative terms can be moved between the two terms, and $F_\theta$ may absorb errors from the Lagrangian, the discretization or the latent representation. The Rayleigh form~\eqref{eq:structureExternalForceNetwork} is expected to reduce this ambiguity. A symmetric force that is linear in $\dot q$ cannot come from any Lagrangian, and it can represent neither a potential nor a magnetic-type force. Neither component can therefore take over the role of the other. 
In Section~\ref{sc:Experiments}, we investigate this separation empirically. After training on data from the forced dynamics, we set the learned force to zero and by Eq.~\eqref{eq:DEL_force} obtain rollouts for comparison with the exact solution of the corresponding unforced system (Figures~\ref{fig:DP_undamped_results}, \ref{fig:CP_undamped_results} and~\ref{fig:pixel_pendulum_undamped}).

%% EXPERIMENTS
\section{Experiments}\label{sc:Experiments}
\begin{figure}[!htb]
    % \hspace{2cm}
    \minipage{0.24\textwidth}
        \centering
        \boxed{\hspace{0.6cm}
        \includegraphics[trim={18cm 7cm 32cm 17cm}, clip, height=2.3cm]{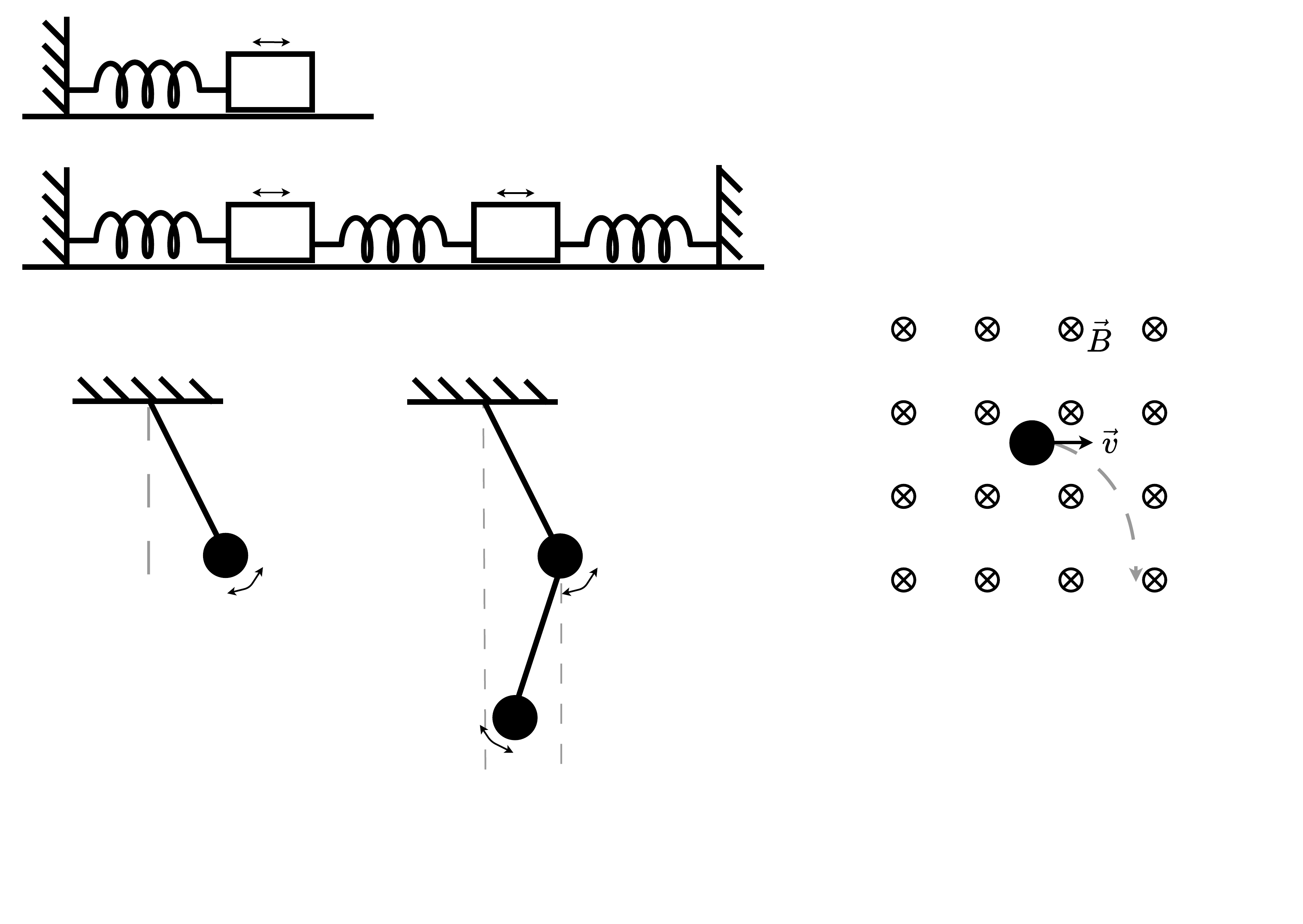}
        \hspace{0.7cm}}
        \subcaption*{Task 1}
    \endminipage\hfill
    \minipage{0.24\textwidth}
        \centering
        \boxed{\includegraphics[trim={40cm 15cm 5cm 14cm}, clip, height=2.3cm]{LNN_datasets.drawio.pdf}}
        \subcaption*{Task 2}
    \endminipage
    % \hspace{2cm}
    \hfill
    \minipage{0.24\textwidth}%
        \centering
        \boxed{
        \includegraphics[trim={4cm 1.5cm 2cm 1.5cm}, clip, height=2.3cm]{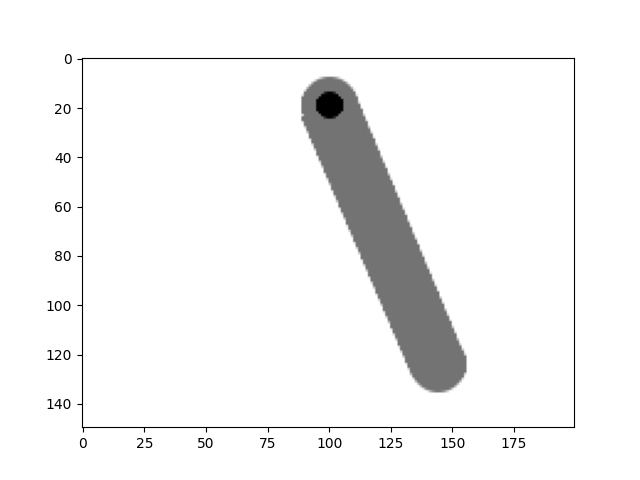}}
        \subcaption*{Task 3}
    \endminipage
    \hfill
    \minipage{0.24\textwidth}%
        \centering
        \boxed{\hspace{0.6cm}
        \includegraphics[trim={0 0cm 18cm 0cm}, clip, height=2.3cm]{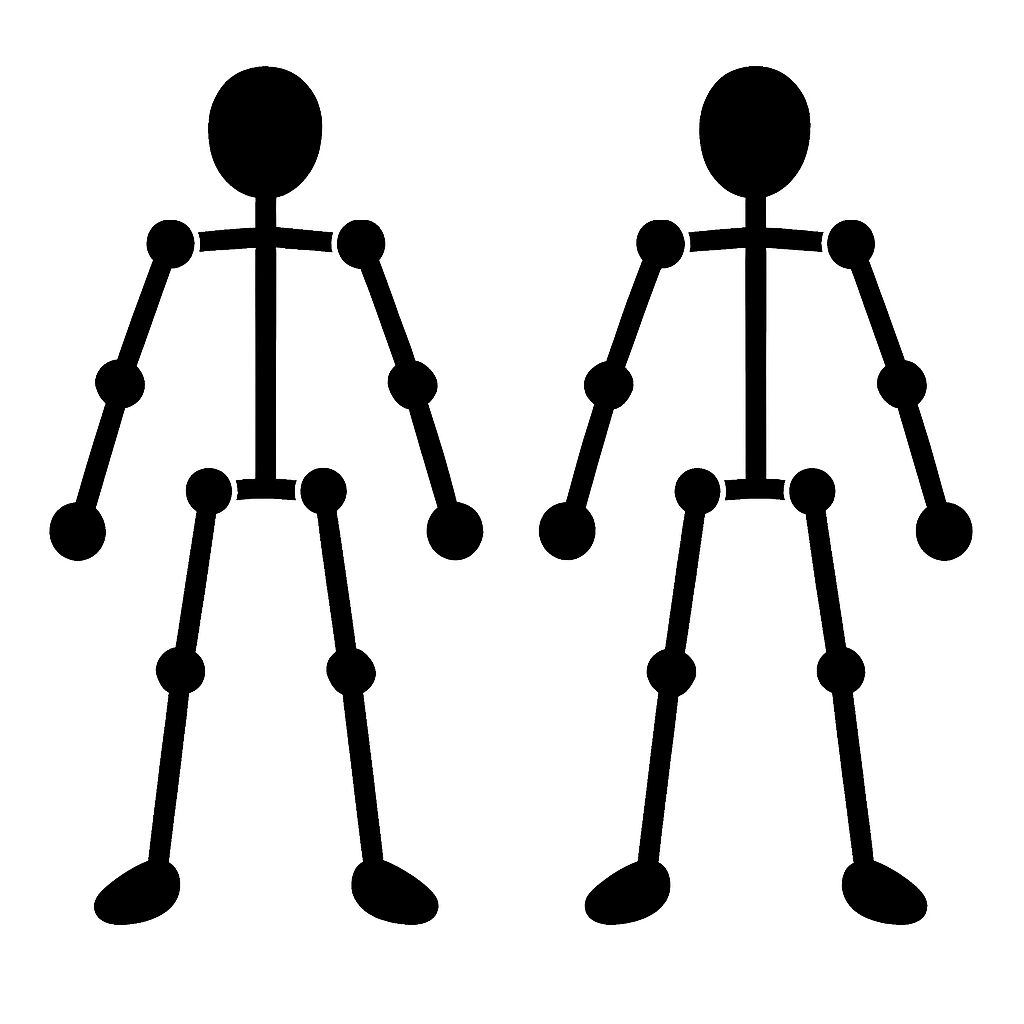}
        \hspace{0.6cm}}
        \subcaption*{Task 4}
    \endminipage
    
    \caption{The four datasets to be considered in the experiments. From the left-panel: a damped double pendulum, a charged particle moving through a magnetic field, a damped pendulum represented with pixel frames, and human motion capture.
    }
    \label{fig:illustraion-datasets}
\end{figure}

To demonstrate that the proposed model can learn and separate the conservative and non-conservative components of the observed system, we perform the following experiments: we train on data obtained from damped, non-conservative systems, but after training, we remove the learned external force term $F_\theta$, and evaluate the rollout predictions for the corresponding conservative system.

The model is evaluated in four learning tasks illustrated in Figure~\ref{fig:illustraion-datasets}. Tasks 1 to 3 involve a damped double pendulum, a dissipative charged particle moving within a magnetic field, and pixel frames from a damped pendulum. Training data for these tasks are generated by solving the respective ODEs with \texttt{scipy.integrate.odeint}.
In task 4, we consider a real-world dataset focused on human motion tracking.

\paragraph{Prediction workflow}
    Unlike training, the prediction workflow involves forward time-stepping. The model is trained by minimizing the residuals of the discrete Lagrange-d’Alembert equations~\eqref{eq:DEL_force} over observed segments to learn $L_\theta$ and $F_\theta$.  
    During the prediction phase, given two initial positions $\{q_0, q_1\}$, the model solves $\mathcal{E}_\Delta(q_0, q_1, \hat{q}_2) = 0$ for $\hat{q}_2$ using the learned $L_\theta$ and $F_\theta$. This process is repeated recursively by feeding predictions as new inputs to generate longer rollouts.
    
    To compare the different approaches, we consider the \textit{relative error} at the $k$-th step for a predicted rollout,
    \begin{equation}
        \text{Relative Error}_k :=
    \frac{\| q_k - \hat{q}_k \|_2}{\| q_k \|_2}.
    \end{equation}

\paragraph{Baseline models}
    In our experiments,  we assume that only position data is available. Thus, all baseline models are adapted to this setting and implemented accordingly.

    To demonstrate the importance of incorporating external forces into the discrete (force) Euler–Lagrange equations, we also provide comparisons with a closely related variant of the Lagrangian Shallow Integrator (LSI) \cite{OberBlobaumVariational2023}. This method uses the discrete Euler-Lagrange equations, and follows a similar idea as the suggested DFLNN in the absence of external forces. We modify the original approach by replacing the Gaussian processes with a neural network architecture similar to the one used in our proposed method. This choice reduces implementation-related differences and allows us to focus primarily on the role of explicitly modeling external force terms. We will refer to this baseline implementation as LSI-NN.
    
    Another natural baseline for comparison is the Generalized Lagrangian Neural Network (GLNN)~\cite{xiao_generalized_2024}, which learns both the Lagrangian and the forcing terms by using the forced Euler–Lagrange equations~\eqref{eq:ForcedEulerLagrange}. In our adaptation, GLNN is trained using only position measurements, and the required values of $\dot{q}$ and $\ddot{q}$ at discrete time points are obtained via finite-difference approximations; see~\ref{apx:GLNN} for further details. We will consider integration schemes of Runge-Kutta 4 (RK4-GLNN).

    Lastly, because the forced Euler--Lagrange equations can be expressed as second-order ordinary differential equations (ODEs), we also include a comparison with Neural ODE~\cite{ChenNeural2018} (NODE). We seek a second order ODE and thereby formulate the problem to learn an equivalent first-order system. The initial velocity is obtained using a finite-difference approximation. We train the model using single-step time updates with Runge–Kutta 4 integration scheme.

\paragraph{Learning on learned latent variables from an autoencoder}\label{sc:Autoencoder}

    \begin{figure}[htbp]
        \centering
        \includegraphics[width=\linewidth]{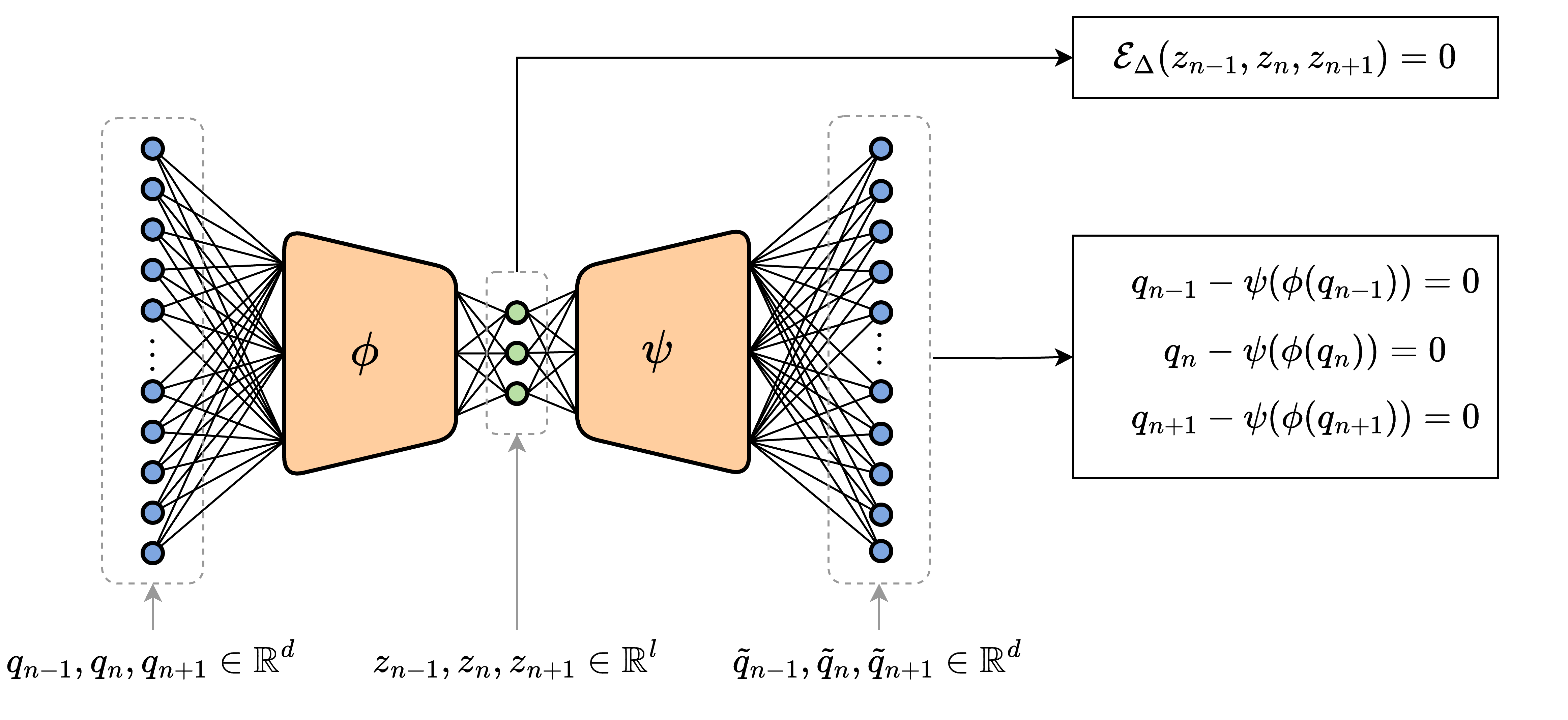}
        \caption{An autoencoder is applied to reduce the high-dimensional input dimension to a lower-dimensional latent space. The proposed model is applied to the dynamics in the latent space.}
        \label{fig:ae-flow}
    \end{figure}
    
    We also train a model on high-dimensional, unstructured pixel data $q \in \mathbb{R}^d$ by training an autoencoder to map to  latent space $z \in \mathbb{R}^l$ where dynamics are modeled, see Figure~\ref{fig:ae-flow}.

    The encoder $\phi_\theta: \mathbb{R}^d \to \mathbb{R}^l$ and decoder $\psi_\theta: \mathbb{R}^l \to \mathbb{R}^d$ are implemented as convolutional neural networks for image data or feedforward neural networks otherwise.
    The autoencoder is trained using a reconstruction loss that per time step reads as follows,
    \begin{equation}
        \mathcal{L}_{\text{AE}}(\phi_\theta, \psi_\theta)(q_{n}) =
        \frac{d}{l}\| q_n - \psi_\theta(\phi_\theta(q_n)) \|_2^2.
    \end{equation}
    In order to recover meaningful latent coordinates \cite{champion2019data, GreydanusHamiltonian2019}, we train the autoencoder together with the proposed model adding a term weighted by the factor $\omega_\text{AE}$ to the loss function:
    \begin{equation}
    \begin{aligned}
        \mathcal{L} &= \frac{\omega_\text{physics}}{N_\mathcal{T}(N-1)}\sum_{\mathcal{T}}  \sum_{n=1}^{N-1} \mathcal{L}_{\text{physics}}(L_\theta, F_\theta)\Big(\phi_\theta(q_{n-1}), \phi_\theta(q_{n}), \phi_\theta(q_{n+1})\Big) \\
        &\quad + \frac{\omega_\text{reg}}{R}\sum_{r=1}^{R} \mathcal{L}_{\text{reg}}(L_\theta)\Big(\phi_\theta(q_{r}), \phi_\theta(q_{r+1})\Big) \\
        &\quad + \frac{\omega_\text{AE}}{N_\mathcal{T}(N+1)}\sum_{\mathcal{T}}  \sum_{n=0}^{N} \mathcal{L}_{\text{AE}}(\phi_\theta, \psi_\theta)(q_{n}).
    \end{aligned}
    \end{equation}
    During prediction, dynamics are rolled out in latent space; the autoencoder is used only to encode initial states and decode the final prediction trajectory.
    
\paragraph{Implementation}
    The experiments were implemented in Python using the PyTorch framework. All models are trained with Adam optimizer with a $\text{lr}=1e-3$. Unless otherwise specified, each model was trained for $10,000$ epochs. See \ref{sc:Hyperparameters} for selected hyperparameters and additional computational notes.
    % Code is available at \href{github}{github}.

\subsection{Task 1: Damped Double Pendulum}\label{sc:Task1}

We simulate a double pendulum system with two identical masses and generate synthetic datasets of angular trajectories 
$\{\theta_1^t, \theta_2^t\}_{t=1}^T$. 
Initial conditions are sampled uniformly from 
$\theta_i^0 \in [-60^\circ, 60^\circ]$ and $\dot{\theta}_i^0 \in [-1, 1]$. 
System parameters are fixed to masses $m_1 = m_2 = 1$, rod lengths $l_1 = l_2 = 1$, and friction coefficients $b_1 = b_2 = 0.5$. 
For all subsequent experiments, unless specified otherwise, we use trajectories of length $T = 40$ time steps with a fixed temporal resolution of $h = 0.1$. The datasets comprises a total of $32$ trajectories, of which $16$ are used for training and $16$ for validation, with a batch size of $256$.
In all Lagrangian-based models, the learned Lagrangian is assumed to have the form \eqref{eq:structuredNetworkEquations} with a potential energy term only depending on the position $U=U(q)$. The external force is assumed to be linear. 

\begin{figure}[htb!]
    \centering
    \begin{minipage}{\textwidth}
        \centering
        \scriptsize 
        Double pendulum: damped dynamics \\
        (training regime)
    \end{minipage}
    
    \begin{minipage}[c]{0.65\textwidth}
        \centering
        \includegraphics[trim={0cm 0cm 0cm 0cm}, clip, width=\textwidth]{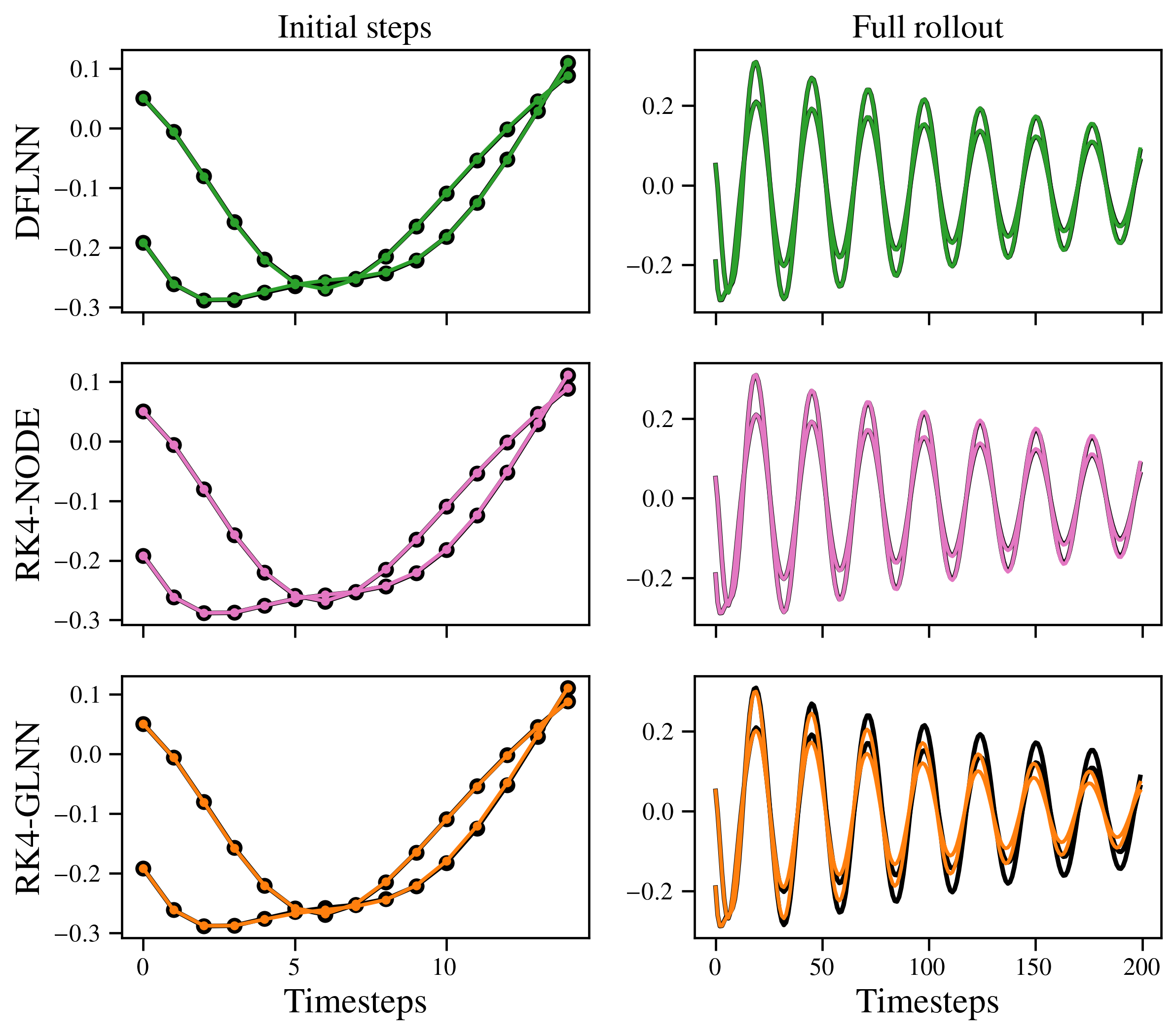}
        \subcaption{Prediction rollout.}
    \end{minipage}
    \hfill
    % Right column with two stacked images
    \begin{minipage}[c]{0.32\textwidth}
        %\vspace{-9.75cm}
        % Top right image
        \centering
        \includegraphics[width=\textwidth]{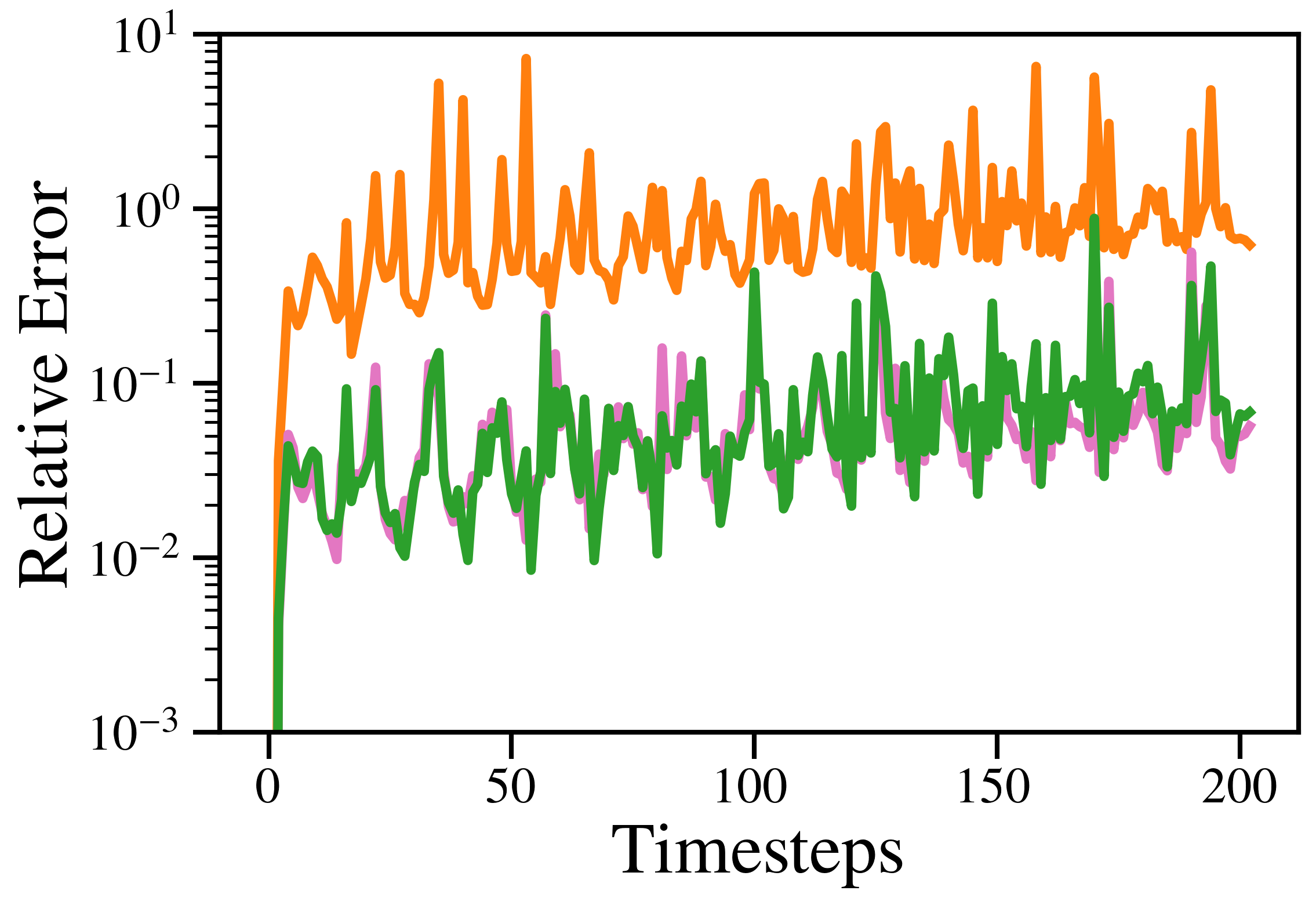} \\
        % \vspace{0.5cm}
        % Bottom right image
        \includegraphics[width=\textwidth]{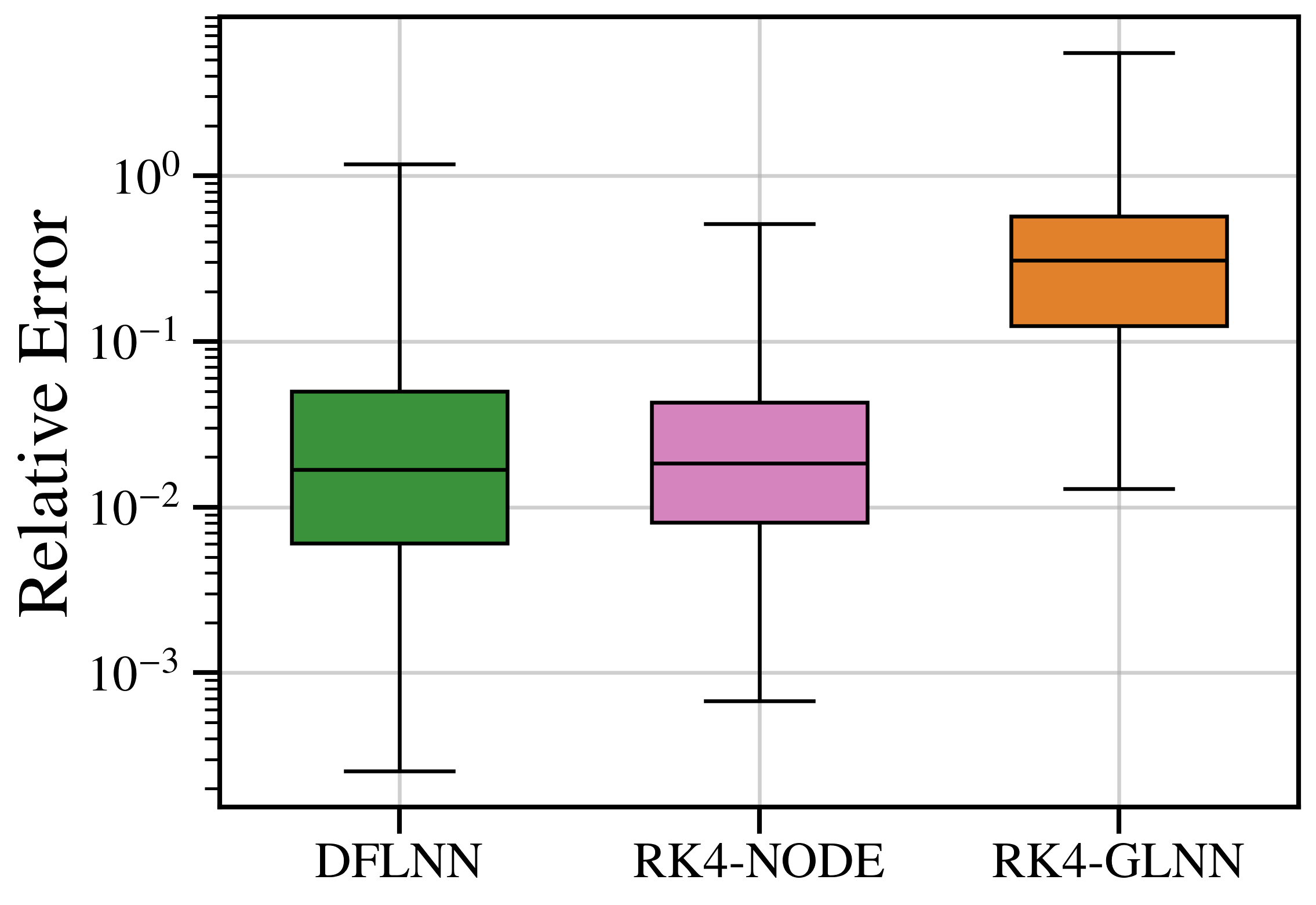}
        \subcaption{Relative error.}
    \end{minipage}
    \caption{Double pendulum with damped dynamics similar to those used during training. Relative error are computed over a set of $20$ test trajectories. Green trajectories represent DFLNN (proposed), orange denote the RK4-GLNN model, and pink indicate RK4-NODE, while the ground truth is plotted in black.}
    \label{fig:DP_damped_results}
\end{figure}

\begin{figure}[htb!]
    \centering
    \begin{minipage}{\textwidth}
        \centering
        \scriptsize 
        Double pendulum: conservative dynamics \\
        (testing regime)
    \end{minipage}
    
    \begin{minipage}[c]{0.65\textwidth}
        \centering
        \includegraphics[trim={0cm 0cm 0cm 0cm}, clip, width=\textwidth]{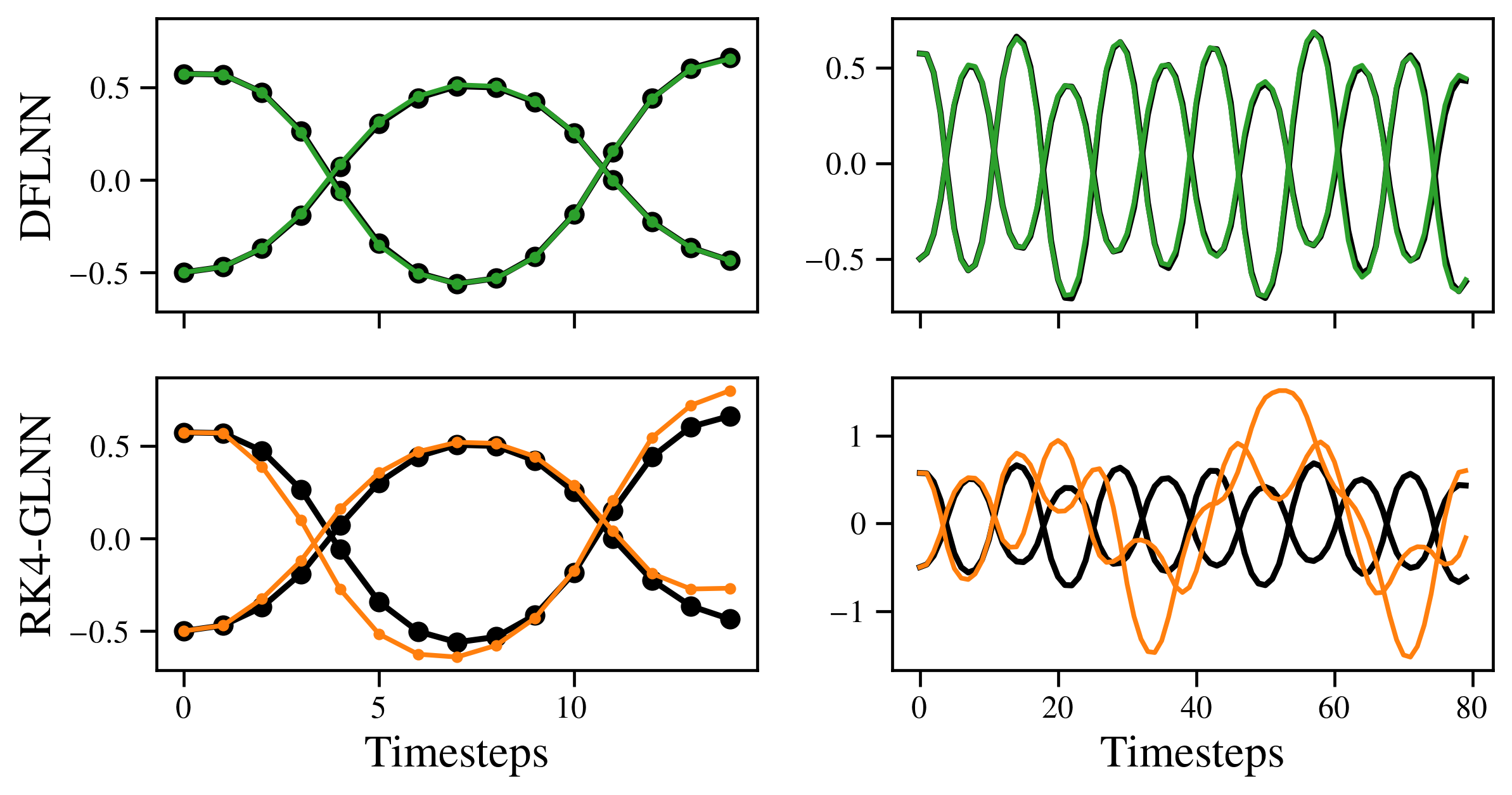}
        \subcaption{Prediction rollout.}
    \end{minipage}
    \hfill
    % Right column with two stacked images
    \begin{minipage}[c]{0.32\textwidth}
        % \vspace{-8.0cm}
        % Top right image
        \centering
        \includegraphics[width=\textwidth]{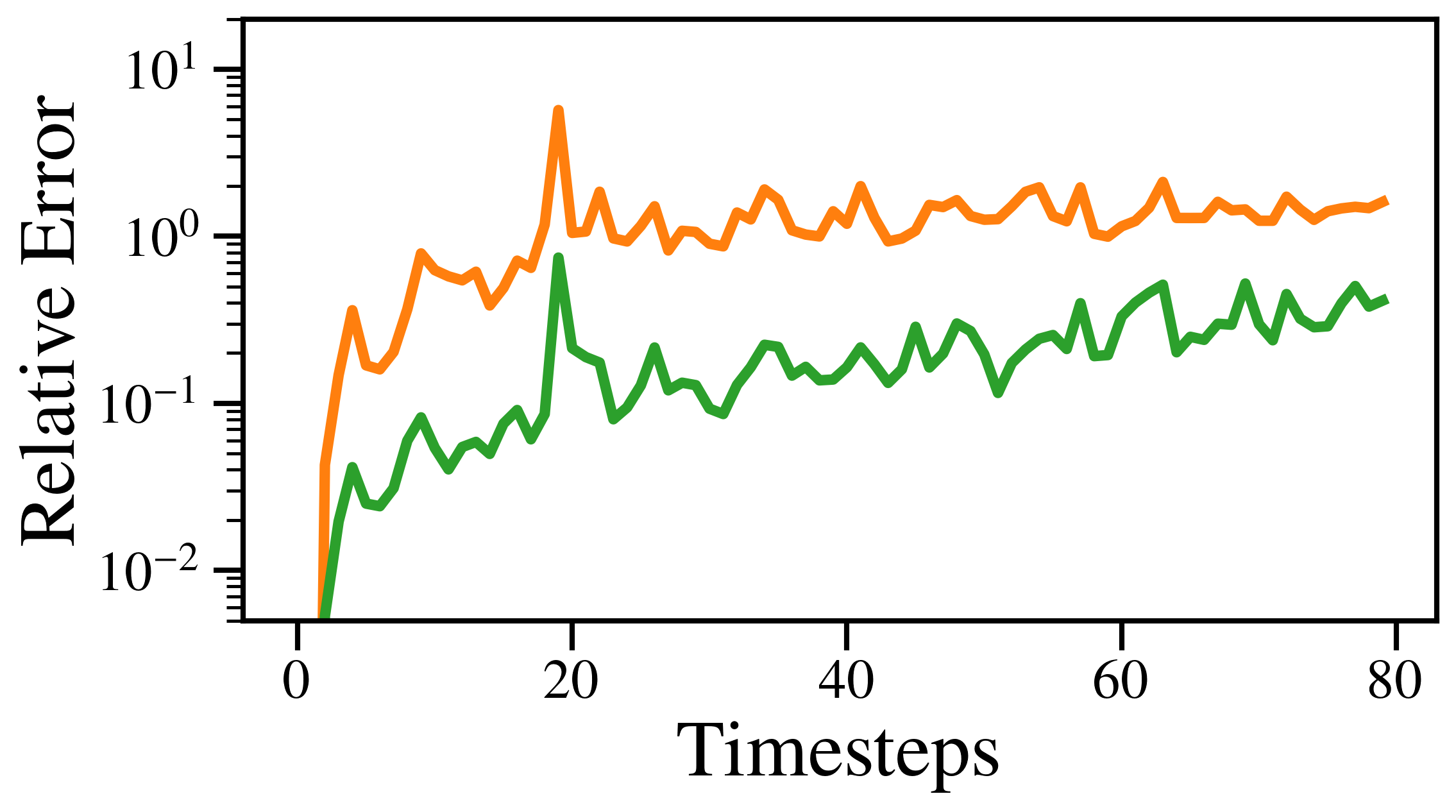}\\
        % \vspace{0.5cm}
        % Bottom right image
        \includegraphics[width=\textwidth]{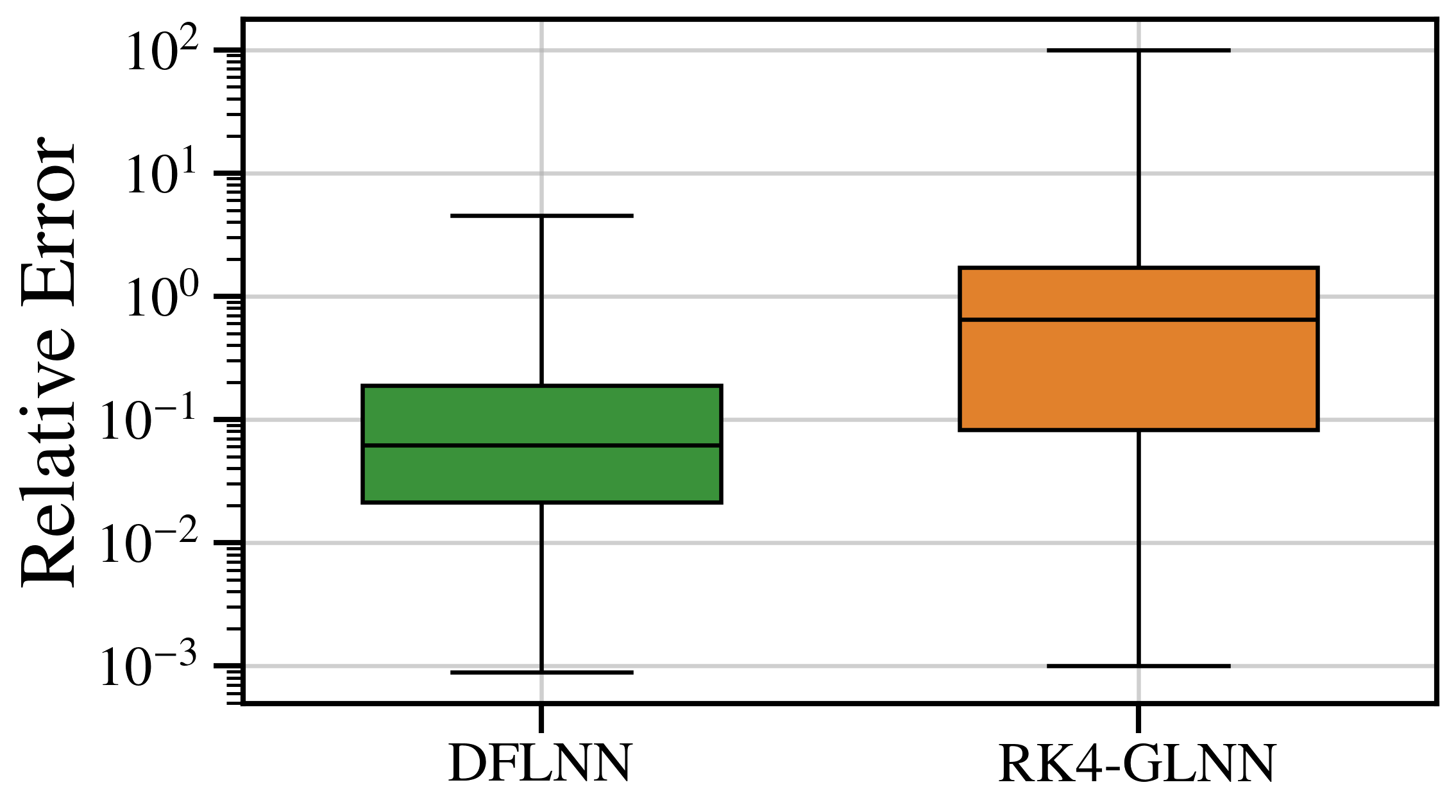}
        \subcaption{Relative error.}
    \end{minipage}
    
    \caption{Double pendulum with conservative, undamped dynamics. Testing regime where we have set the learned force to zero, as explained in the end of Section~\ref{sc:TheForceTerm}. Relative error are computed over a set of $20$ test trajectories. Green trajectories represent DFLNN (proposed) and orange denote the RK4-GLNN model, while the ground truth is plotted in black.
    }
    \label{fig:DP_undamped_results}
\end{figure}
The performance of the proposed model when performing inference on the damped double pendulum is presented in Figure~\ref{fig:DP_damped_results}. The system parameters match those of the training data, but the predictions are rolled out from new, unseen initial conditions and over $200$ time steps, five times longer than the $40$-step trajectories seen during training. These results show that DFLNN achieves a performance comparable to that of RK4-NODE, whereas RK4-GLNN accumulates error at each time step, leading to dynamics with an excessively strong damping effect.

In Figure \ref{fig:DP_undamped_results}, we set the learned force to zero---as explained in the end of Section~\ref{sc:TheForceTerm}---to evaluate the learned Lagrangian. This places us in a previously unseen testing regime where we compare to corresponding unforced dynamics. The DFLNN follows the conservative dynamics closely over the full horizon, showing that the learned Lagrangian carries the conservative part of the dynamics on its own. The RK4-GLNN rollouts, in contrast, deviate from the ground truth also in this setting, so its errors in the damped regime cannot be attributed to the force term alone and the learned Lagrangian itself is inaccurate.

\begin{figure}[htb!]
    \centering
    \begin{minipage}{\textwidth}
        \centering
        \scriptsize 
        Double pendulum: Varying time step size $h$
    \end{minipage}
    
    \begin{minipage}[c]{0.65\textwidth}
        \centering
        \includegraphics[trim={0cm 0cm 0cm 0cm}, clip, width=\textwidth]{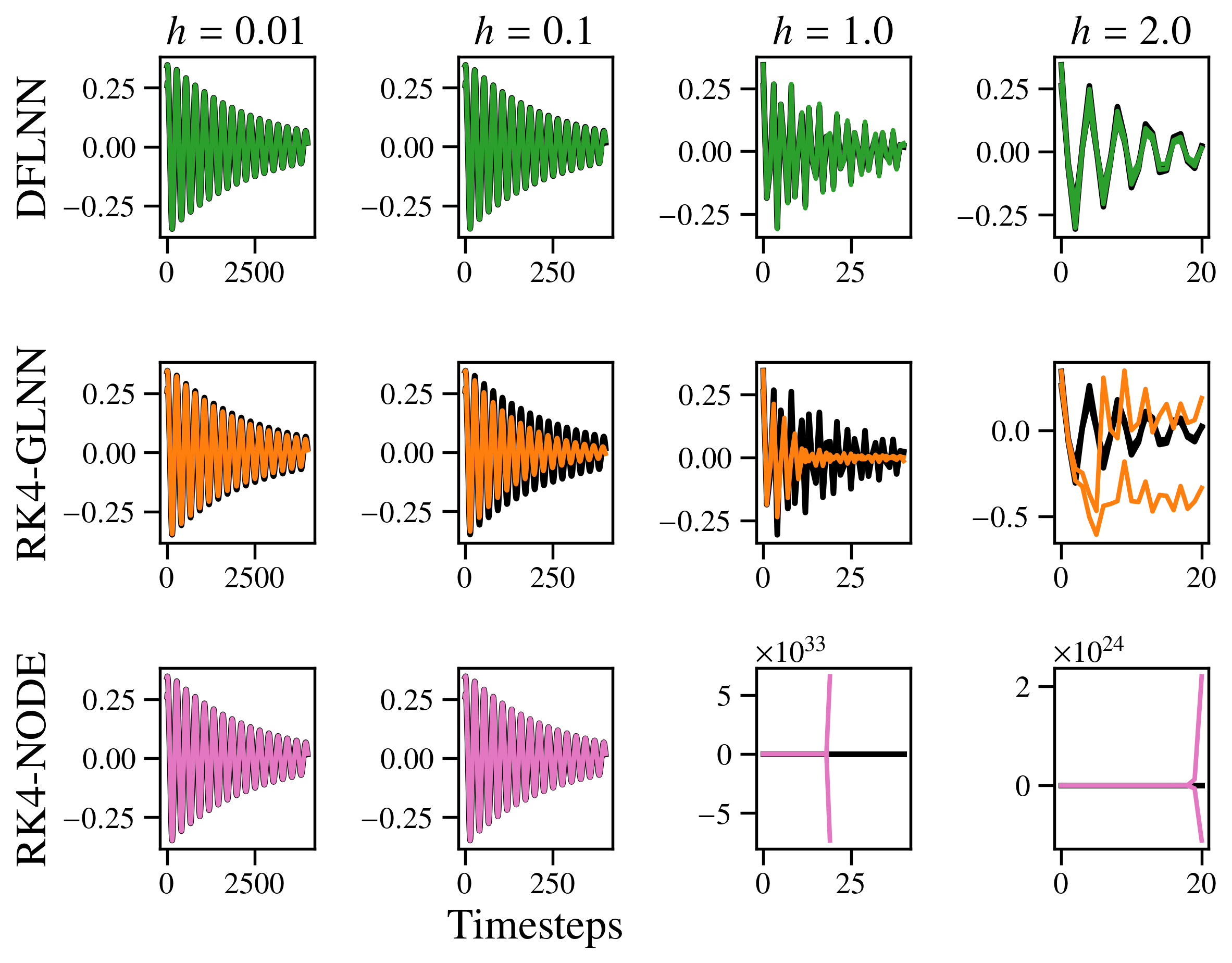}
        \subcaption{Prediction rollout.}
    \end{minipage}
    \hfill
    % Right column with two stacked images
    \begin{minipage}[c]{0.32\textwidth}
        %\vspace{-9.75cm}
        % Top right image
        \centering
        \includegraphics[width=\textwidth]{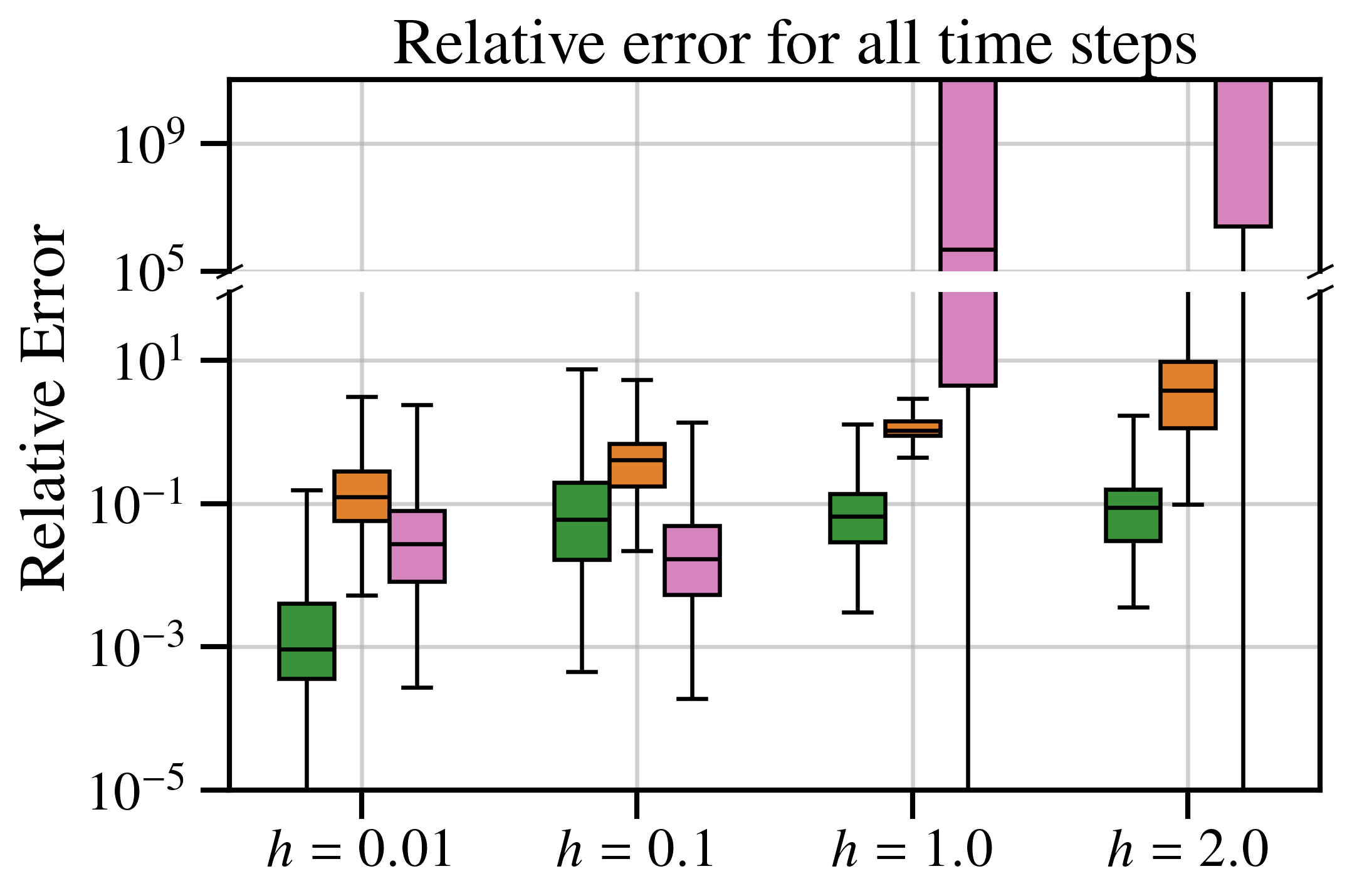} \\
        \vspace{0.5cm}
        % Bottom right image
        \includegraphics[width=\textwidth]{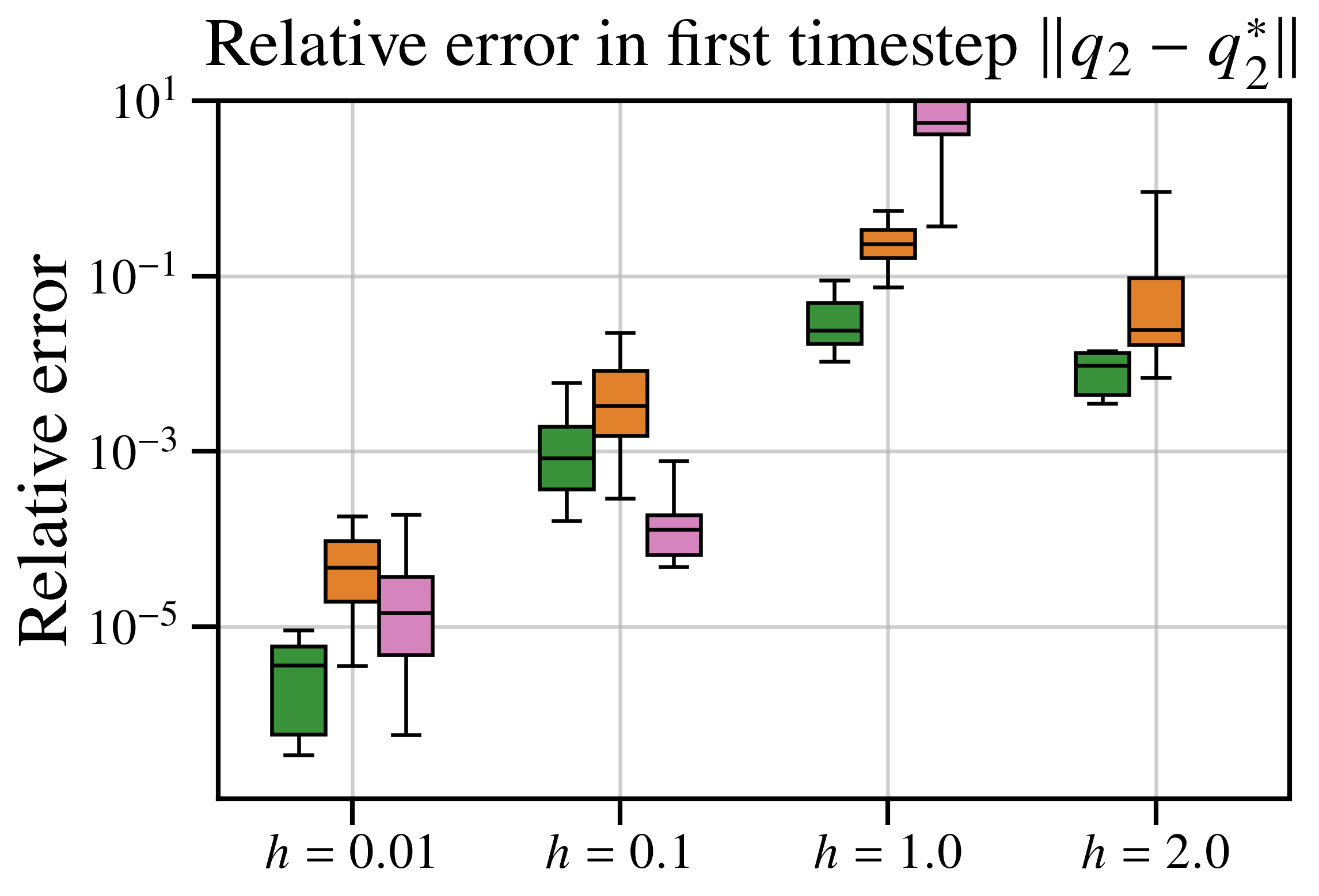}
        \subcaption{Relative error.}
    \end{minipage}
    \caption{Double pendulum with damped dynamics over varying time step size $h$. Relative error are computed over a set of $10$ test trajectories, each $10$ times longer than the trajectories used during training. %, i.e. $4,000$, $400$, $40$ and $20$ time steps per trajectory with step size $h={0.01, 0.1, 1.0, 2.0}$, respectively. 
    Left panel: A single example of predicted trajectory. Right: Relative error of all points along all $10$ predicted trajectories (upper), and relative error for only the first time step (lower). Green trajectories represent DFLNN (proposed), orange denote the RK4-GLNN model, and pink indicate RK4-NODE, while the ground truth is plotted in black.}
    \label{fig:DP_step_size}
\end{figure}

Next, we investigate how robust the model is with respect to the choice of time step size. For training, we generate trajectories of the damped double pendulum over the fixed time interval $[0, 4]$ and consider four temporal resolutions, $h \in \{0.01, 0.1, 1.0, 2.0\}$, corresponding to $400$, $40$, $4$ and $2$ time steps per trajectory, respectively. Since coarser sampling yields fewer points per trajectory, we vary the number of trajectories (i.e.\ initial conditions) inversely with $h$ so that the total amount of training data stays fixed, using $N_\mathcal{T} \in \{16, 160, 1600, 3200\}$ for $h \in \{0.01, 0.1, 1.0, 2.0\}$, respectively. 
The models were trained with a batch size of $256$. For each temporal resolution, we also generate $16$ additional trajectories for validation during training.

We evaluate the models for varying time step size by rolling out trajectories from $10$ new, unseen initial conditions, each $10$ times longer than the trajectories used during training. The results are shown in Figure~\ref{fig:DP_step_size}. RK4-NODE captures the dynamics at the finest resolutions but fails completely at the coarsest ones. RK4-GLNN remains to produce plausible trajectories as $h$ increases, but gradually overestimates the damping as $h$ grows. DFLNN retains its long-term accuracy across all resolutions, with only a mild degradation at the largest step size.

\begin{figure}[htb!]
    \centering
    
    \begin{minipage}{\textwidth}
        \centering
        \scriptsize 
        Double pendulum: effect of raising the damping coefficient
    \end{minipage}
    
    \begin{subfigure}[t]{0.58\textwidth}
        \centering
        \includegraphics[trim={0 0 0 0}, clip, width=\linewidth]{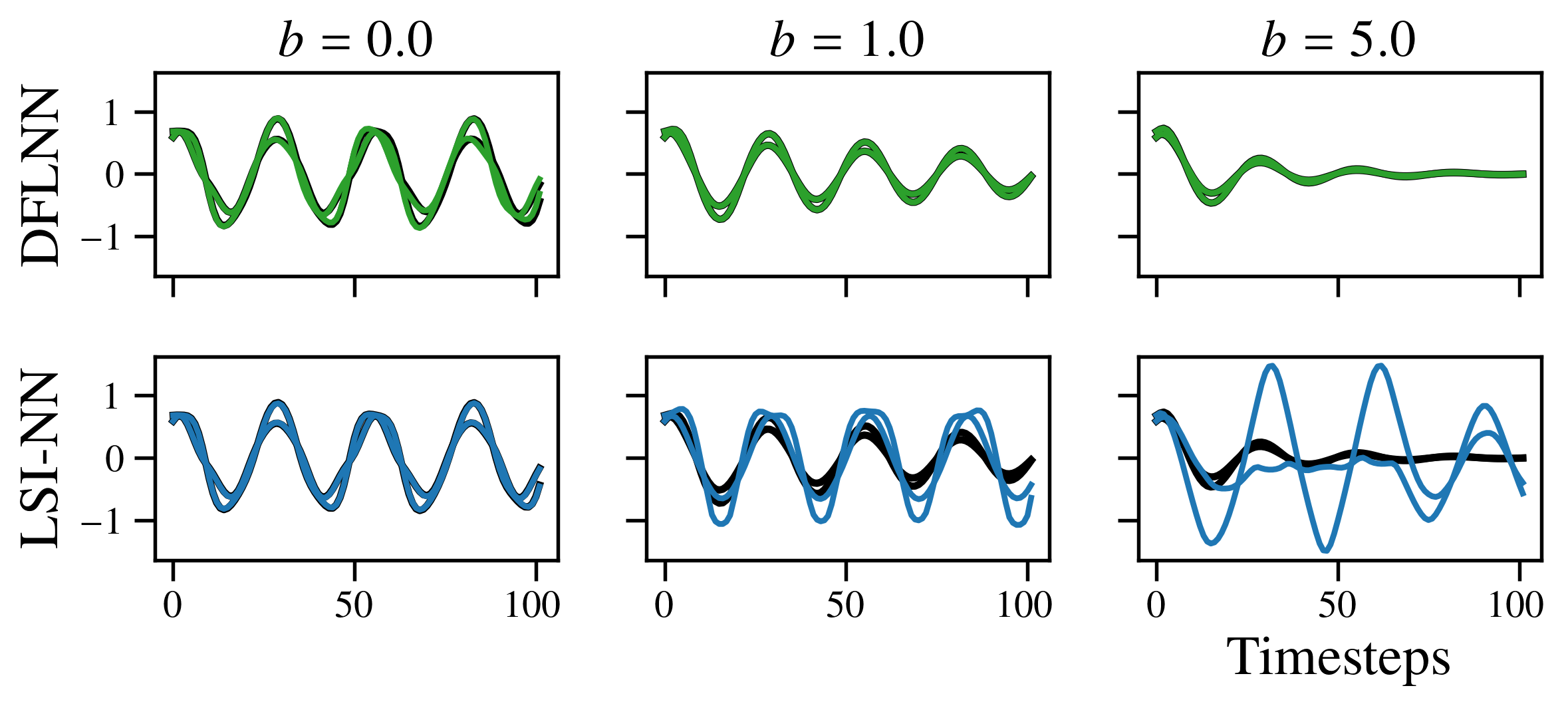}
        \caption{Prediction rollout.}
    \end{subfigure}
    \hfill
    \begin{subfigure}[t]{0.38\textwidth}
        \centering
        \includegraphics[trim={0 0 0 0}, clip, width=\linewidth]{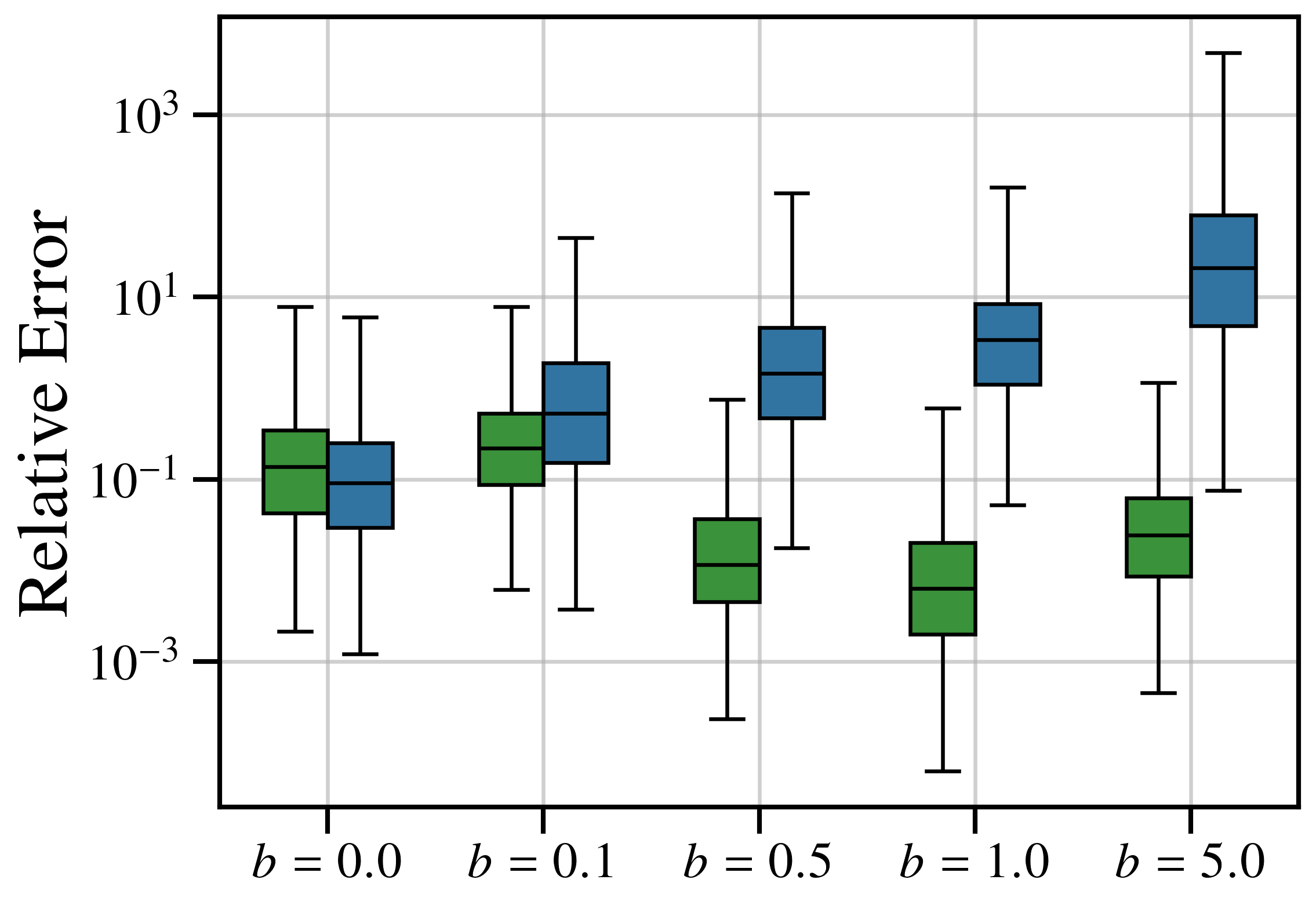}
        \caption{Relative error.}
    \end{subfigure}
    
    \caption{Investigate the significance of including the external force term in the discrete Lagrangian formulation of the double pendulum system. We contrast the proposed DFLNN model with the LSI-NN, which employs a comparable approach but omits an explicit treatment of external forces when learning the dynamics from data. Left: A single example of the predicted trajectories. Right: Relative error evaluated across $20$ predicted trajectories.}
    \label{fig:DP_b_results}
\end{figure}
To examine the effect of the external force term in the DFLNN architecture, Figure \ref{fig:DP_b_results} presents a comparison of DFLNN models trained with damping coefficients ranging from $b=0.0$ to $b=5.0$. We benchmark against LSI-NN, a related architecture that differs in that it includes only a Lagrangian term and omits any external force term. For this comparison, we employ the same training dataset size as before. The results indicate that both models achieve comparable performance when no external forces are present ($b=0$). In contrast, as the damping magnitude increases, the performance of LSI-NN gradually deteriorates, highlighting that the advantage of incorporating an external force term in the DFLNN architecture becomes more pronounced under stronger external forcing.

\begin{figure}[htb!]
    \centering
    \begin{minipage}{\textwidth}
        \centering
        \scriptsize 
        Double pendulum: Increasing noise level $\mu$
    \end{minipage}
    
    \begin{minipage}[c]{0.65\textwidth}
        \centering
        \includegraphics[trim={0cm 0cm 0cm 0cm}, clip, width=\textwidth]{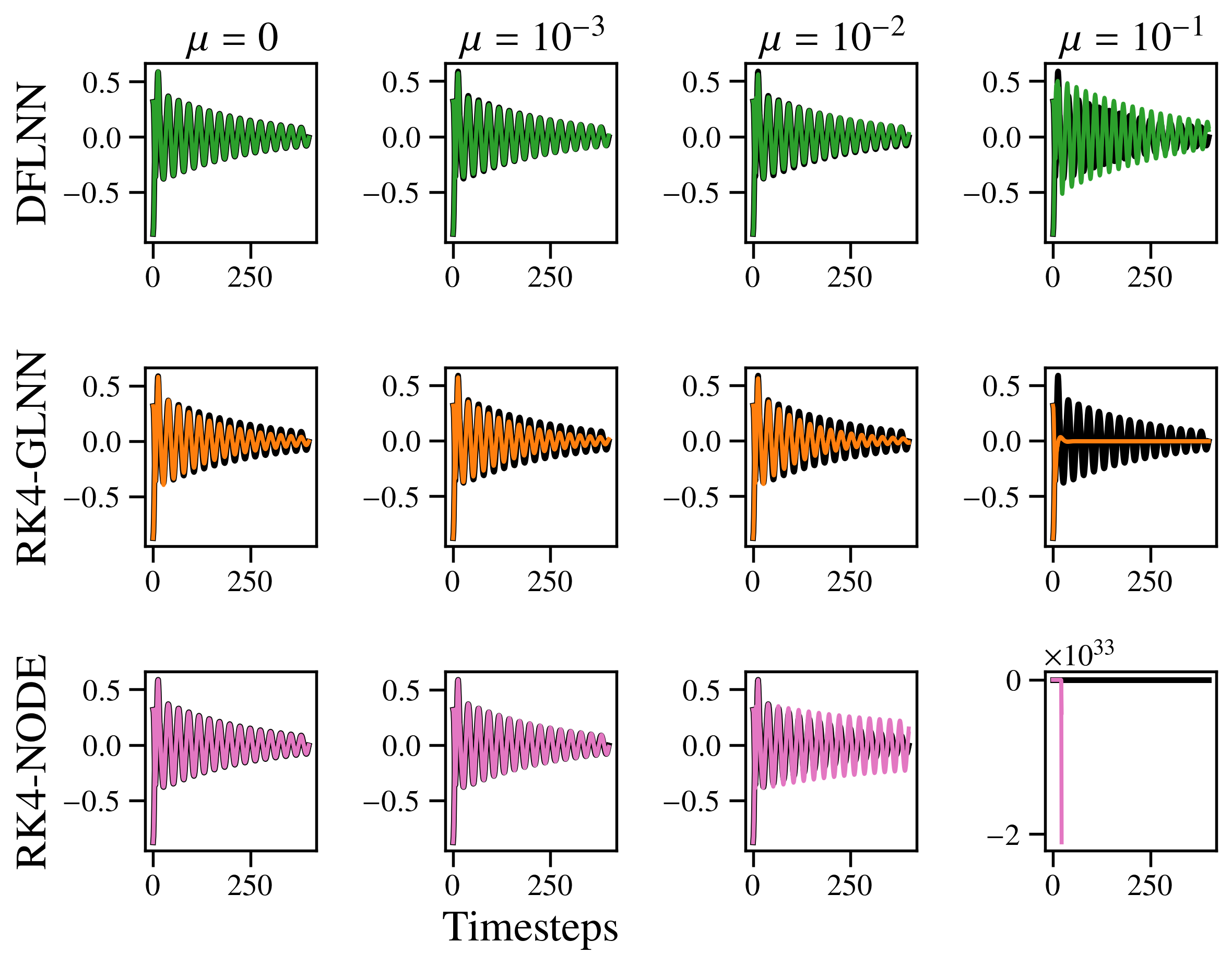}
        \subcaption{Prediction rollout.}
    \end{minipage}
    \hfill
    % Right column with two stacked images
    \begin{minipage}[c]{0.32\textwidth}
        %\vspace{-9.75cm}
        % Top right image
        \centering
        \includegraphics[width=\textwidth]{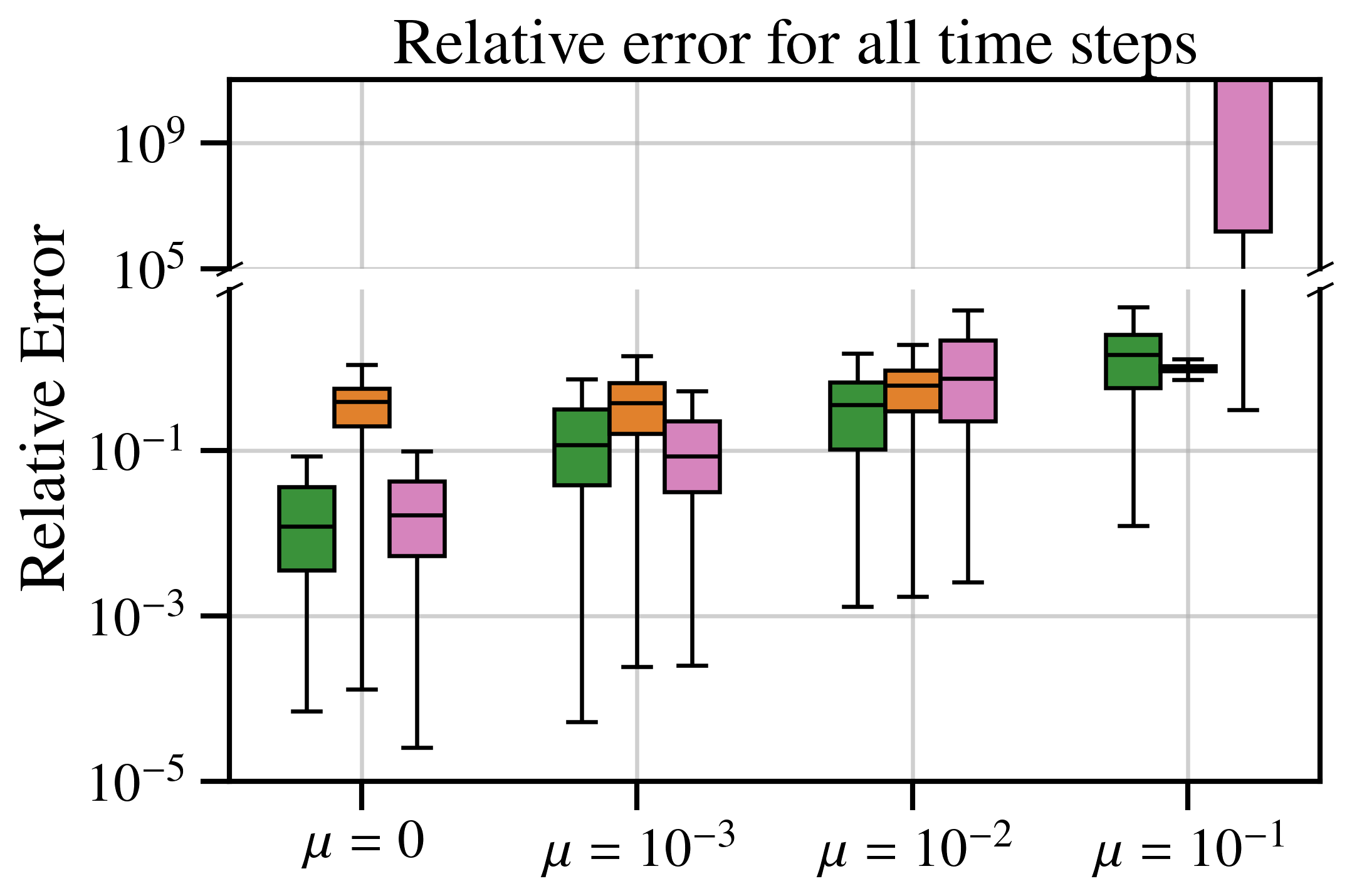} \\
        \vspace{0.5cm}
        % Bottom right image
        \includegraphics[width=\textwidth]{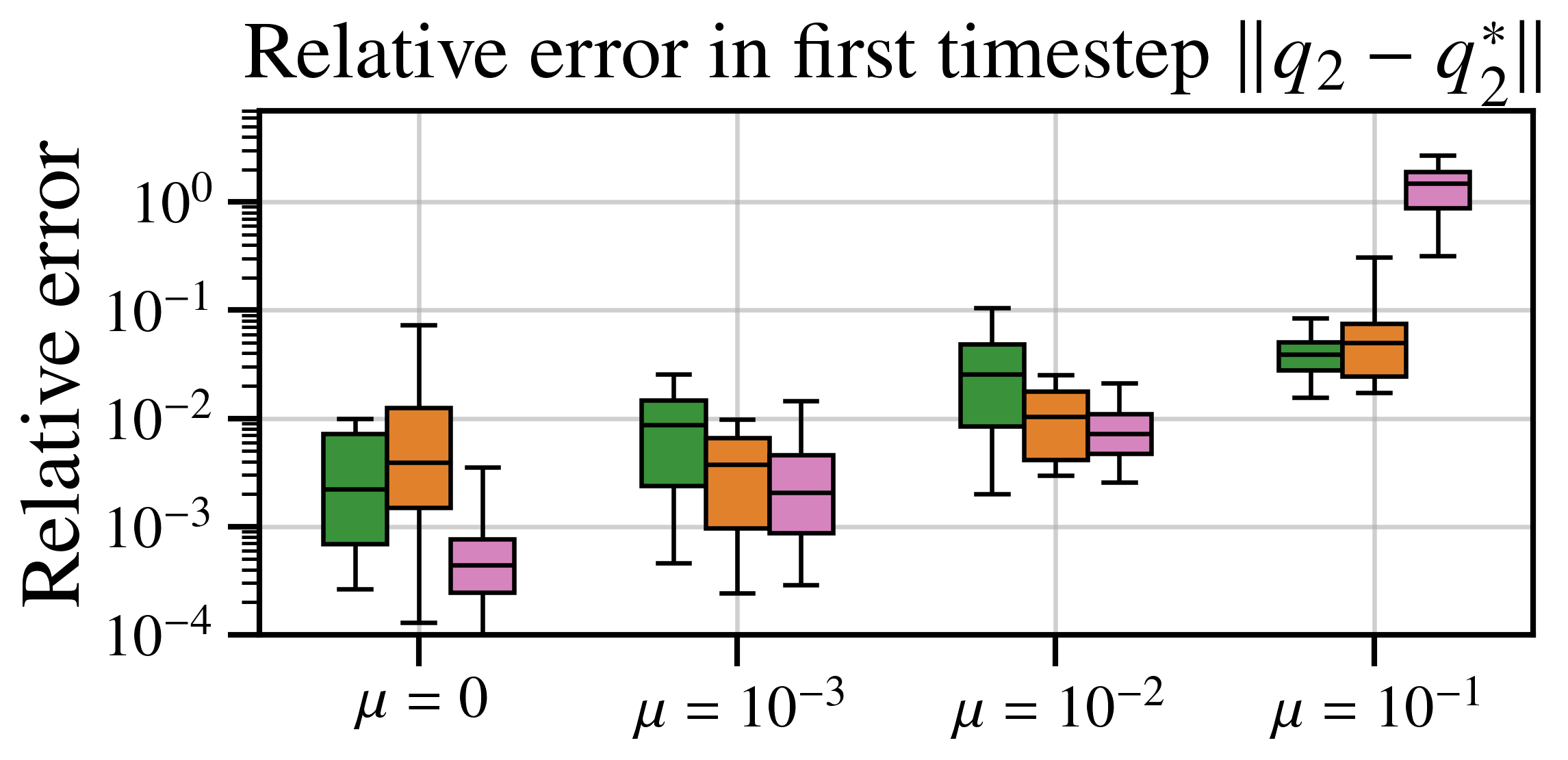} \\
        \vspace{0.5cm}
        \includegraphics[width=\textwidth]{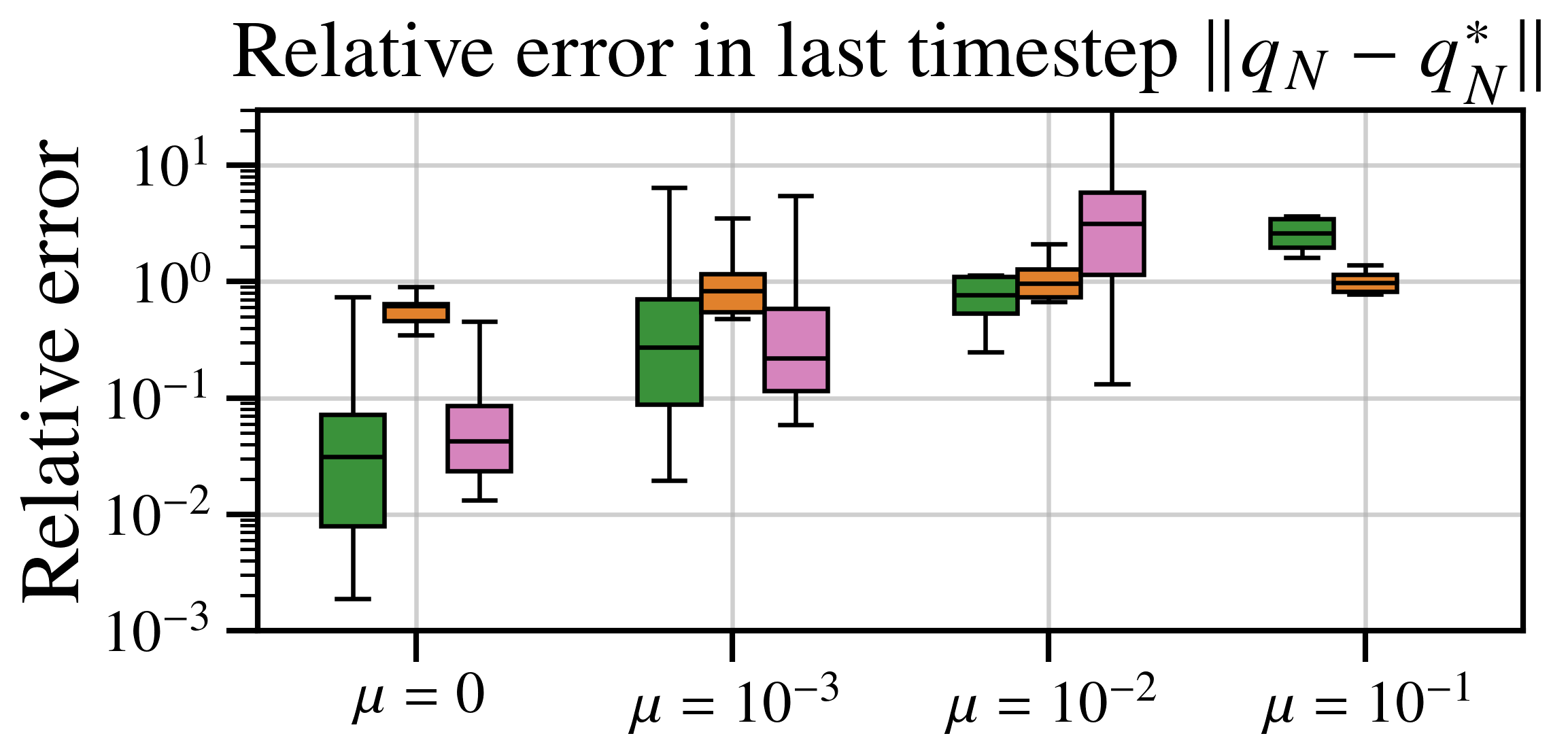}
        \subcaption{Relative error.}
    \end{minipage}
    \caption{Double pendulum with damped dynamics under varying Gaussian noise level $\mu$.
    The models are trained on the noisy data and evaluated against the clean ground truth. Relative error are computed over a set of $10$ test trajectories each twice as long as the trajectories used during training. Left panel: a single example of a predicted trajectory. Right: relative error of all points along all $10$ predicted trajectories (upper), and relative error for only the last time step (lower). Green trajectories represent DFLNN (proposed), orange denote the RK4-GLNN model, and pink indicate RK4-NODE, while the ground truth is plotted in black.}
    \label{fig:DP_vaying_noise}
\end{figure}
Lastly, we examine how robust the model is with respect to measurement noise. We generate a training and validation dataset of $64$ and $16$ trajectories, respectively. Each integrated over $200$ time steps of step size $h=0.1$. We corrupt the trajectories with additive Gaussian noise,
\begin{equation*}
    \tilde{q}_n = q_n + \mu\,\operatorname{std}(q)\,\varepsilon_n,
    \qquad \varepsilon_n \sim \mathcal{N}(0, I),
\end{equation*}
where $\operatorname{std}(q)$ denotes the standard deviation of the clean position trajectory, and $\mu$ is a dimensionless scalar that sets this relative noise level. We consider $\mu \in \{0, 10^{-3}, 10^{-2}, 10^{-1}\}$, and train each model on the noisy training dataset while evaluating performance over $10$ unseen, clean ground-truth trajectories. The models were trained with a batch size of $256$.

The results for varying noise level are shown in Figure~\ref{fig:DP_vaying_noise}. As expected, all methods degrade as the noise level $\mu$ increases, but to very different degrees. As $\mu$ increases, the baseline methods degrade most rapidly. At the largest noise value considered, $\mu = 10^{-1}$, RK4-NODE completely diverges, attaining values on the order of $10^{33}$, whereas RK4-GLNN almost immediately decay toward a resting state instead of following the oscillation. Across the entire range of $\mu$ values considered, DFLNN remains closest to the true dynamics at all noise levels. The final-step error should be read with care in this damped setting. Since the true trajectory decays toward rest, a low final-step error can reflect the system having come to rest rather than an accurate prediction, which is why we also report the error over all time steps.

\subsection{Task 2: Dissipative Charged Particle in a Magnetic Field}\label{sc:Task2}

\begin{figure}[htb!]
    \centering

    \begin{minipage}{\textwidth}
        \centering
        \scriptsize 
        Charged particle: model comparison
    \end{minipage}

    \begin{minipage}{0.45\textwidth}
        \centering
        \hspace{0.03\textwidth} 
        \scriptsize  Initial position $q_0^{i}\in(0, 0)$ \\ (training regime)
    \end{minipage}%
    \hfill
    \begin{minipage}{0.45\textwidth}
        \centering
        \hspace{0.04\textwidth} 
        \scriptsize Initial position between $q_0^{i} \in (0, 1)$ \\ (testing regime)
    \end{minipage}\\
    
    \begin{minipage}[c]{0.22\textwidth}
        \centering
        \includegraphics[trim={0 0 0 0}, clip, width=\linewidth]{Charged_Particle_rollout_Origo_damped.png}
    \end{minipage}
    \begin{minipage}[c]{0.22\textwidth}
        \centering
        \includegraphics[trim={0 0 0 0}, clip, width=\linewidth]{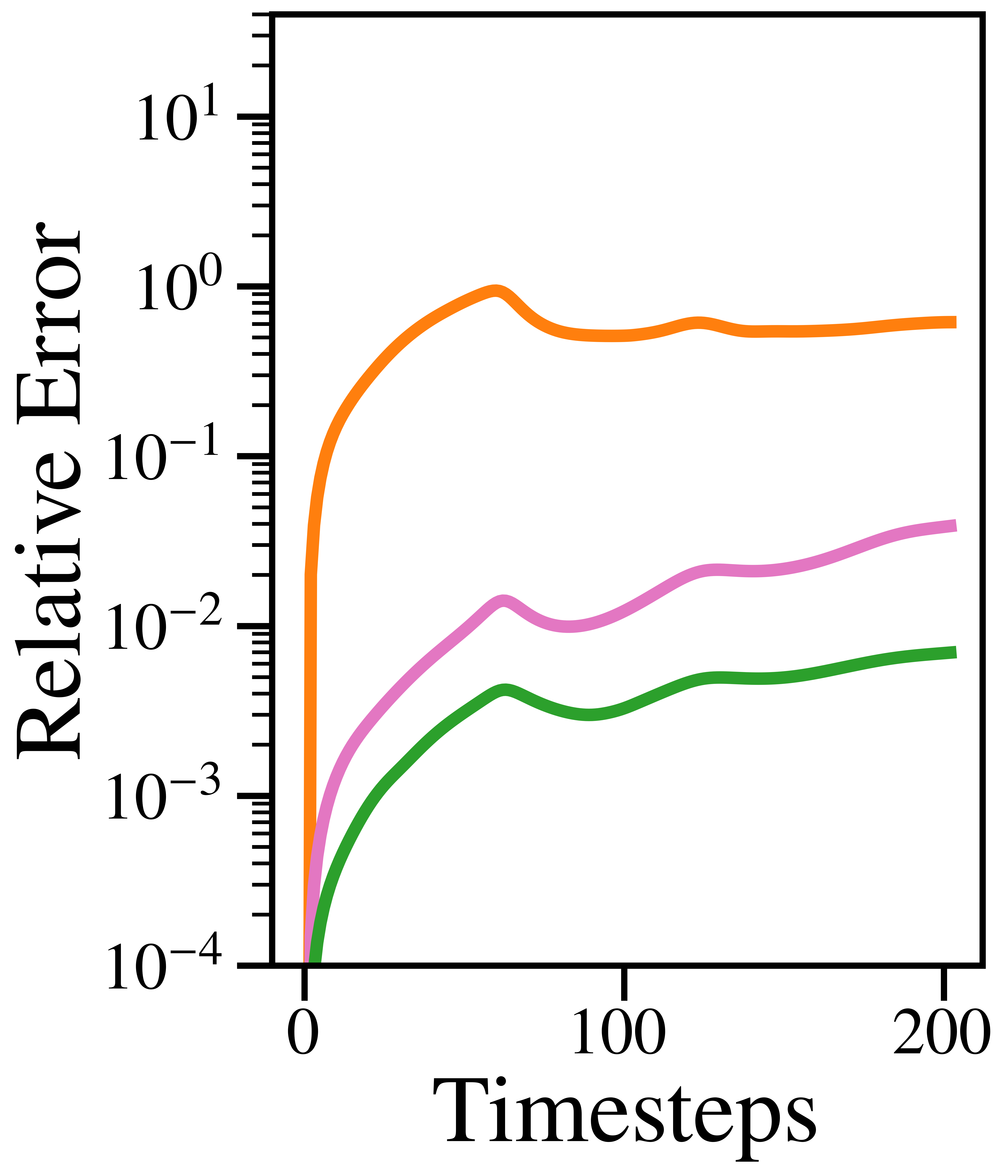} \\
        \includegraphics[trim={0 1.3cm 0 0}, clip, width=\linewidth]{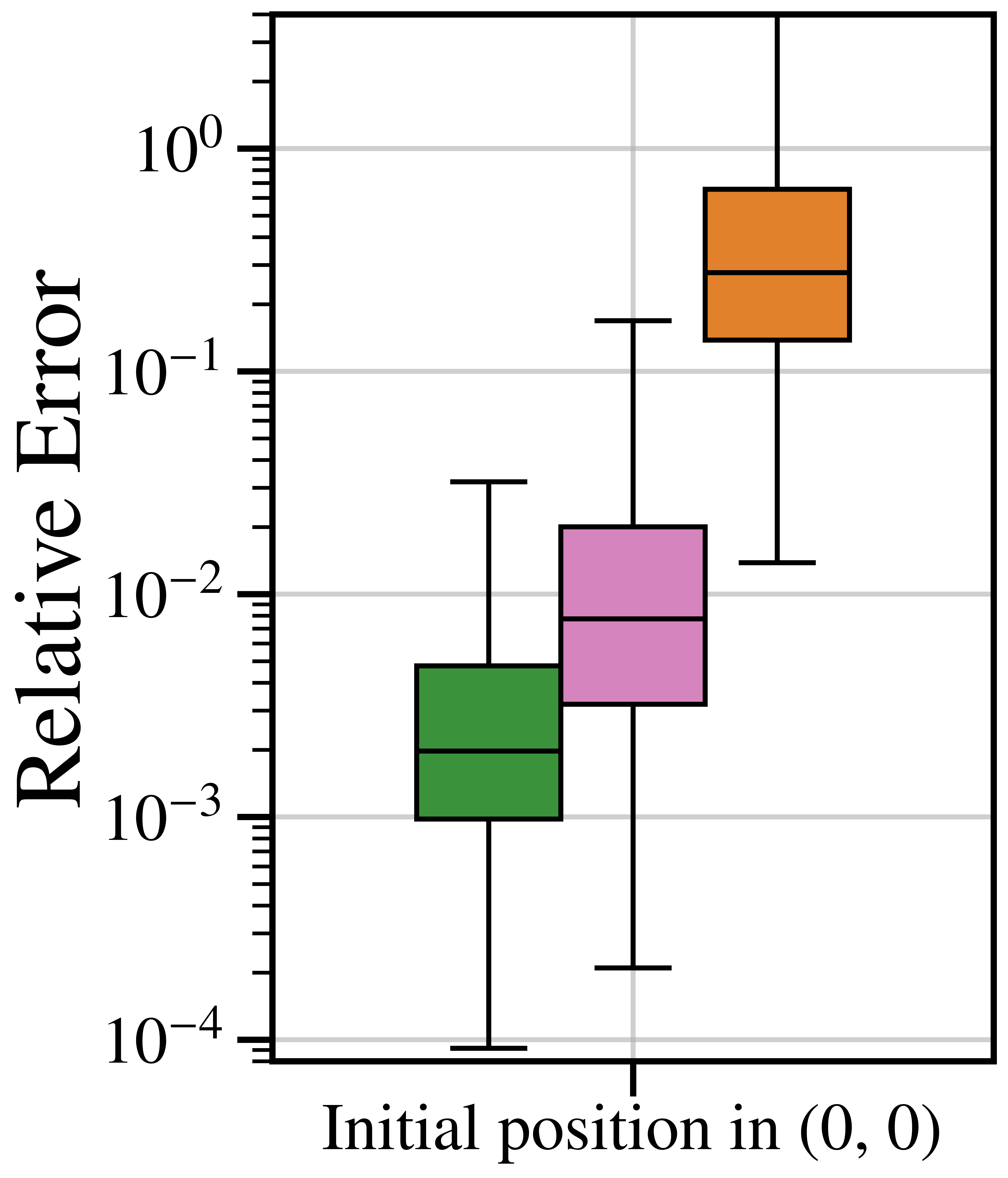}
    \end{minipage}
    \hfill
    \begin{minipage}[c]{0.22\textwidth}
        \centering
        \includegraphics[trim={0 0 0 0}, clip, width=\linewidth]{Charged_Particle_rollout_OrigoInRand_damped.png}
    \end{minipage}
    \begin{minipage}[c]{0.22\textwidth}
        \centering
        \includegraphics[trim={0 0 0 0}, clip, width=\linewidth]{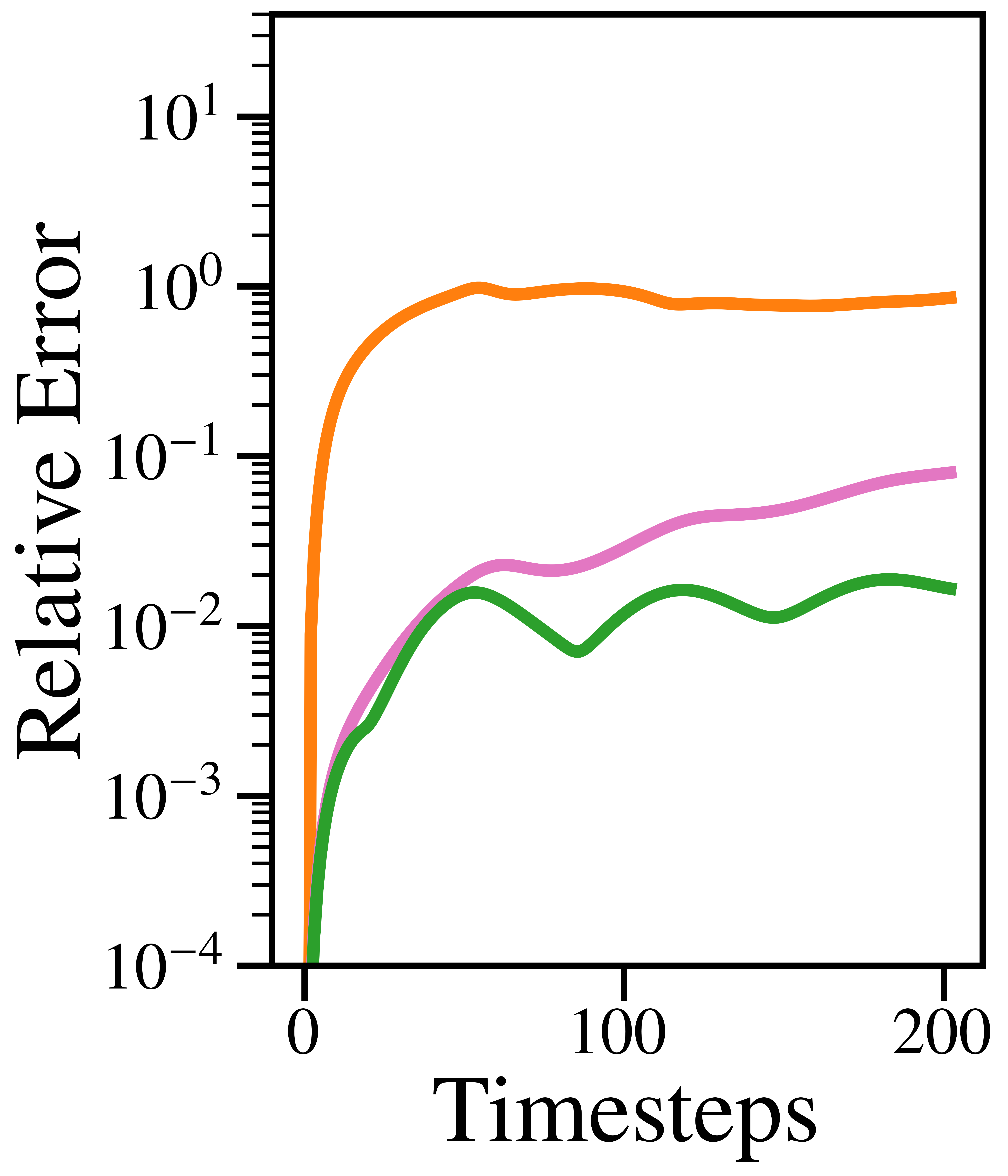} \\
        \includegraphics[trim={0 1.3cm 0 0}, clip, width=\linewidth]{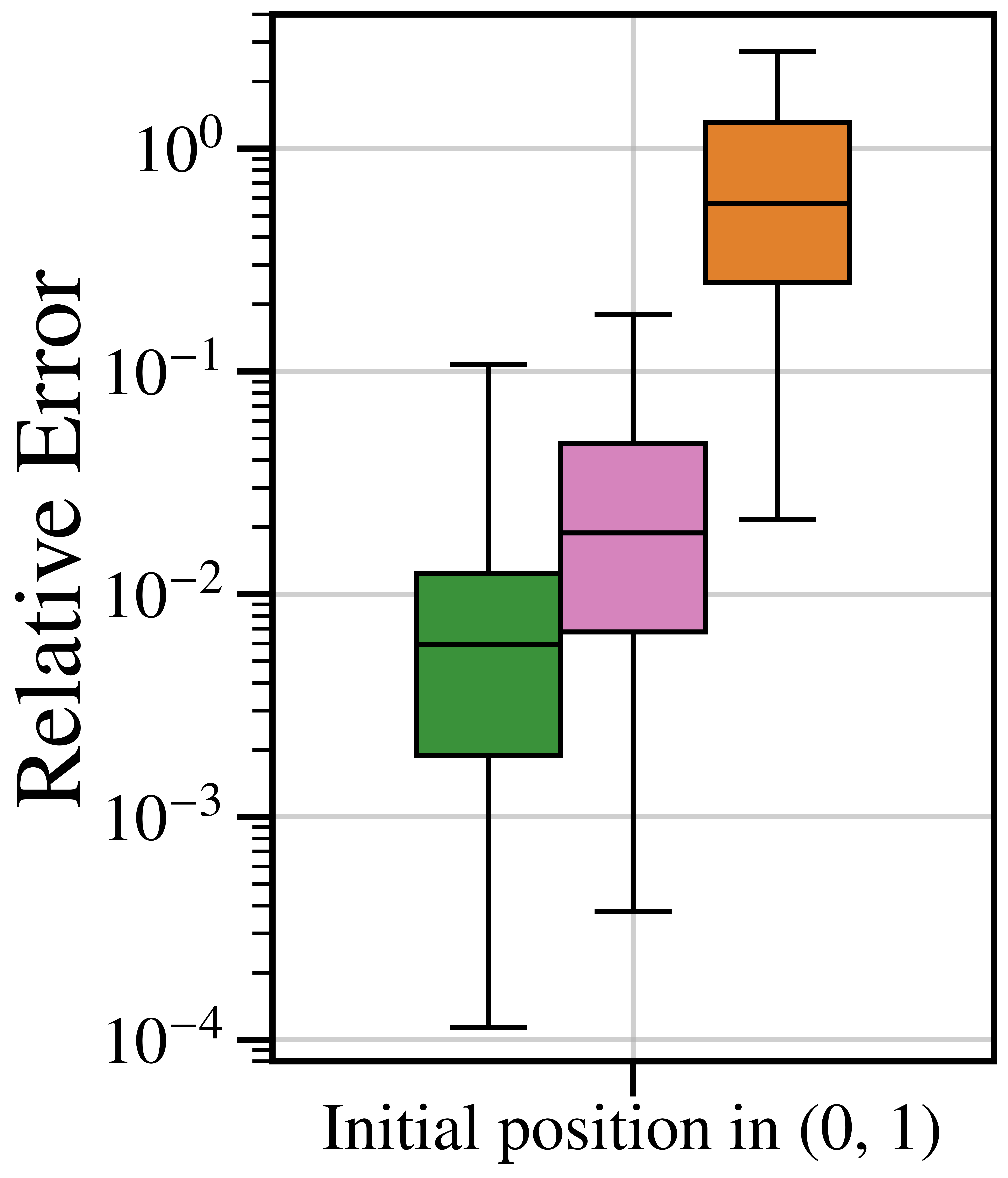}
    \end{minipage}
    \\
    %\vspace{-10cm}
    \begin{minipage}{0.45\textwidth}
        \centering
        \subcaption{Predictions of trajectories with initial position $q_0^{i}\in(0, 0)$, consistent with the training regime.}
    \end{minipage}%
    \hfill
    \begin{minipage}{0.45\textwidth}
        \centering
        \subcaption{Predicted trajectories on a test setup for initial positions $q_0^{i} \in (0, 1)$ that were not included in the training data.}
    \end{minipage}
    
    \caption{Result for a charged particle. (a) Performance of the trained models on predicted trajectories with initial conditions similar to those used in training. (b) Prediction of trajectories for initial positions lying outside the training range. Each subfigure is composed of three panels: a single example of predicted trajectories (left); the relative error over 20 predicted trajectories shown as a boxplot of all points (top right); and the mean relative error as a function of the timestep (bottom right). Green trajectories correspond to DFLNN (proposed), orange to the RK4-GLNN model, and pink to RK4-NODE, with the ground truth plotted in black.}
    \label{fig:CP_results}
\end{figure}

We study the dynamics of a charged particle moving in a magnetic field with linear dissipation, expressed in Cartesian coordinates. Synthetic datasets of particle trajectories are generated by numerical integration of the governing equations of motion, see~\ref{apx:dataset-chargedparticle}. 
For a charged particle traveling in a magnetic field, the potential energy depends on both the position $q$ and the velocity $\dot{q}$. Thus, we assume that the Lagrangian has the form \eqref{eq:structuredNetworkEquations} with $U=U(q, \dot{q})$, and the external force is assumed to be linear \eqref{eq:structureExternalForceNetwork}.

In this first experiment, we examine each model's ability to generate plausible rollouts outside its training regime. All models are trained on trajectories starting from the origin, $q_0 = (0,0,0)^\top$, with initial velocities sampled from $\dot q_0 \in [-1, 1]$. After training, we test on trajectories whose initial positions are instead sampled uniformly from $[0, 1]$, outside the training distribution. System parameters are fixed to mass $m = 1$, charge $q = 1$, magnetic field vector $B = [0, 0, 0.5]^\top$, and friction coefficient $b = 0.1$. Each training trajectory spans $40$ time steps with temporal resolution $h = 0.1$. In total, $32$ trajectories are generated, with $16$ used for training and $16$ for validation, and a batch size of $256$. We evaluate on $20$ trajectories of $200$ time steps, five times longer than those seen during training. The corresponding results are shown in Figure~\ref{fig:CP_results}. They again show that RK4-GLNN fails to capture the underlying dynamics accurately, while the proposed model outperforms the baselines both within and beyond the training regime.

\begin{figure}[htb!]
    \centering
    
    \begin{minipage}{\textwidth}
        \centering
        \scriptsize 
        Charged particle: conservative dynamics \\
        (testing regime)
    \end{minipage}

    % --- Bottom Row ---
    \begin{minipage}[t]{\textwidth}
    \begin{subfigure}[c]{0.58\textwidth}
        \centering
        \includegraphics[trim={0 0 0 0}, clip, width=\linewidth]{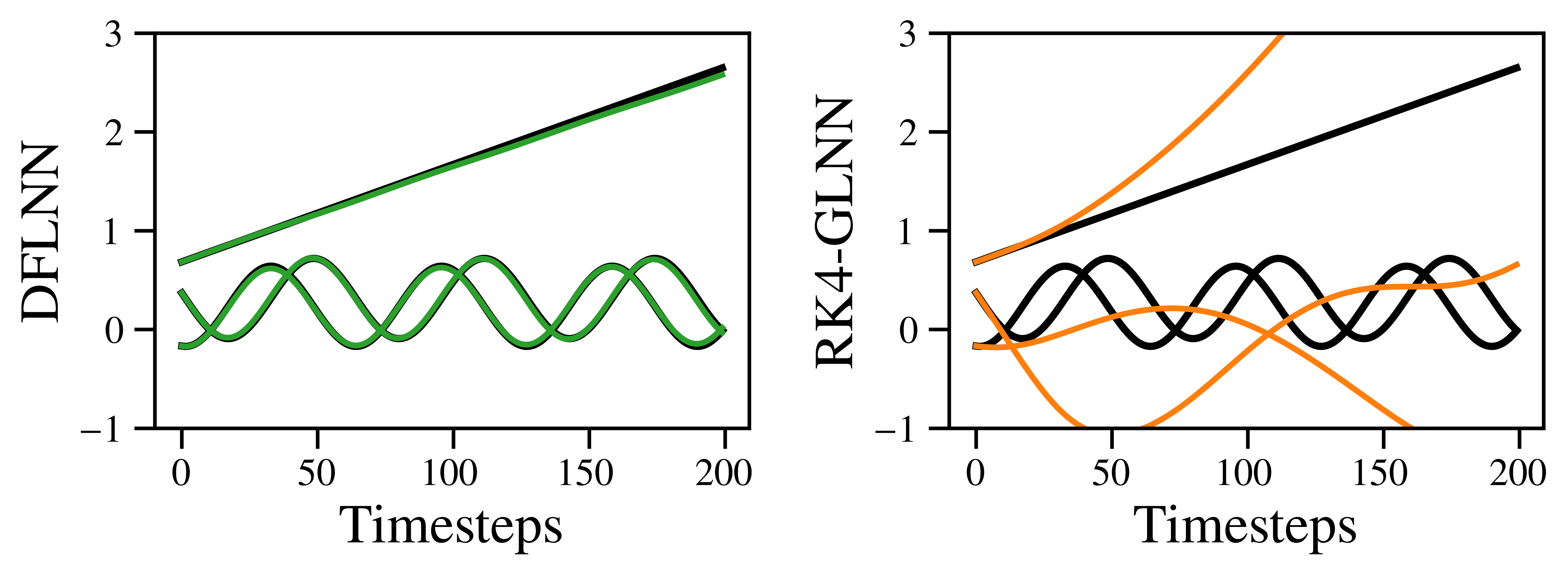}
        \subcaption{Predicted rollout.}
    \end{subfigure} 
    \hfill
    \begin{subfigure}[c]{0.35\textwidth}
        \centering
        \includegraphics[trim={0 0 0 0}, clip, width=\linewidth]{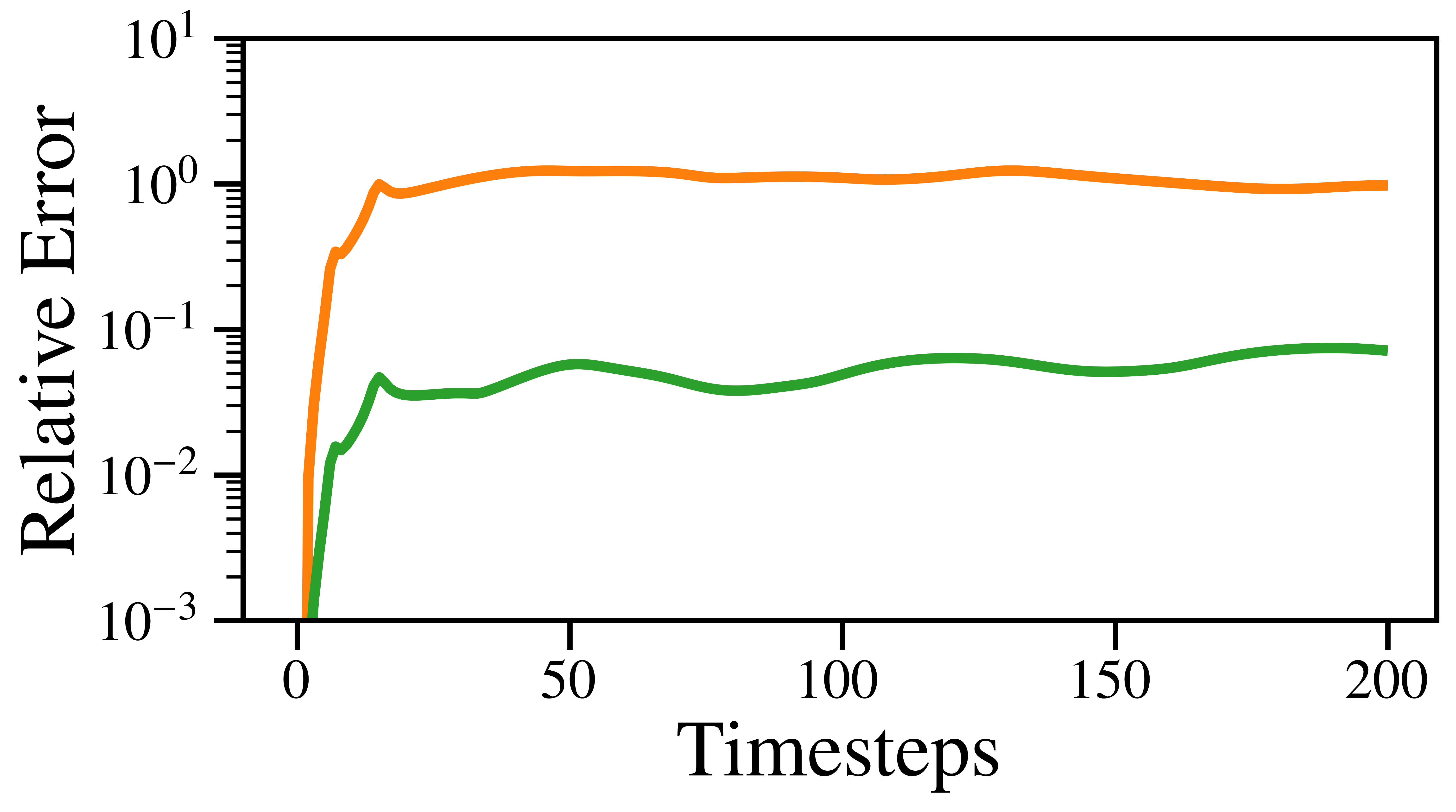}
        \subcaption{Relative error.}
    \end{subfigure} 
    \end{minipage}
    
    \caption{Charged particle with conservative, undamped dynamics. Testing regime where we have set the learned force to zero, as explained in the end of Section~\ref{sc:TheForceTerm}. Left: a single example of a predicted trajectory. Right: mean relative error as a function of the time step over $20$ test trajectories. Green trajectories represent DFLNN (proposed) and orange denote the RK4-GLNN model, while the ground truth is plotted in black.}
    \label{fig:CP_undamped_results}
\end{figure}
Similarly to Task 1, we set the learned force to zero--as explained in the end of Section~\ref{sc:TheForceTerm}---to evaluate the learned Lagrangian. We roll out predictions from initial positions outside the training distribution, sampled on the interval $q_0 \in [0,1]$. The results in Figure~\ref{fig:CP_undamped_results} show the same picture as for the double pendulum. DFLNN reproduces the undamped dynamics, indicating that the magnetic force has been absorbed into the velocity-dependent potential $U_\theta(q,\dot q)$, while the dissipation has been absorbed into the learned force $F_\theta$. The RK4-GLNN rollouts deviate from the ground truth also in this conservative setting, confirming that its learned Lagrangian, and not only its force term, is inaccurate.

\begin{figure}[htb!]
    \centering
    
    \begin{minipage}{\textwidth}
        \centering
        \scriptsize 
        Charged particle: effect of raising the damping coefficient
    \end{minipage}

    % --- Bottom Row ---
    \begin{minipage}[t]{\textwidth}
    \begin{subfigure}[c]{0.58\textwidth}
        \centering
        \includegraphics[trim={0 0 0 0}, clip, width=\linewidth]{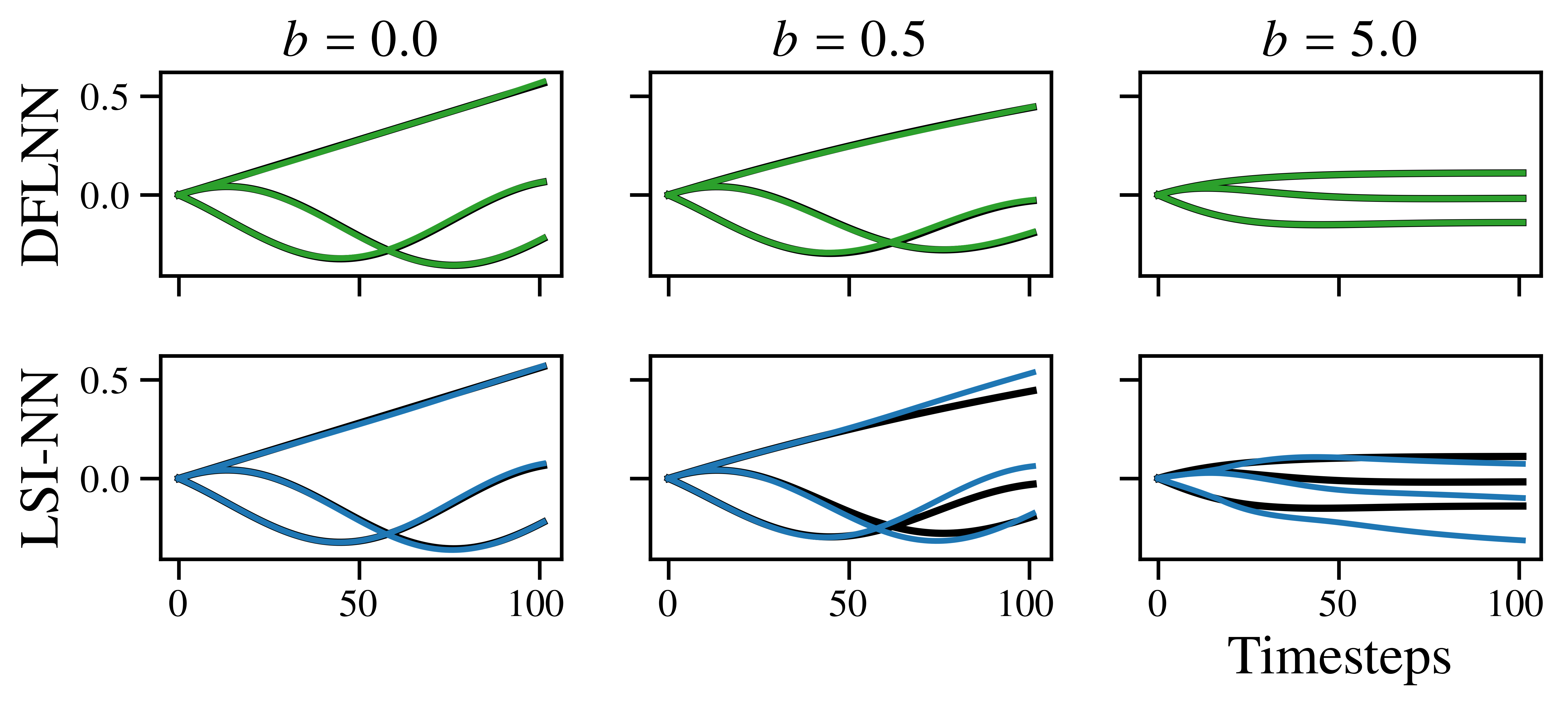}
        \subcaption{Predicted rollout.}
    \end{subfigure} 
    \hfill
    \begin{subfigure}[c]{0.35\textwidth}
        \centering
        \includegraphics[trim={0 0 0 0}, clip, width=\linewidth]{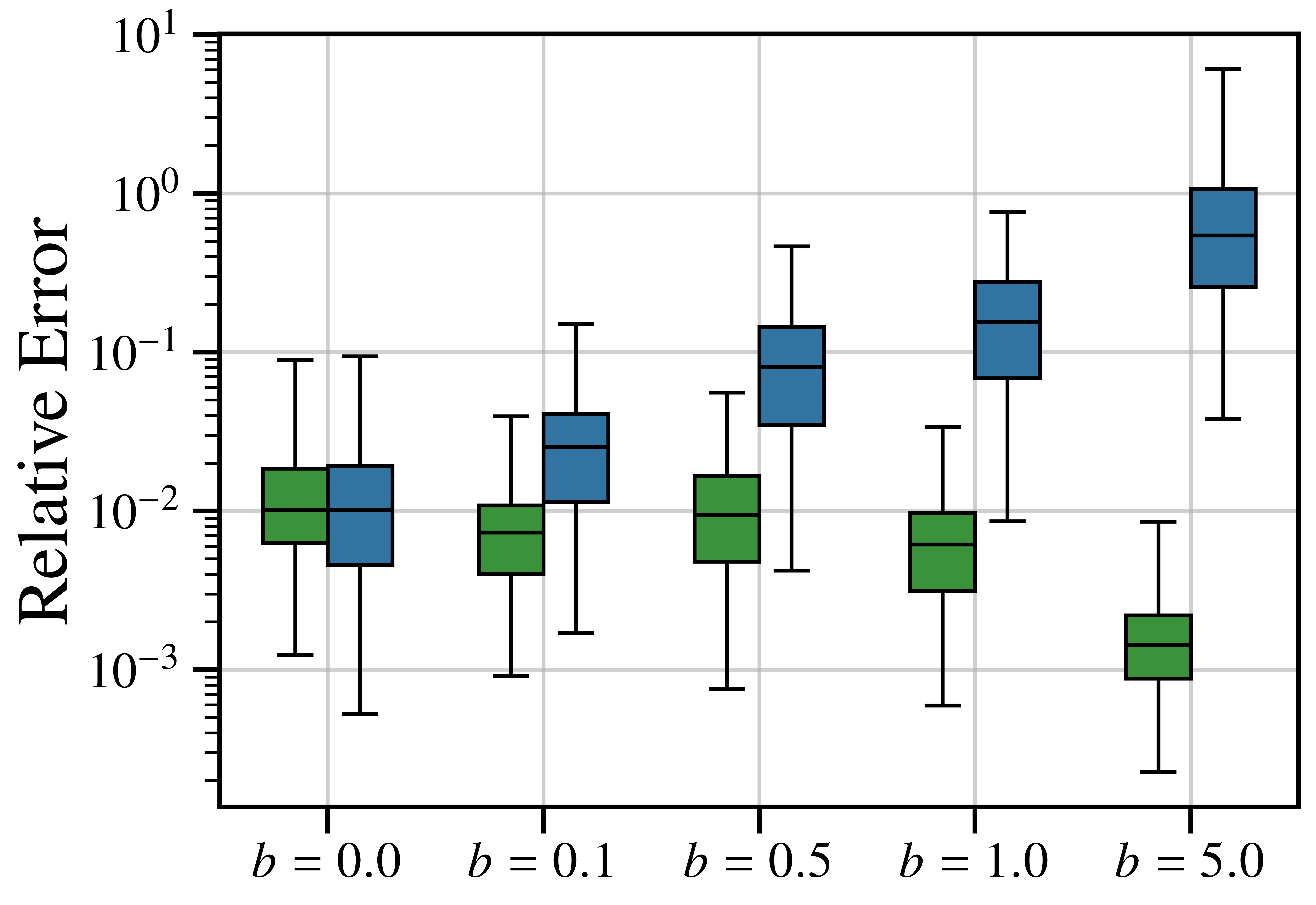}
        \subcaption{Relative error.}
    \end{subfigure} 
    \end{minipage}
    
    \caption{Examine how crucial it is to incorporate the external force term in the discrete Lagrangian formulation for the charged particle problem. We compare the proposed DFLNN model with the LSI-NN, which adopts a similar methodology but does not explicitly account for external forces when learning the dynamics from data. Left: A single example of the predicted trajectories. Right: Relative error evaluated across $20$ predicted trajectories.}
    \label{fig:CP_b_results}
\end{figure}

To investigate how explicitly integrating the external force term into the DFLNN architecture affects performance---rather than employing a DFLNN variant that models only the Lagrangian part---we report in Figure \ref{fig:CP_b_results} the results obtained for an increased damping coefficient, raised from $b=0.0$ to $b=5.0$. For these experiments, the training data were generated from dynamics with parameters $m=1$, $q=10$, and $B=[0, 0, 0.5]^\top$. In total, we used $288$ trajectories, each with $200$ timesteps and a stepsize of $h=0.01$, of which $256$ trajectories were used for training and $32$ for validation. The models were trained for $2,000$ epochs with a batch size of $4096$. As in Task \ref{sc:Task1}, we observe that the advantages of using DFLNN become increasingly significant as the damping coefficient increases.

\begin{figure}[htb!]
    \centering
    \begin{minipage}{\textwidth}
        \centering
        \scriptsize 
        Charged particle: Varying time step size $h$
    \end{minipage}
    
    \begin{minipage}[c]{0.65\textwidth}
        \centering
        \includegraphics[trim={0cm 0cm 0cm 0cm}, clip, width=\textwidth]{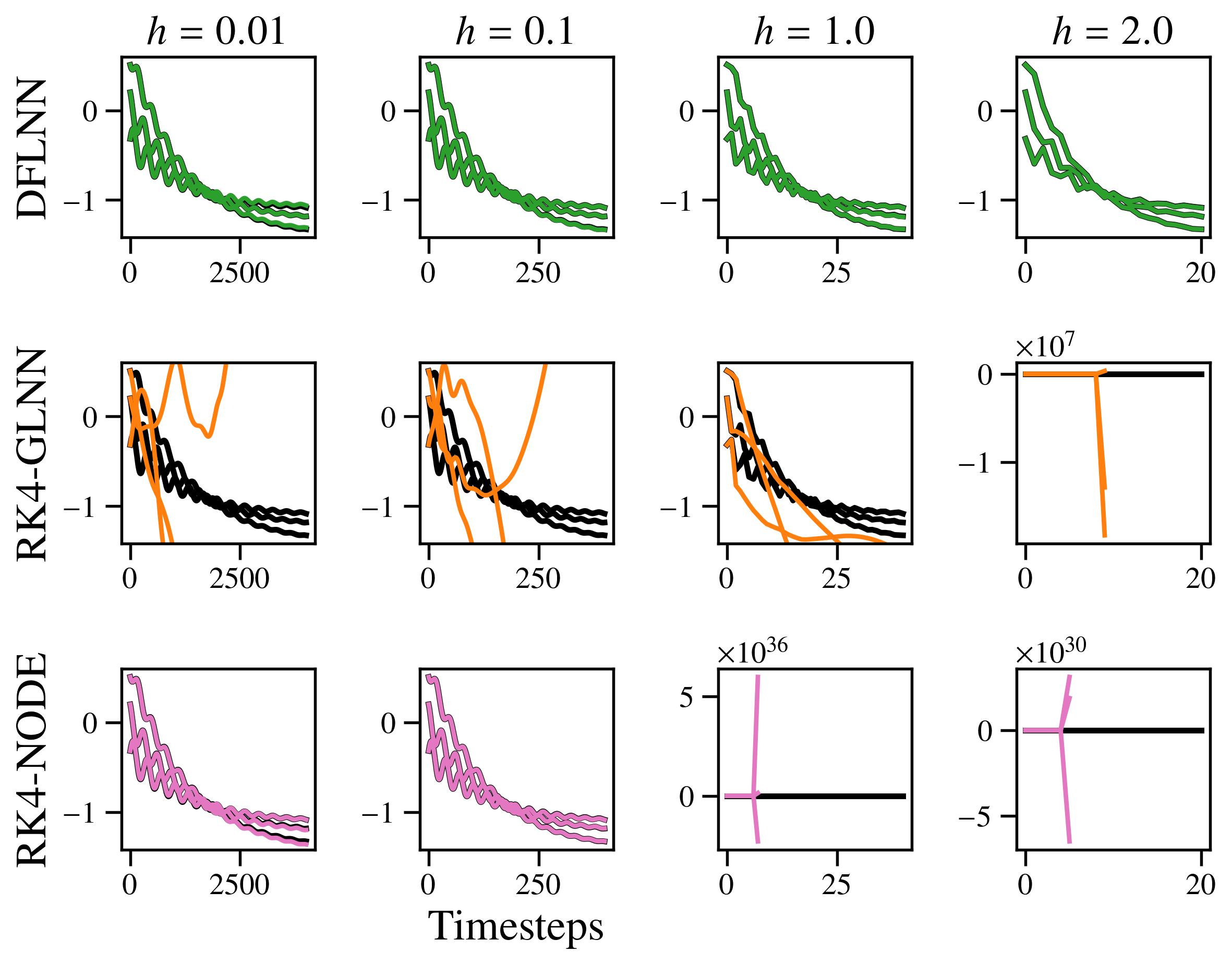}
        \subcaption{Prediction rollout.}
    \end{minipage}
    \hfill
    % Right column with two stacked images
    \begin{minipage}[c]{0.32\textwidth}
        \centering
        \includegraphics[width=\textwidth]{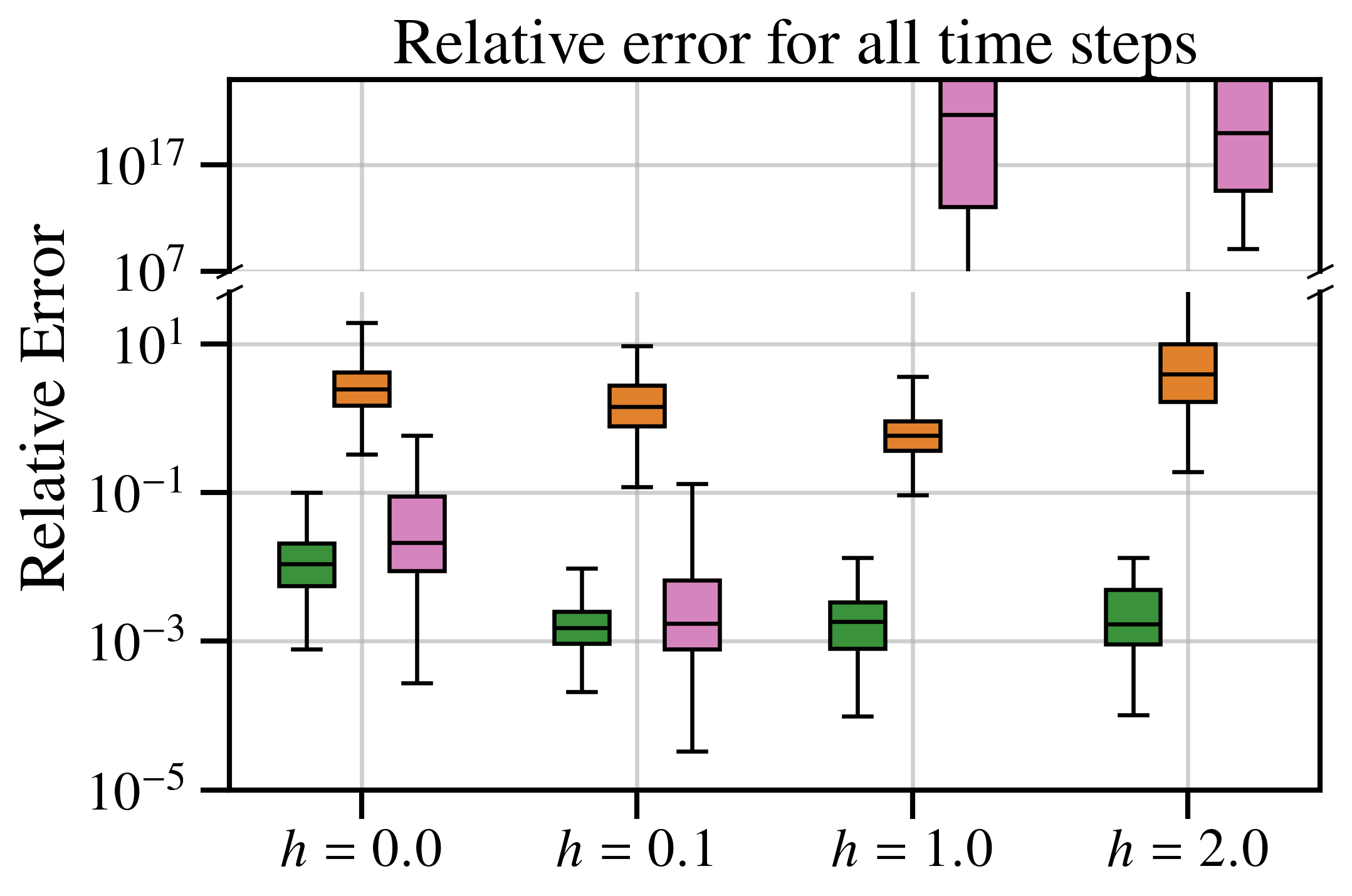} \\
        \vspace{0.5cm}
        \includegraphics[width=\textwidth]{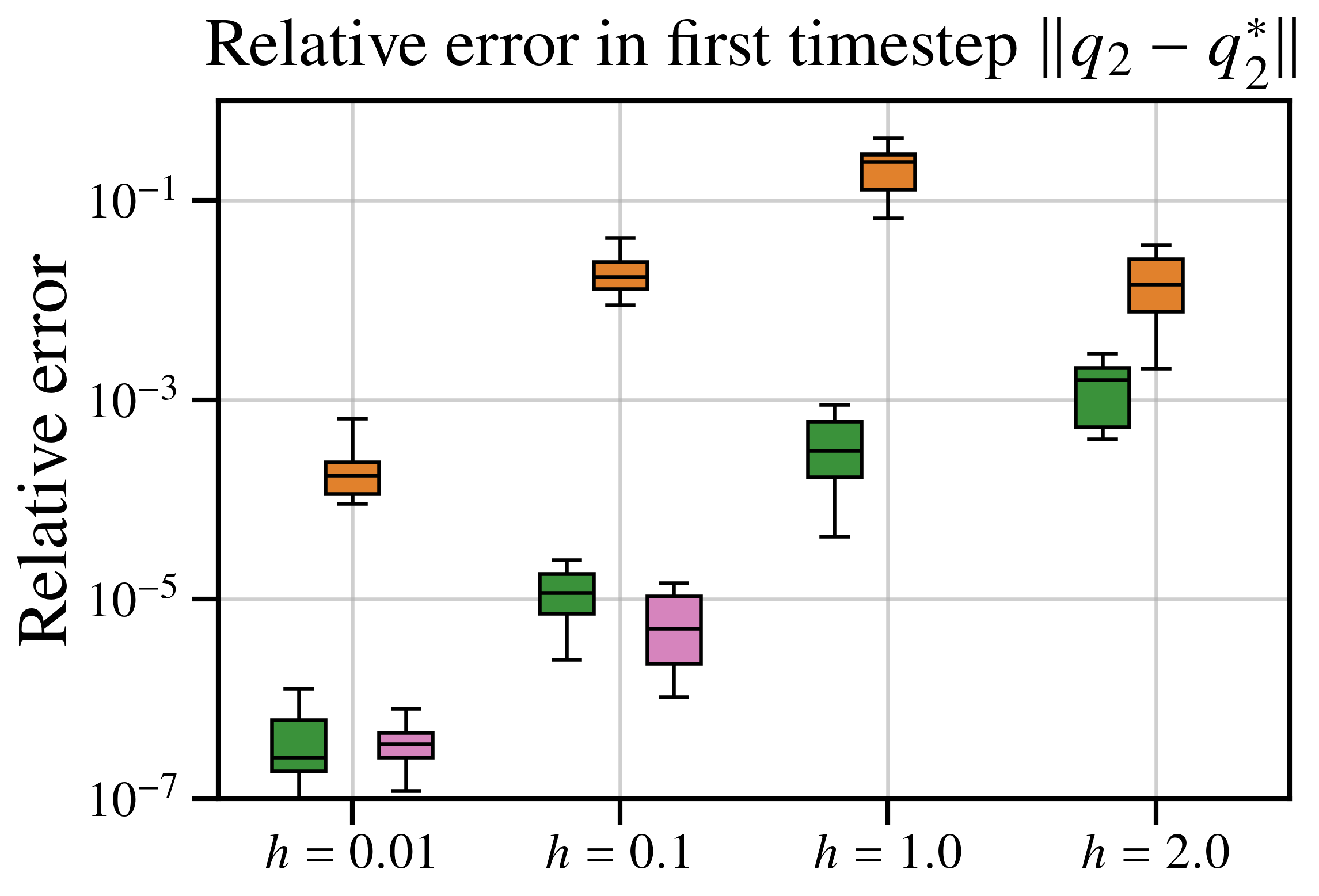}
        \subcaption{Relative error.}
    \end{minipage}
    \caption{Charged particle with dissipative dynamics over varying time step size $h$. Relative error are computed from $10$ test trajectories of $200$ time steps each, four times longer than those seen during training. Left panel: a single example of a predicted trajectory. Right: relative error of all points along all $10$ predicted trajectories (upper), and relative error for only the first time step (lower). Green trajectories represent DFLNN (proposed), orange denote the RK4-GLNN model, and pink indicate RK4-NODE, while the ground truth is plotted in black.}
    \label{fig:CP_step_size}
\end{figure}

Lastly, we repeat the time-step study. We consider unit mass and unit charge, magnetic field $B=[0.6, 1.0, 1.6]^\top$, and damping coefficient $b=0.08$. Initial positions are sampled from from the interval $q\in[0, 1]$ and $\dot{q}\in[-1, 1]$ for both training and testing. For training, we integrate trajectories over the fixed time interval $[0,10]$ and consider the temporal resolutions $h \in \{0.1, 0.5, 1.0, 2.0\}$, corresponding to $100$, $20$, $10$ and $5$ time steps per trajectory, respectively. As previously, since coarser sampling yields fewer points per trajectory, we vary the number of trajectories (i.e.\ initial conditions) inversely with $h$ so that the total amount of training data stays fixed, using $N_\mathcal{T} \in \{640, 3200, 6400, 12800\}$ for $h \in \{0.1, 0.5, 1.0, 2.0\}$, respectively. The models were trained with a batch size of $256$. For each temporal resolution, we also generate $16$ additional trajectories for validation during training. 

We evaluate the models by rolling out trajectories from $10$ new, unseen initial conditions which we integrate four times longer than those seen during training. The results are shown in Figure~\ref{fig:CP_step_size}. The overall picture resembles the damped double pendulum, with DFLNN as the only model that captures the dynamics at every resolution, including the largest step size $h=2.0$. RK4-NODE is accurate at the two finest resolutions, where it performs comparably to DFLNN, but diverges within a few steps for $h \geq 1.0$, reaching magnitudes of order $10^{30}$--$10^{36}$. For RK4-GLNN, the predicted rollouts depart from the ground truth already at the finest resolution.

\subsection{Task 3: Pixel data for a Damped Simple Pendulum}\label{sc:Task3}
In contrast to the previous experiments, this experiment employs an autoencoder with a learned latent space. This setup is included to illustrate the broader applicability of the proposed model, by showing that the measurement approach is not restricted to position sensors, but can also be applied to e.g. video recordings.

To evaluate the model’s capacity to learn from the latent embeddings produced by an autoencoder, we employ synthetic image time series representing a damped simple pendulum. The data are generated with the Gymnasium library~\cite{towers2024gymnasium}, the successor to OpenAI Gym~\cite{brockman2016openai}, yielding RGB frames of size $500 \times 500 \times 3$ at each time step. Damping is implemented by modifying the \texttt{Pendulum-v1} environment to incorporate a linear dissipative force $\vec{F}=-0.02\dot{\vec{q}}$.
Each frame is cropped around the pendulum, converted to grayscale, and downsampled by a factor of $5$, resulting in $q_t \in \mathbb{R}^{56 \times 56}$. The initial angular displacement is constrained to $120$ degrees. The model is trained on a collection of $522$ trajectories, each of duration $T = 200$ time steps with step size $h = 0.1$, using $512$ trajectories for training and $10$ for validation. Training is carried out for $4,000$ epochs with a batch size of $2048$.

\begin{figure}[htb!]
    \centering

    \begin{minipage}{\textwidth}
        \centering
        \scriptsize 
        Pixel frames of a pendulum: damped dynamics\\
        (training regime)
    \end{minipage}
    \\
    % Right column with two stacked images
    \begin{minipage}[c]{\textwidth}
        \centering
        \includegraphics[trim={0cm, 0.1cm, 0cm, 0.2cm}, clip, width=\textwidth]{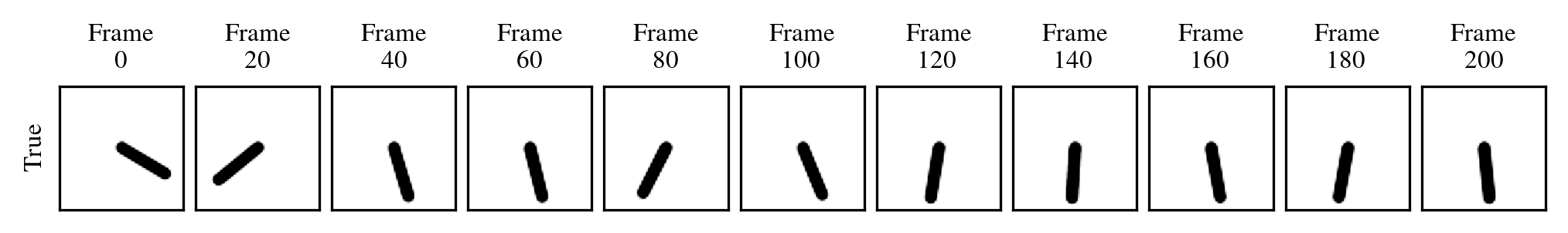}
        \\
        \includegraphics[trim={0cm, 0.1cm, 0cm, 0.8cm}, clip, width=\textwidth]{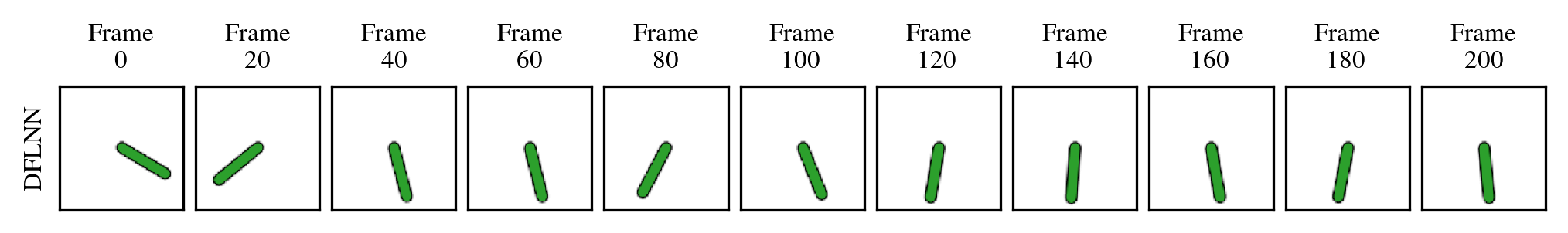}
        \
        \includegraphics[trim={0cm, 0.1cm, 0cm, 0.8cm}, clip, width=\textwidth]{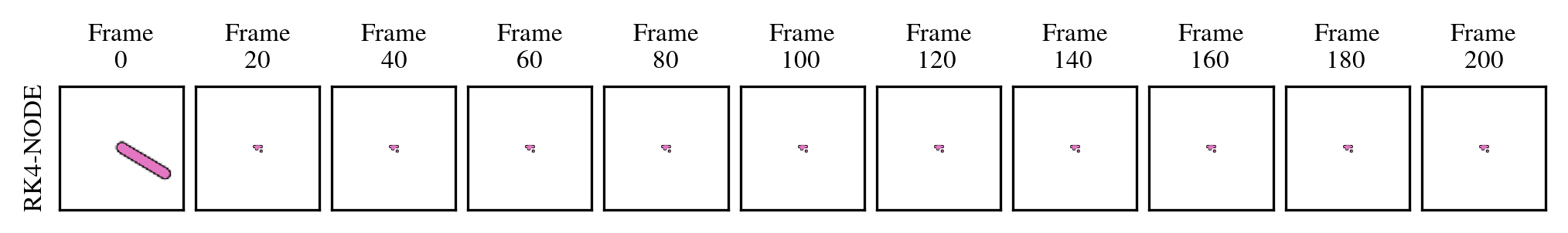}
        \\
        \includegraphics[trim={0cm, 0.1cm, 0cm, 0.8cm}, clip, width=\textwidth]{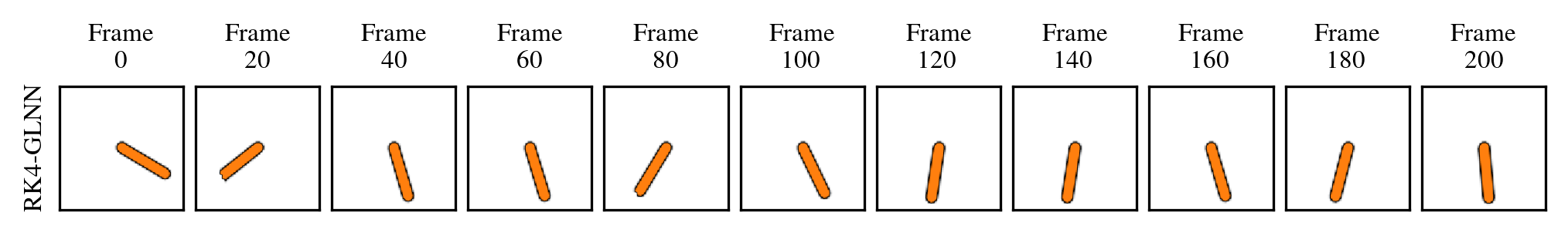}
        \subcaption{Pixel frame prediction}
    \end{minipage}
    \vspace{0.1cm}
    \\
    \begin{minipage}[c]{0.3\textwidth}
        % Top right image
        % \vspace{1.0cm}
        \centering
        \includegraphics[width=0.8\textwidth]{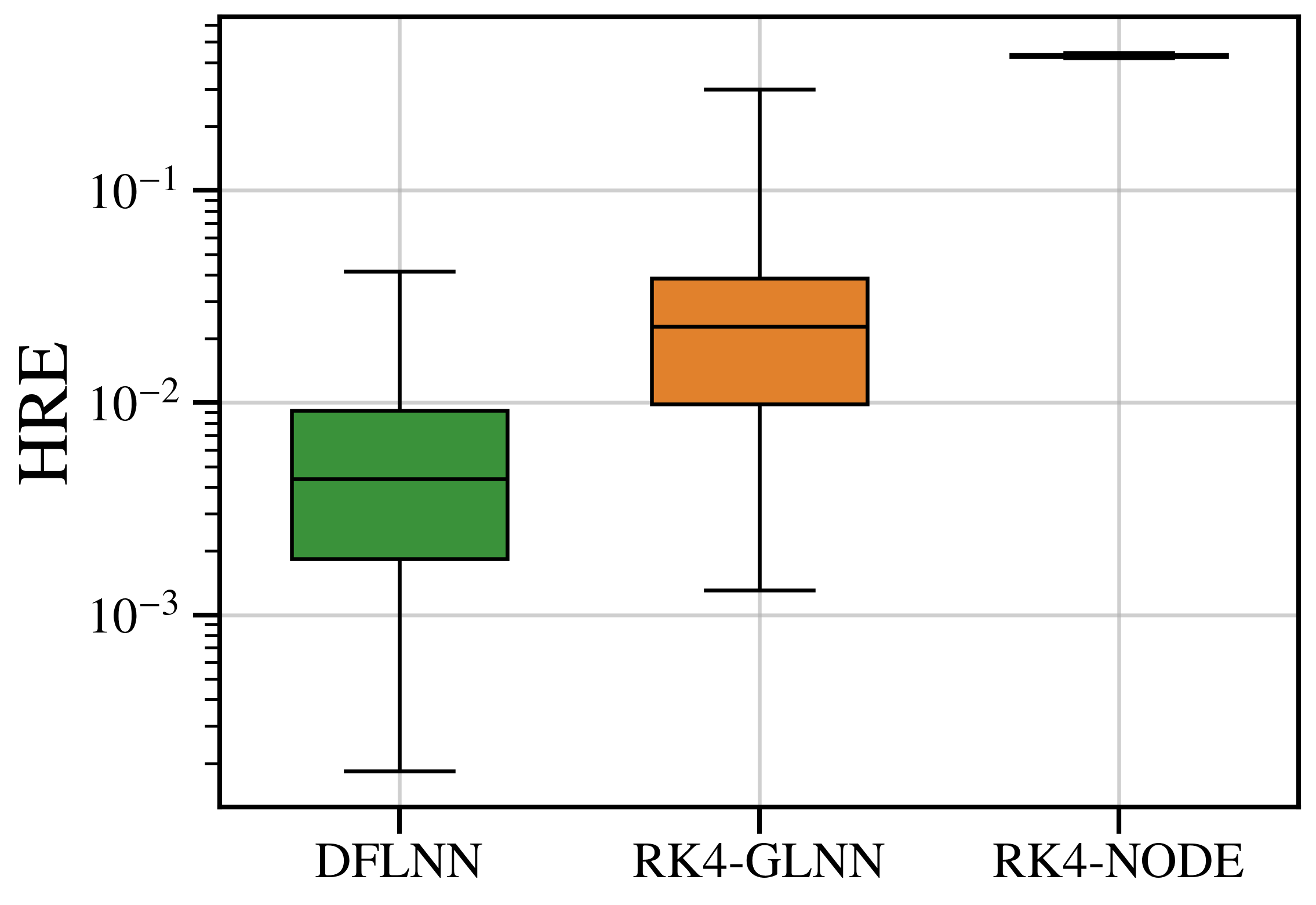}
        % \subcaption{\centering Relative error}

        \includegraphics[width=0.8\textwidth]{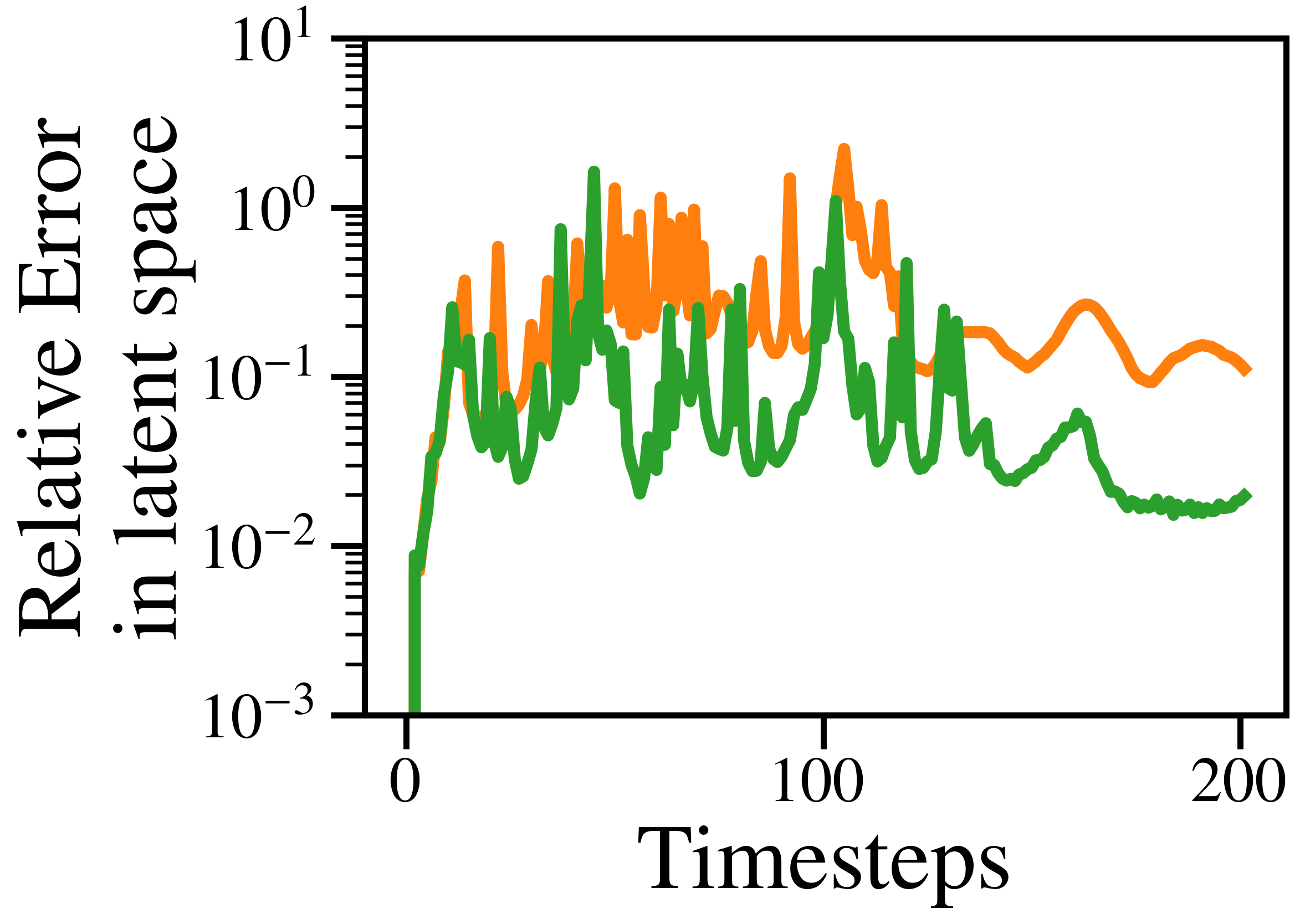}
        \subcaption{\centering Prediction error}
    \end{minipage}
    \begin{minipage}{0.6\textwidth}
        \centering
        \includegraphics[trim={0 0 0 0}, clip, width=\linewidth]{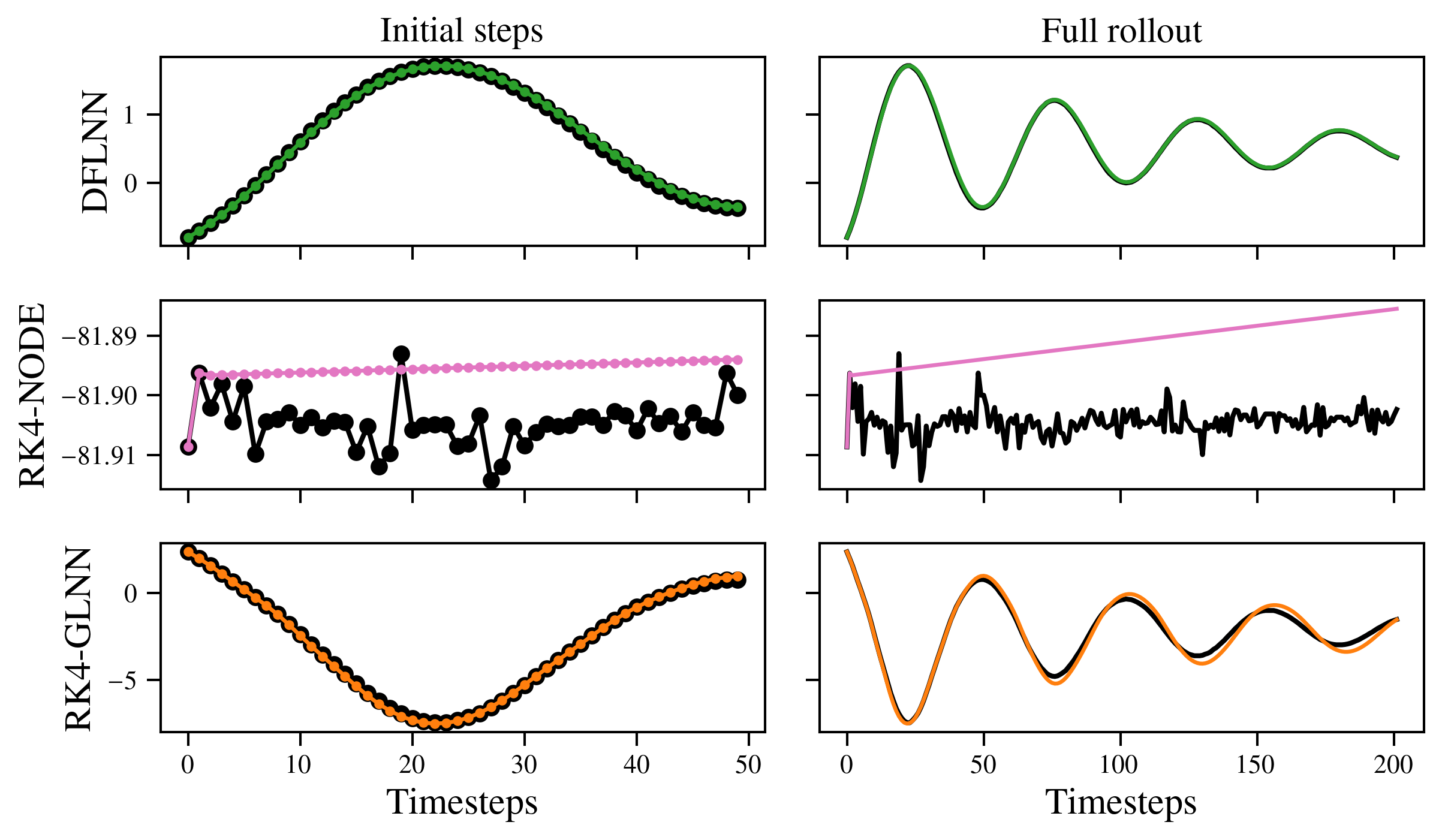}
        \subcaption{Latent projection}
    \end{minipage}

    \caption{Predictions for image sequences of a simple pendulum with damped dynamics, using conditions comparable to those in training. Green traces correspond to the DFLNN (proposed) model, orange to the RK4-GLNN model, and pink to RK4-NODE. The black curves represent the ground truth. Top panel: one sequence of predicted image frames. Bottom left: Harris relative error in image space evaluated over all predicted timesteps ($200$ forecast steps for $20$ trajectories), along with the relative error of the latent trajectories. The ground-truth latent trajectories are obtained by encoding the true pixel frames with the trained autoencoder.  Bottom right: rollout predictions in latent space. }
    \label{fig:pixel_pendulum_damped}
\end{figure}

\begin{figure}[htb!]
    \centering

    \begin{minipage}{\textwidth}
        \centering
        \scriptsize 
        Pixel frames of a pendulum: conservative dynamics\\
        (testing regime)
    \end{minipage}
    \\
    % Right column with two stacked images
    \begin{minipage}[c]{\textwidth}
        \centering
        \includegraphics[trim={0cm, 0.1cm, 0cm, 0.2cm}, clip, width=\textwidth]{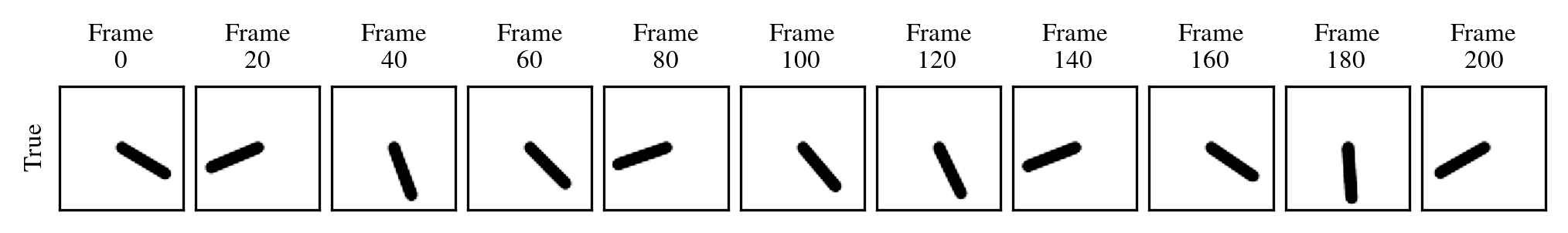}
        \\
        \includegraphics[trim={0cm, 0.1cm, 0cm, 0.8cm}, clip, width=\textwidth]{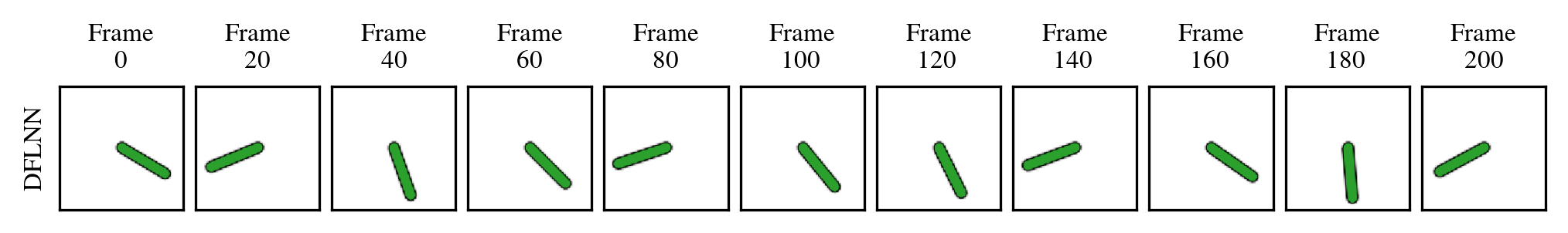}
        \\
        \includegraphics[trim={0cm, 0.1cm, 0cm, 0.8cm}, clip, width=\textwidth]{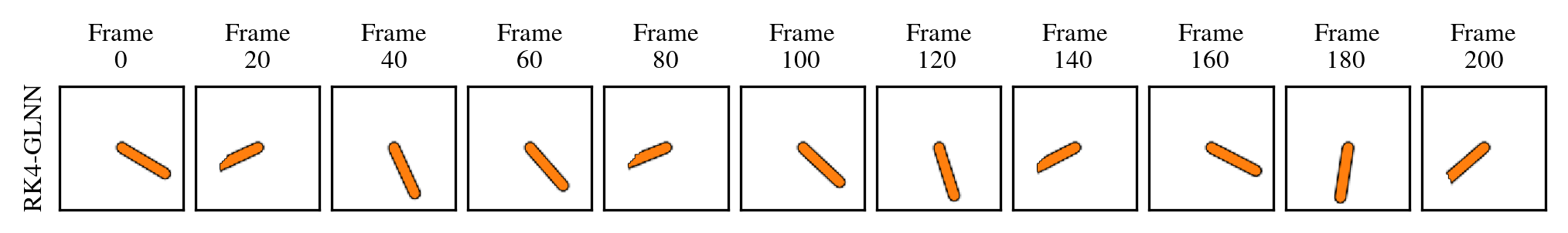}
        \subcaption{Pixel frame prediction}
    \end{minipage}
    \vspace{0.1cm}
    \\
    \begin{minipage}[c]{0.3\textwidth}
        % Top right image
        % \vspace{1.0cm}
        \centering
        \includegraphics[width=\textwidth]{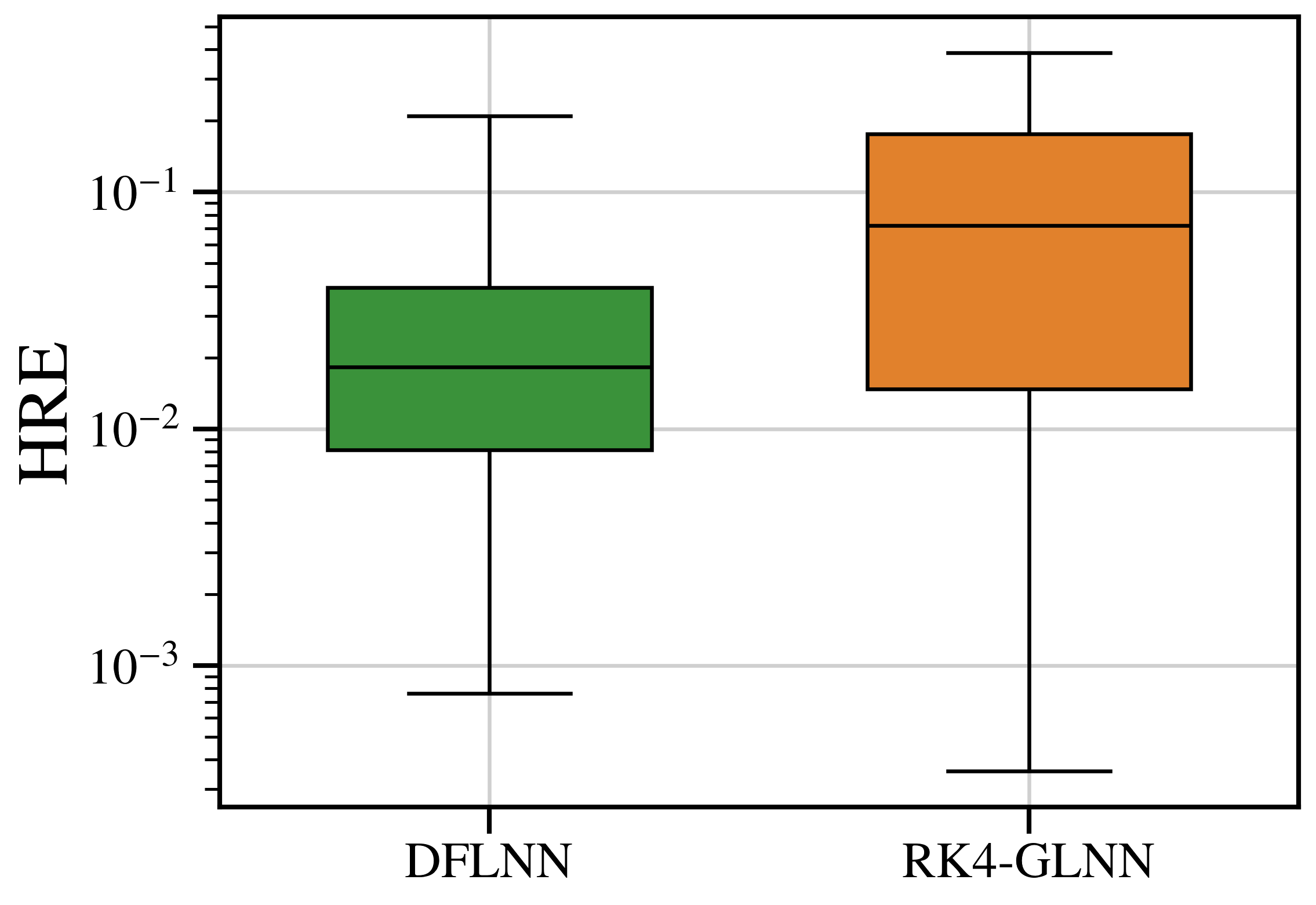}
        
        \includegraphics[width=0.8\textwidth]{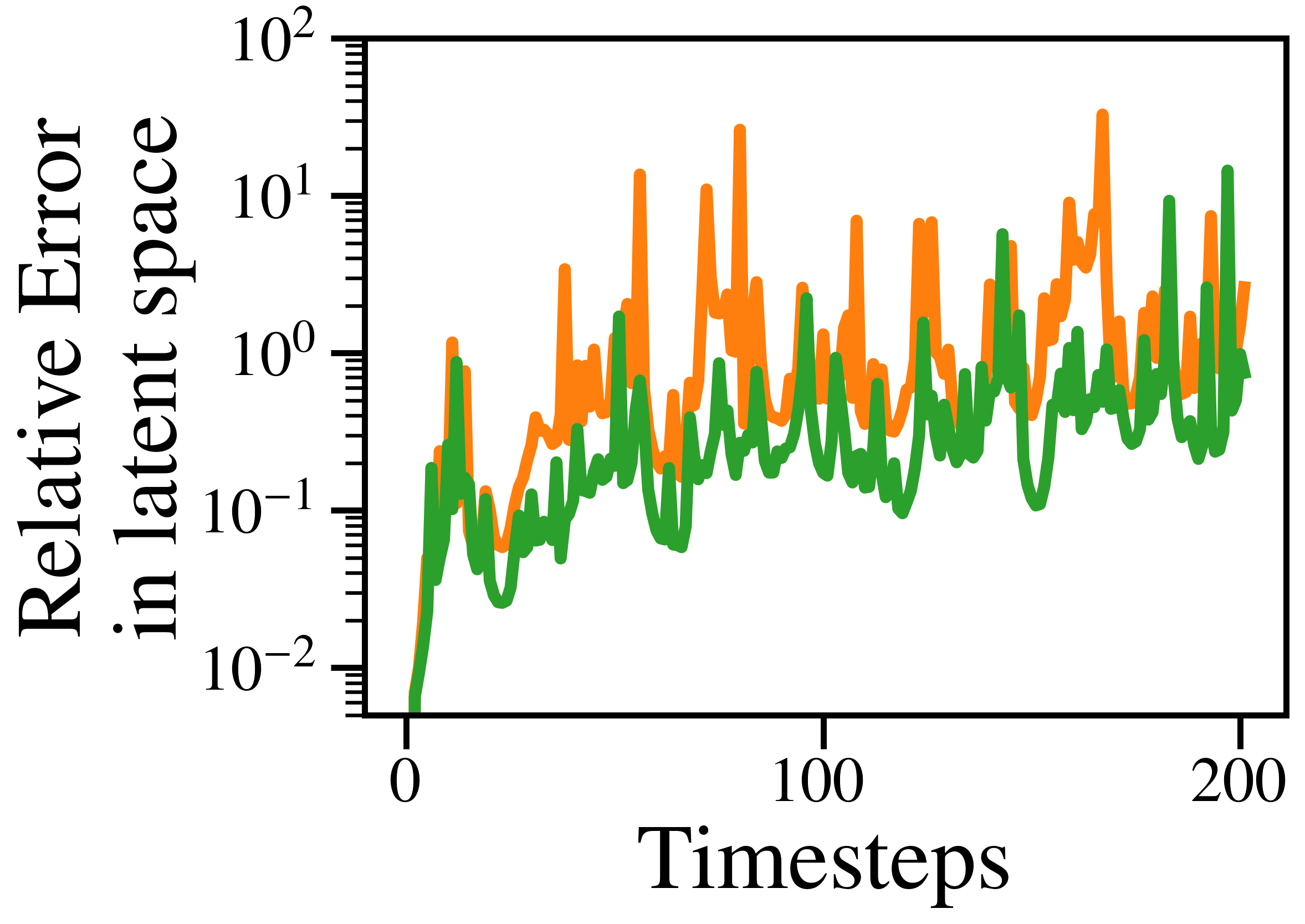}
        \subcaption{\centering Prediction error}
    \end{minipage}
    \begin{minipage}{0.6\textwidth}
        \centering
        \includegraphics[trim={0 0 0 0}, clip, width=\linewidth]{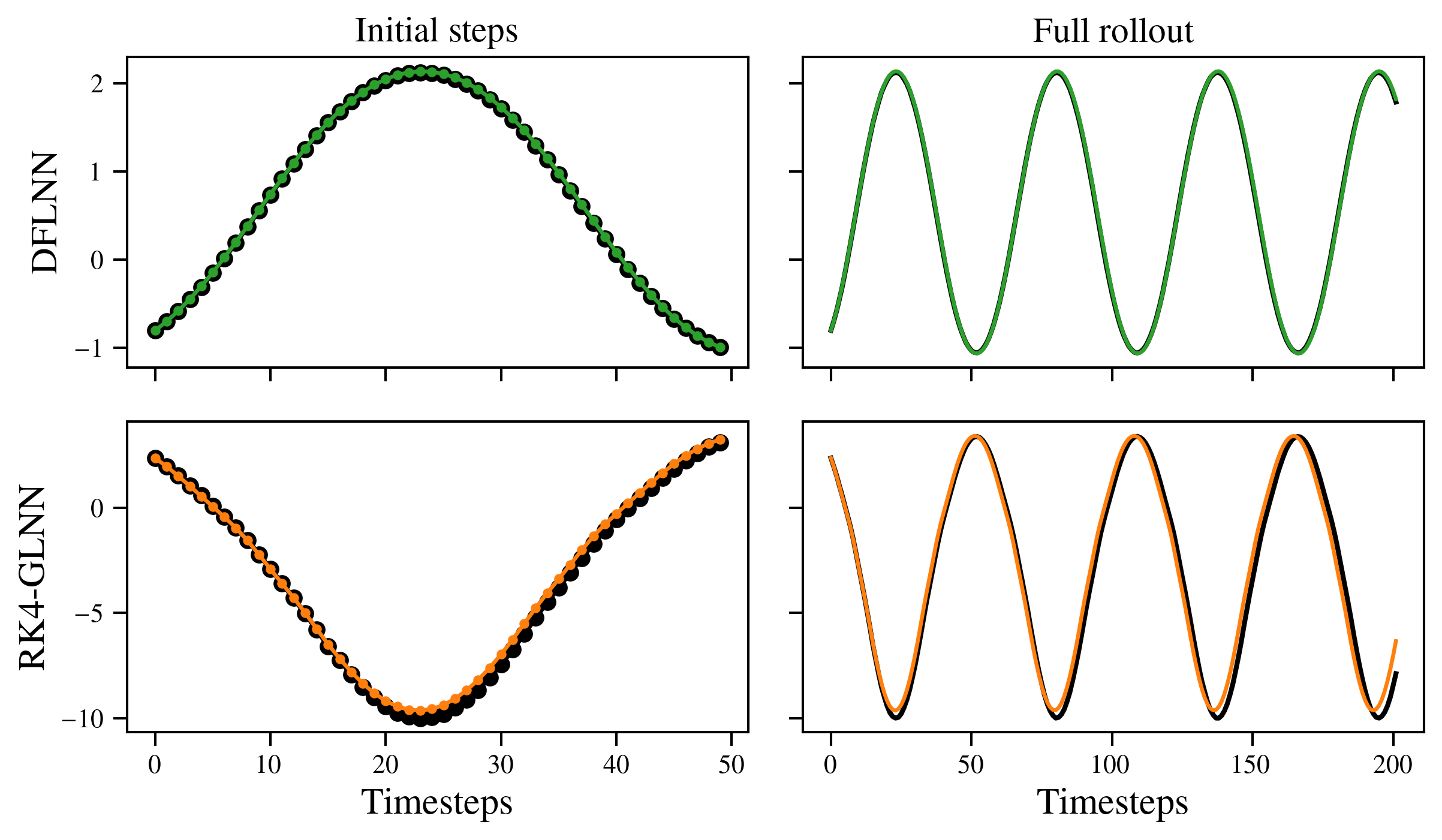}
        \subcaption{Latent projection}
    \end{minipage}

    \caption{Predicted image sequences for a simple pendulum under conservative, undamped dynamics. Testing regime where we have set the learned force to zero, as explained in the end of Section~\ref{sc:TheForceTerm}. Green curves correspond to the DFLNN (proposed) model, and orange to the RK4-GLNN model. Black curves represent the ground truth. Top panel: one sequence of predicted image frames. Bottom left: Harris relative error in image space evaluated over all predicted timesteps ($200$ forecast steps for $20$ trajectories), along with the relative error of the latent trajectories. The ground-truth latent trajectories are obtained by encoding the true pixel frames with the trained autoencoder. Bottom right: rollout predictions in latent space. 
    }
    \label{fig:pixel_pendulum_undamped}
\end{figure}

The use of an autoencoder is essential in this setting, converting the data format from $56 \times 56$ pixel frame to a reduced latent space in which the model can learn the dynamics; see Section~\ref{sc:Autoencoder}. When assessing the model’s precision on the validation dataset, we use a Harris-based Relative Error (HRE)~\cite{HarrisCorner1988} metric to extract structural features and prevent the large white background of each image from dominating the validation metric~\cite{wang2004image}, see~\ref{apx:HarrisRelativeError} for a more detailed description.

In Figure~\ref{fig:pixel_pendulum_damped}, we show the relative error over $200$ time steps for $20$ unseen trajectories, along with an illustrative rollout in the damped regime, consistent with the training setup.
Since performance depends both on accurately capturing the dynamics and on the autoencoder’s ability to embed pixel frames into a meaningful latent space, we also evaluate and compare rollouts directly in latent space. The latent trajectory for the ground-truth data is obtained by passing the true pixel frames through the trained autoencoder and plotting the resulting latent representations. Inspecting these trajectories, we can quickly see that the latent space learned by RK4-NODE is largely unstructured, so rollouts predicted in this latent space fail to produce meaningful decoded image sequences. In contrast, both Lagrangian-based models successfully co-train with the autoencoder and induce structure in the latent space. Moreover, from the latent trajectories we observe that the proposed DFLNN model tracks the true latent dynamics more accurately than the baseline models.

As in the previous experiments, we set the learned force to zero---as explained in the end of Section~\ref{sc:TheForceTerm}---to evaluate the learned Lagrangian. Figure~\ref{fig:pixel_pendulum_undamped} shows that the conservative dynamics is correctly captured by the Lagrangian even in a learned latent space.

\begin{figure}[htb!]
    \centering

    \begin{minipage}{\textwidth}
        \centering
        \scriptsize 
        Pixel frames of a pendulum: Evaluate in pixel space
    \end{minipage}

    \begin{minipage}{0.45\textwidth}
        \centering
        \hspace{0.03\textwidth} 
        \scriptsize  Damped dynamics \\ (training regime)
    \end{minipage}%
    \hfill
    \begin{minipage}{0.35\textwidth}
        \centering
        \hspace{0.03\textwidth} 
        \scriptsize Conservative dynamics \\ (testing regime)
    \end{minipage}
    \hspace{6cm}
    \\
    \vspace{-0.05cm}
    \begin{minipage}[c]{0.45\textwidth}
        \centering
        \includegraphics[trim={0 0 0 0}, clip, width=\linewidth]{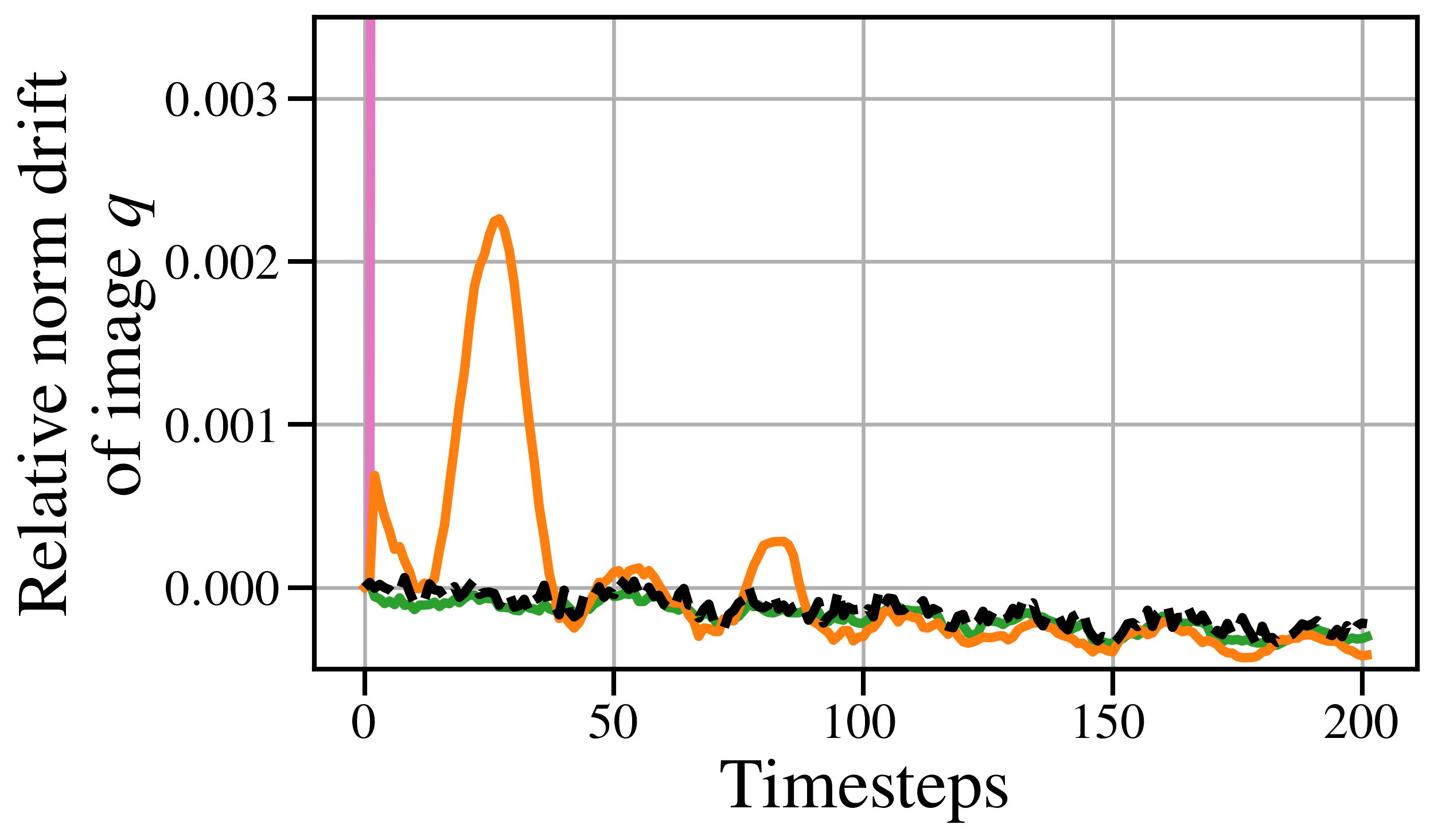}
    \end{minipage}
    \hfill
    \begin{minipage}[c]{0.45\textwidth}
        \centering
        \includegraphics[trim={0cm 0 0 0}, clip, width=\linewidth]{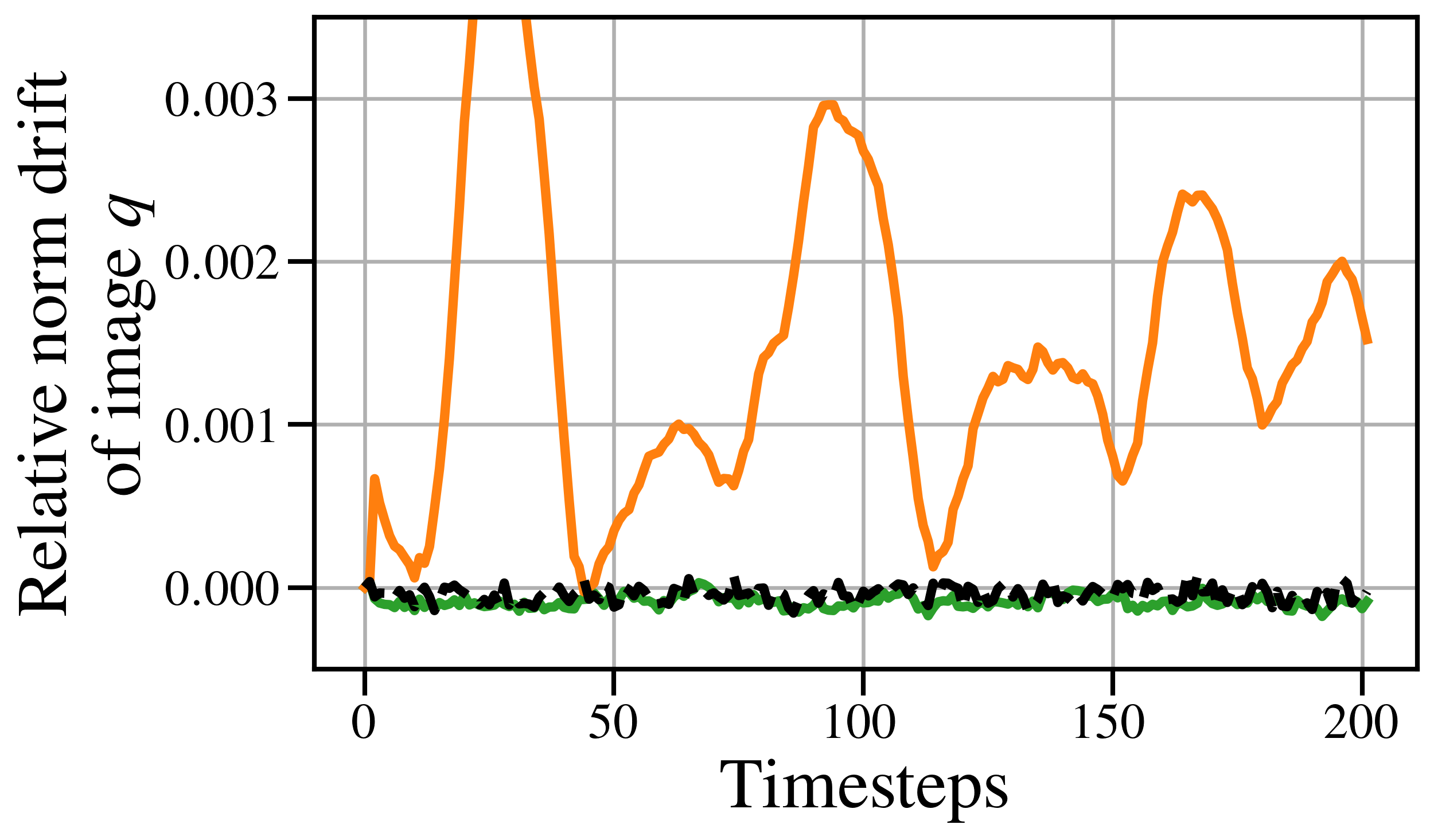}
    \end{minipage}
    \hspace{1cm}
    \\
    %\vspace{-10cm}
    \begin{minipage}{0.45\textwidth}
        \centering
        \subcaption{}
    \end{minipage}%
    \hfill
    \begin{minipage}{0.45\textwidth}
        \centering
        \subcaption{}
    \end{minipage}
    
    \caption{Mean relative drift of the frame norm, $(\|q_t\| - \|q_0\|)/\|q_0\|$, of the predicted pixel sequences for $200$ time steps, averaged over $20$ test trajectories. (a) Damped regime, matching the training conditions. (b) Conservative regime, obtained by setting the force to zero. Green curves represent DFLNN (proposed), orange the RK4-GLNN model, and pink the RK4-NODE model, while the ground truth is shown as a black dashed line. The DFLNN prediction tracks the ground-truth norm drift closely in both regimes, whereas RK4-GLNN shows large deviations. RK4-NODE is shown only in the damped regime, since the force is not captured in a separate term.}

    \label{fig:PP_pixel_normdiff_results}
\end{figure}
As an additional quantitative validation in image space, we monitor the mean relative drift of the frame norm, $(\|q_t\| - \|q_0\|)/\|q_0\|$, averaged across $20$ previously unseen trajectories over the rollout (Figure~\ref{fig:PP_pixel_normdiff_results}). In both the damped and the manipulated conservative regime, the DFLNN prediction (green) matches the ground-truth norm drift (black) closely throughout the full $200$ steps-step horizon, while RK4-GLNN (orange) shows substantial deviations in the frame norm. This suggests that, even if the RK4-GLNN latent trajectory appears plausible, its rollout may drift into states outside the region the autoencoder encountered during training. In contrast, DFLNN produces latent trajectories that remain sufficiently aligned with the true image representation, preventing these issues for the autoencoder.

\subsection{Task 4 (a-b): Human Motion Capture}\label{sc:Task4}

\begin{figure}[htb!]
    \centering

    \begin{minipage}{\textwidth}
        \centering
        \scriptsize 
        Human swinging from a bar
    \end{minipage}

    % Trial 2
    \begin{minipage}[c]{0.14\textwidth}
        % Top right image
        % \vspace{1.0cm}
        \centering
        \includegraphics[width=\textwidth]{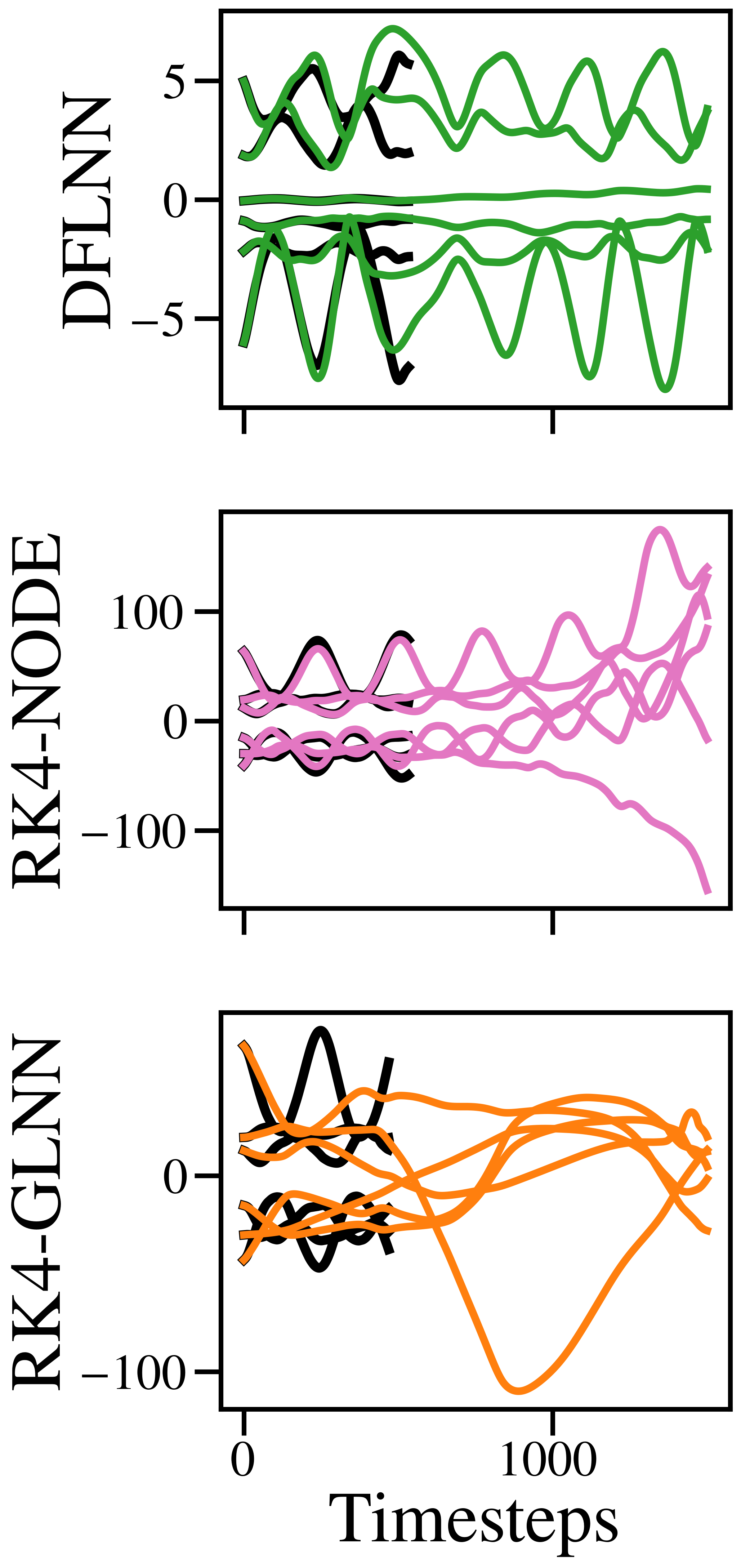}
        \subcaption*{\centering \hspace{0.4cm} Latent\\ \hspace{0.4cm} projection}
    \end{minipage}
    \hfil
    % Right column with two stacked images
    \begin{minipage}[c]{0.85\textwidth}
        \centering
        \includegraphics[trim={0cm, 0.5cm, 0cm, 0.2cm}, clip, width=\textwidth]{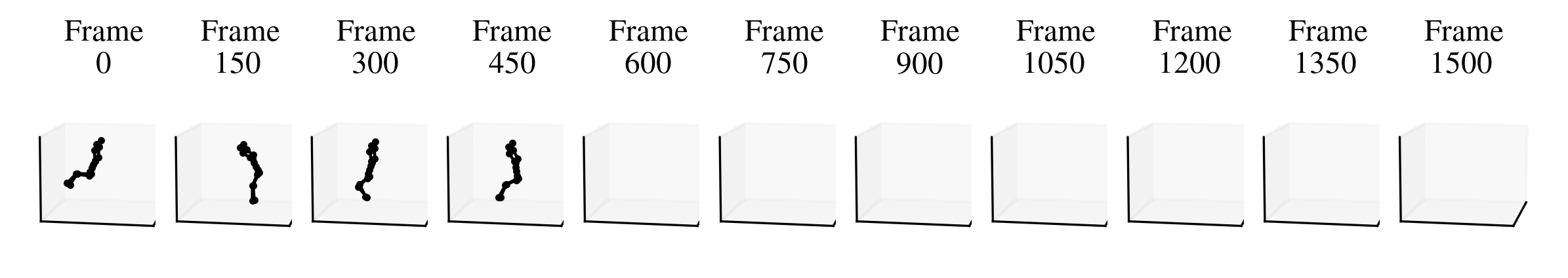}
        \\
        \includegraphics[trim={0cm, 0.5cm, 0cm, 1.6cm}, clip, width=\textwidth]{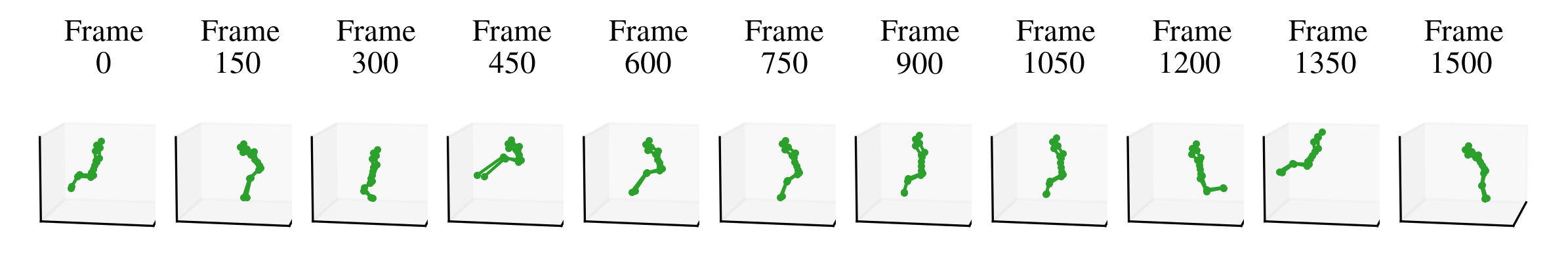}
        \
        \includegraphics[trim={0cm, 0.5cm, 0cm, 1.6cm}, clip, width=\textwidth]{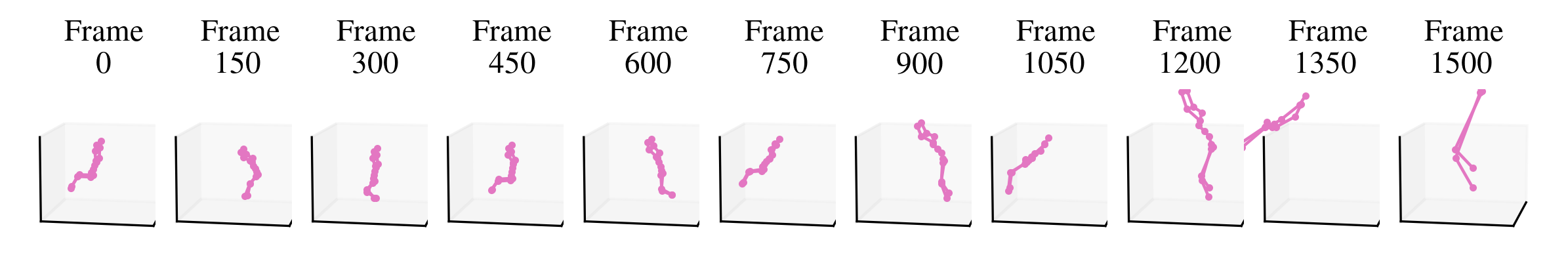}
        \\
        \includegraphics[trim={0cm, 0.5cm, 0cm, 1.6cm}, clip, width=\textwidth]{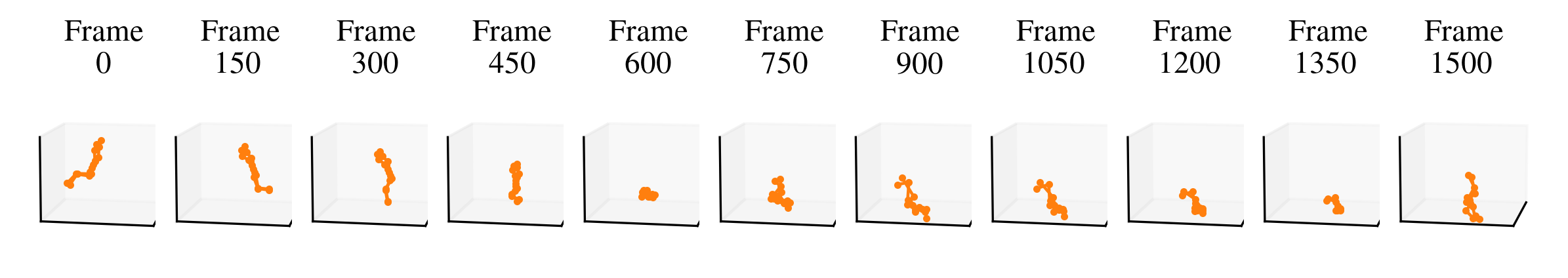}
        \subcaption*{Motion prediction}
    \end{minipage}
    \begin{minipage}{\textwidth}
        \centering
        \vspace{-0.2cm}
        \subcaption{Trial $2$}
    \end{minipage}
    \\
    \vspace{0.1cm}

    % Trial 3
    \begin{minipage}[c]{0.14\textwidth}
        % Top right image
        % \vspace{1.0cm}
        \centering
        \includegraphics[width=\textwidth]{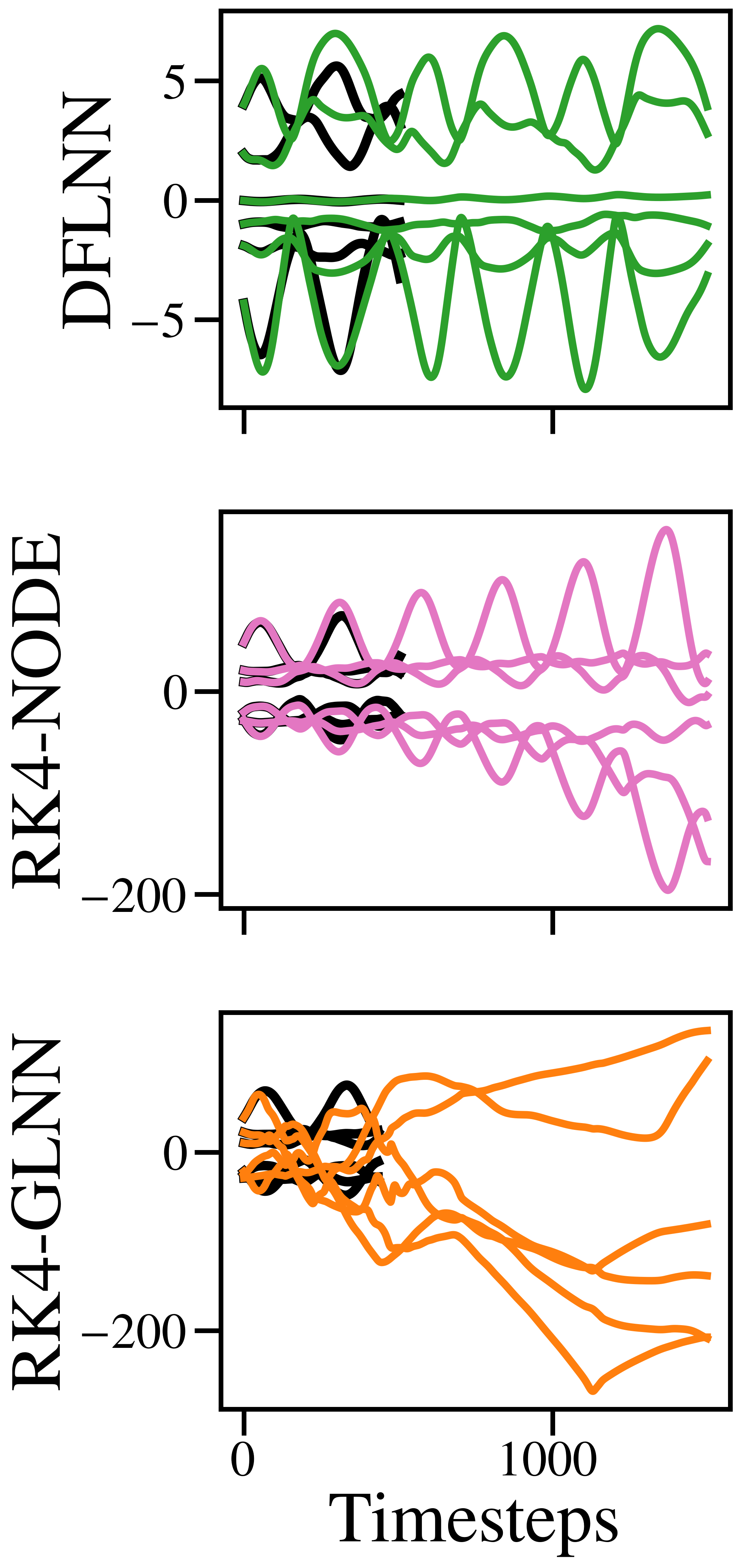}
        \subcaption*{\centering \hspace{0.4cm} Latent\\ \hspace{0.4cm} projection}
    \end{minipage}
    \hfill
    % Right column with two stacked images
    \begin{minipage}[c]{0.85\textwidth}
        \centering
        \includegraphics[trim={0cm, 0.5cm, 0cm, 0.2cm}, clip, width=\textwidth]{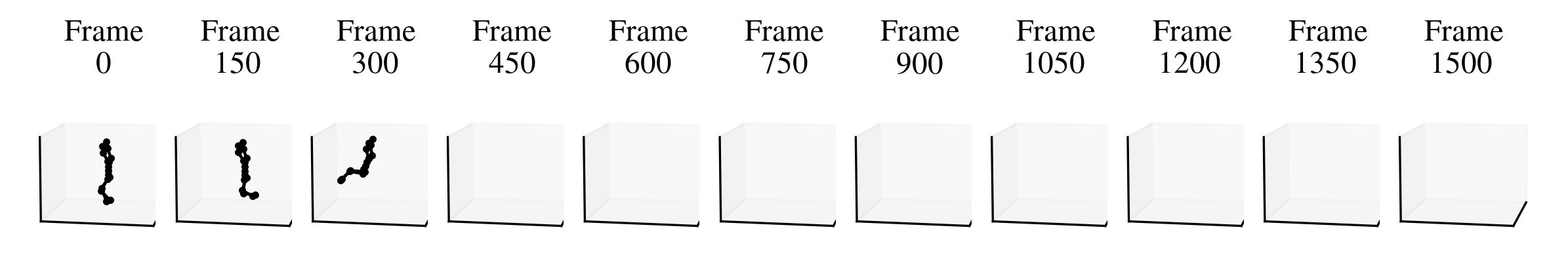}
        \\
        \includegraphics[trim={0cm, 0.5cm, 0cm, 1.6cm}, clip, width=\textwidth]{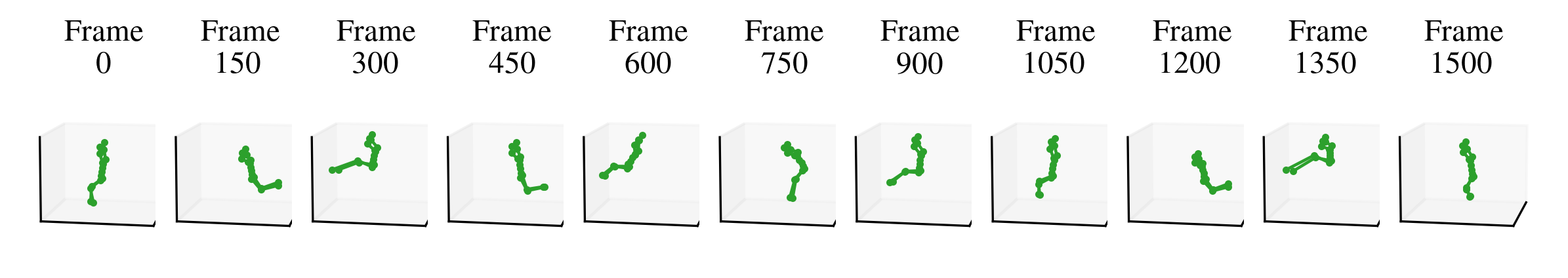}
        \
        \includegraphics[trim={0cm, 0.5cm, 0cm, 1.6cm}, clip, width=\textwidth]{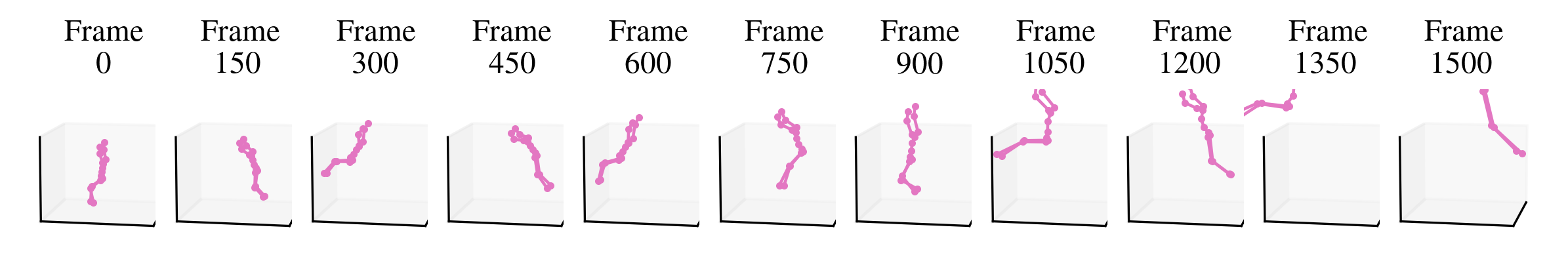}
        \\
        \includegraphics[trim={0cm, 0.5cm, 0cm, 1.6cm}, clip, width=\textwidth]{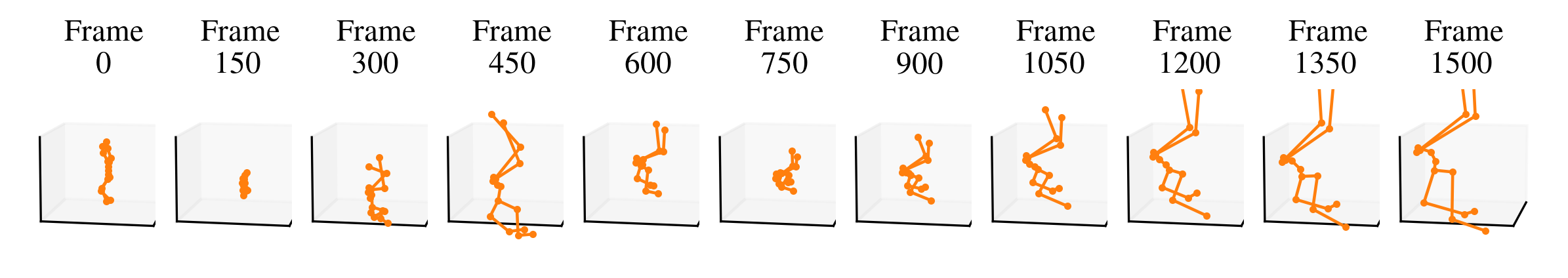}
        \subcaption*{Motion prediction}
    \end{minipage}
    \begin{minipage}{\textwidth}
        \centering
        \vspace{-0.2cm}
        \subcaption{Trial $3$}
    \end{minipage}

    \caption{Reconstruction and extrapolation beyond the training data for a subject swinging from a bar. Green trajectories represent DFLNN (proposed), orange denote the RK4-GLNN model, and pink indicate RK4-NODE, while the ground truth is plotted in black. Left: rollout predictions in the latent space. Right: corresponding decoded motion predictions in the original motion space.}
    \label{fig:motion_capture_swinging}
\end{figure}

\begin{figure}[htb!]
    \centering

    \begin{minipage}{\textwidth}
        \centering
        \scriptsize 
        Human motion running
    \end{minipage}
    
    \begin{minipage}[c]{0.8\textwidth}
        \centering
        \includegraphics[trim={0cm, 0.3cm, 0cm, 0cm}, clip, width=\textwidth]{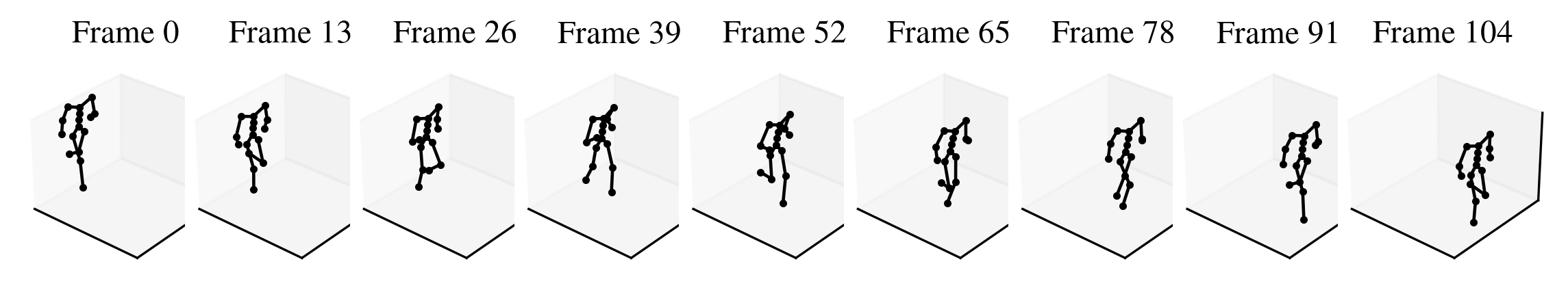}
        \\
        \includegraphics[trim={0cm, 0.3cm, 0cm, 0.8cm}, clip, width=\textwidth]{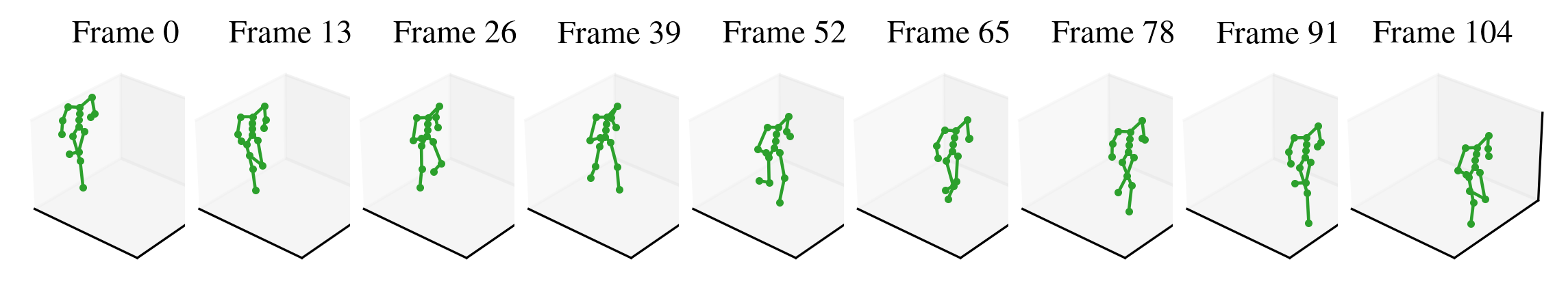}
        \\
        \includegraphics[trim={0cm, 0.3cm, 0cm, 0.8cm}, clip, width=\textwidth]{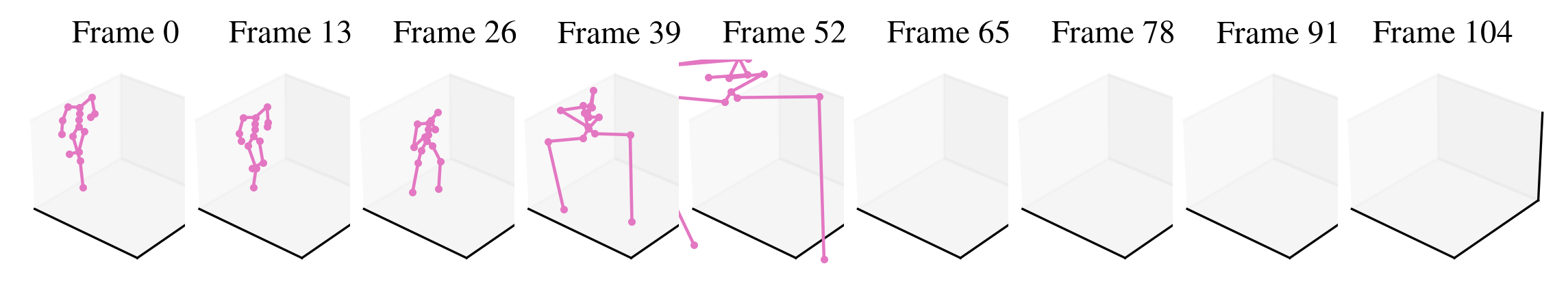}
        \\
        \includegraphics[trim={0cm, 0.3cm, 0cm, 0.8cm}, clip, width=\textwidth]{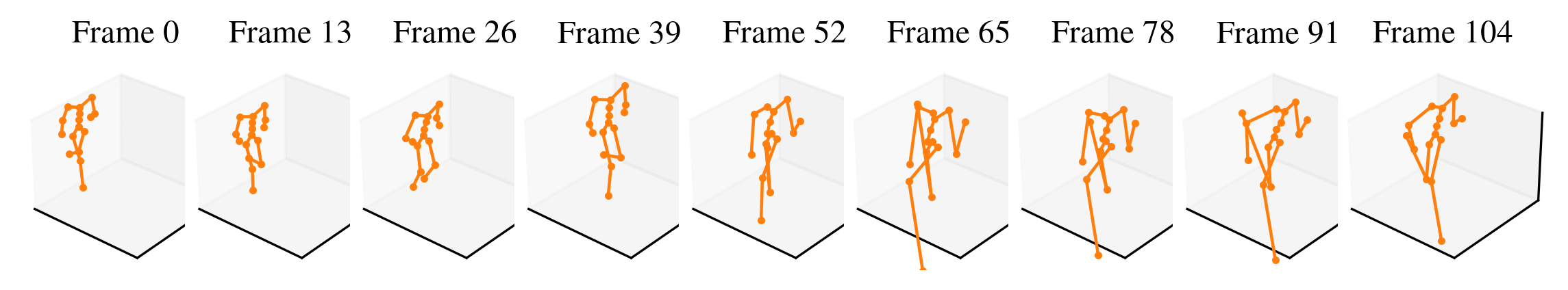}
        \subcaption{Motion prediction}
    \end{minipage}
    \hfill
    % Right column with two stacked images
    \begin{minipage}[c]{0.18\textwidth}
        % Top right image
        % \vspace{-0.8cm}
        \centering
        \includegraphics[width=\textwidth]{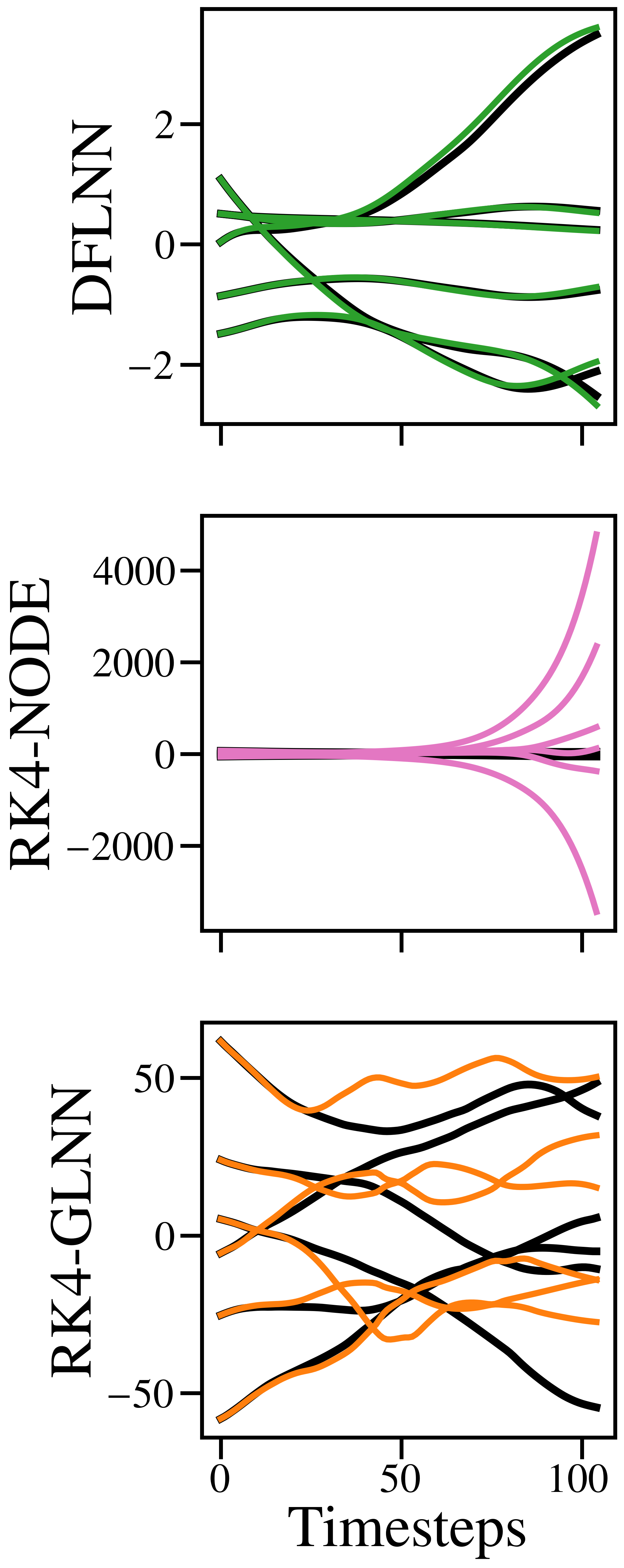}
        \subcaption{\centering Latent\\ \hspace{0.4cm} projection}
    \end{minipage}\\
    \vspace{0.5cm}
    \begin{minipage}[c]{0.5\textwidth}
        % Top right image
        % \vspace{-0.8cm}
        \centering
        \includegraphics[width=\textwidth]{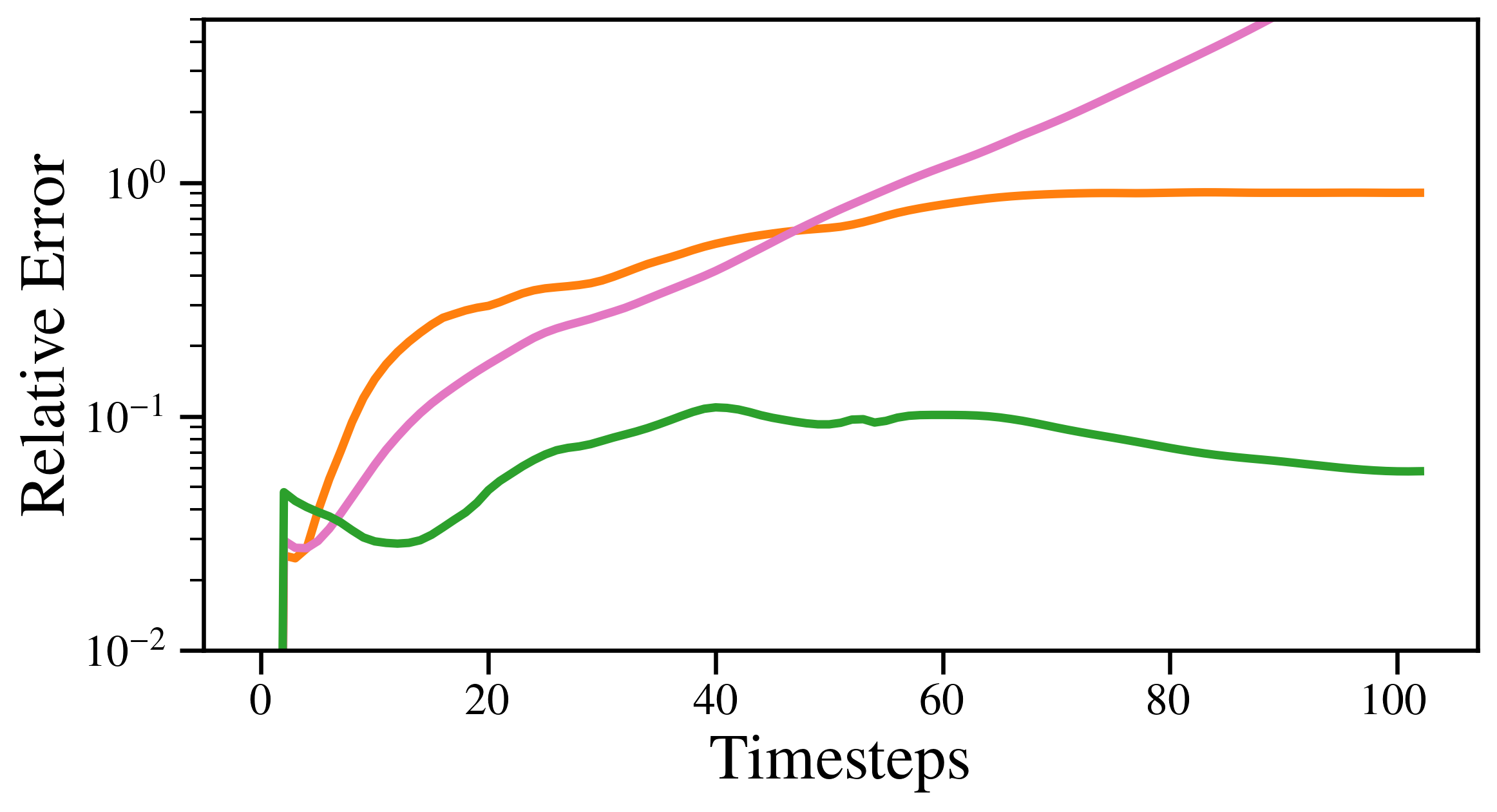}
        \subcaption{Relative error}
    \end{minipage}\\
    
    \caption{Predicted outcome for an unseen running instance (trial $10$) of human running motion. Green curves correspond to DFLNN (proposed), orange to the RK4-GLNN model, and pink to RK4-NODE. The ground truth is shown in black. The top-left panel illustrates the complete motion as a skeletal representation. The top-right panel presents the latent rollout prior to decoding with the autoencoder. The bottom panel shows the relative error.}
    \label{fig:motion_capture_running}
\end{figure}

We evaluate our method on real-world data of human motion, using motion capture recordings from the CMU Graphics Lab Motion Capture Database~\cite{CMU-Database, de2009guide}. We consider the motions of 16 linked joints on human body---from the tibia to the radius---including \texttt{rtibia}, \texttt{ltibia}, \texttt{rfemur}, \texttt{lfemur}, \texttt{rhipjoint}, \texttt{lhipjoint}, \texttt{root}, \texttt{lowerback}, \texttt{upperback}, \texttt{thorax}, \texttt{rclavicle}, \texttt{lclavicle}, \texttt{rhumerus}, \texttt{lhumerus}, \texttt{rradius} and \texttt{lradius}. Joint angles, recorded at $120$~Hz relative to a fixed reference, are converted to Cartesian coordinates and smoothed using a Savitzky-Golay filter~\cite{savitzky_smoothing_1964}.
    
We hypothesize that a person body behaves similarly to a multi-pendulum system and aim to identify a Lagrangian with mechanical structure with potential $U(q)$. Due to the likely low intrinsic dimensionality of the motion, we apply an autoencoder to reduce the observed dimension to $l=6$. The external force is assumed to comprise both frictional dissipation and active energy input from the subject. Accordingly, we model it as the sum of a nonlinear dissipative term plus a neural network $F_\theta^\text{Free}$ to model the forces exerted by the human motion $F_\theta = - K_\theta(q)\dot{q} + F_\theta^\text{Free}$.

\subsubsection{Task 4a: Human swinging from a bar}

From the CMU dataset, we use two recordings of a person swinging from a bar (subject $43$, trials $2$ and $3$). For each recording, we train on the first $80\%$ of the timesteps and validate on the remaining $20\%$. In Figure~\ref{fig:motion_capture_swinging}, we assess model performance both in reconstructing the training trajectories and in rolling out beyond the training horizon. The results show that all models track the motion for a few time steps; RK4-GLNN quickly drifts away, whereas RK4-NODE and DFLNN closely follow the motion over the entire training interval. However, when extrapolating past the training window (after roughly $500$ frames), only the DFLNN model continues to yield plausible trajectories, while the baseline models diverge much earlier.

\subsubsection{Task 4b: Human running}

Next we consider a human running. The CMU dataset includes eleven recordings (subject $9$, trials $1-11$), in which we train on the first $8$ samples (validate on trials $1-2$ and train on trials $3-8$) and reserve the last $3$ samples for testing (trials $9-11$).
 
The results are shown in Figure~\ref{fig:motion_capture_running}, which reports the mean relative error over the three test trials. We additionally display an example prediction for sample $10$ from the test set. It can be seen that all models initially continue the running motion for the first few timesteps. However, both RK4-NODE and RK4-GLNN rapidly diverge from the true trajectory, whereas the proposed model successfully follows the complete motion.

%% CONCLUSION
\section{Conclusions and future work}\label{sc:Conclusion}
In this work, we introduced the Discrete Forced Lagrangian Neural Network (DFLNN), a structure-preserving framework for learning non-conservative dynamics directly from discrete position data. By grounding our architecture in the discrete Lagrange-d'Alembert principle, we circumvent the need for accurate velocity estimates while ensuring that the learned model adheres to fundamental physical laws. 

Our results demonstrate that DFLNN offers significant advantages over both unstructured baselines and continuous-time physics-informed models that rely on finite-difference approximations. We showcased the model's versatility by effectively disentangling conservative and dissipative forces in the latent space of high-dimensional video data and by achieving state-of-the-art long-term prediction on real-world human motion datasets.

Looking ahead, several promising directions emerge. First, extending this framework to control settings could allow for energy-efficient planning in robotics. Second, incorporating uncertainty quantification would further robustify the model for medical and biomechanical applications where data is often noisy or sparse. Third, a systematic study of learning from irregularly sampled position data is an interesting and challenging avenue for future research. Finally, exploring the application of DFLNN to graph-based representations could unlock new capabilities for modeling complex, interacting multi-body systems.

\section*{CRediT authorship contribution statement}
\textbf{Martine Dyring Hansen}: Writing – original draft, Visualization, Validation, Software, Methodology, Investigation, Formal analysis, Data curation; \textbf{Elena Celledoni}: Writing – review and editing, Supervision, Resources, Funding acquisition, Conceptualization; \textbf{Benjamin Kwanen Tapley}: Writing – review and editing, Supervision, Resources, Funding acquisition, Conceptualization.

% Corresponding authors are required to acknowledge co-author contributions using CRediT (Contributor Roles Taxonomy) roles:
% Conceptualization
% Data curation
% Formal analysis 
% Funding acquisition
% Investigation
% Methodology
% Project administration
% Resources
% Software
% Supervision
% Validation
% Visualization
% Writing – original draft
% Writing – review and editing
%% Not all CRediT roles will apply to every manuscript and some authors may contribute through multiple roles.

\section*{Declaration of competing interest}
The authors declare that they have no known competing financial interests or personal relationships that could have appeared to
influence the work reported in this paper.

\section*{Acknowledgments}
The authors would like to express gratitude to Marta Ghirardelli, Gijs Bellaard, Nicky van den Berg, Lars Ruthotto, Molei Tao, and Christian Offen for their valuable discussions during the preparation of this paper.
This research was supported by the Research Council of Norway, through the project PhysML (No. 338779), and EU through MSCA-SE: REMODEL (Project ID: 101131557). The data used in this project was obtained from mocap.cs.cmu.edu. The database was created with funding from NSF EIA-0196217.

\appendix
\section{Baseline: Adaptive Generalised Lagrangian Neural Networks (adapted GLNN)}
\label{apx:GLNN}

 We briefly describe here the approach of Xiao et al. \cite{xiao_generalized_2024}, see also \cite{CranmerLagrangian2020}. This methodology requires both position and velocity data. Starting from \eqref{eq:ForcedEulerLagrange}, by expanding the term $\frac{d}{dt}(\frac{\partial L}{\partial \dot{q}}(q,\dot{q}))$ differentiating with respect to $t$, and assuming $\frac{\partial^2 L}{\partial \dot{q}^2}$ to be invertible, the authors derive the following explicit expression of the forced Euler-Lagrange equations 
 \begin{equation}
 \label{EL-GLNN}
 \ddot{q}=\left(\frac{\partial^2 L}{\partial \dot{q}^2}\right)^{-1}(\frac{\partial L}{\partial q}-\dot{q}\cdot \frac{\partial^2 L}{\partial q\partial\dot{q}}+F),
 \end{equation}
 then use a numerical integration to approximate $(q,\dot{q})\approx (\hat{q},\hat{\dot{q}})$ at the times where the data is observed, and obtain a loss function as the mean of terms of the form $\|q-\hat{q}\|^2+\|\dot{q}-\hat{\dot{q}}\|^2$. In the absence of the external force $F$, the method reduces to the Lagrangian Neural Network (LNN) \cite{CranmerLagrangian2020}. Both approaches rely on $\frac{\partial^2 L}{\partial \dot{q}^2}$ being at least locally invertible resulting in a so-called regular Lagrangian $L$, \cite[p. 379]{Marsden_West_2001}.  However, no specific technique for regularizing the Lagrangian is described in the paper.
   A notable drawback of this method is its reliance on instantaneous position and velocity data.

Without access to observed velocities, we use their midpoint approximation. Denoting with $x(t):=[q(t), \dot{q}(t)]$, we approximate $x(t_{i+\frac{1}{2}})$ using $(q_i,q_{i+1})$: 
$$x(t_{i+\frac{1}{2}})\approx \bar{x}(q_{i}, q_{i+1}),\qquad \bar{x}(q_{i}, q_{i+1}) := [\bar{q}_{i+\frac{1}{2}},\bar{\dot{q}}_{i+\frac{1}{2}}],$$ here we use  
 the definitions of Section~\ref{Learning-Physics} for $\bar{q}_{i+\frac{1}{2}},\bar{\dot{q}}_{i+\frac{1}{2}}$. 
 
The loss function we use is
    \begin{equation}\label{eq:loss-baseline}
        \mathcal{L} = \frac{1}{N_\mathcal{T}(N-1)}\sum_{\mathcal{T}} \sum_{n=1}^{N-1} \| \hat{\bar{x}}(q_{n-1}, q_{n}) - \bar{x}(q_{n}, q_{n+1}) \|_2^2,
    \end{equation}
where $\hat{\bar{x}}(q_{n-1}, q_{n})$ is the prediction to the model. 

One significant limitation of this approach lies in the requirement to compute the inverse of the Hessian $\left(\frac{\partial^2L}{\partial\dot{q}^2}\right)^{-1}$ for each data point, as noted by \cite{xiao_generalized_2024}. This process is significantly resource-intensive. In DFLNN, the Hessian is only necessary to evaluate when regularizing the model, thereby avoiding the need to determine the Hessian at every data point. This offers a notable computational advantage. Furthermore, the GLNN requires that the Hessian matrix is invertible at all points to avoid invalid computations, which makes the training process challenging.

We remark that applying a symplectic integrator to the second order equations \eqref{EL-GLNN} written as a first order system, does not necessarily amount to a syplectic discretization of the corresponding Hamiltonian equations.
Assuming, for example, to use an implicit midpoint method to integrate the second order equations \eqref{EL-GLNN}, and comparing with the midpoint integration of the corresponding Hamiltonian equations (which is equivalent to our method as shown in \cite[p. 208, example 6.3]{HairerLubichWanner2006}), one can easily see that the two methods are not equivalent unless we also require $\left(\frac{\partial^2L}{\partial\dot{q}^2}\right)$ is a constant matrix. In particular, when $F=0$, while our method is symplectic the the other method isn't.

\begin{figure}[htb!]
    \centering
    \includegraphics[width=0.9\linewidth]{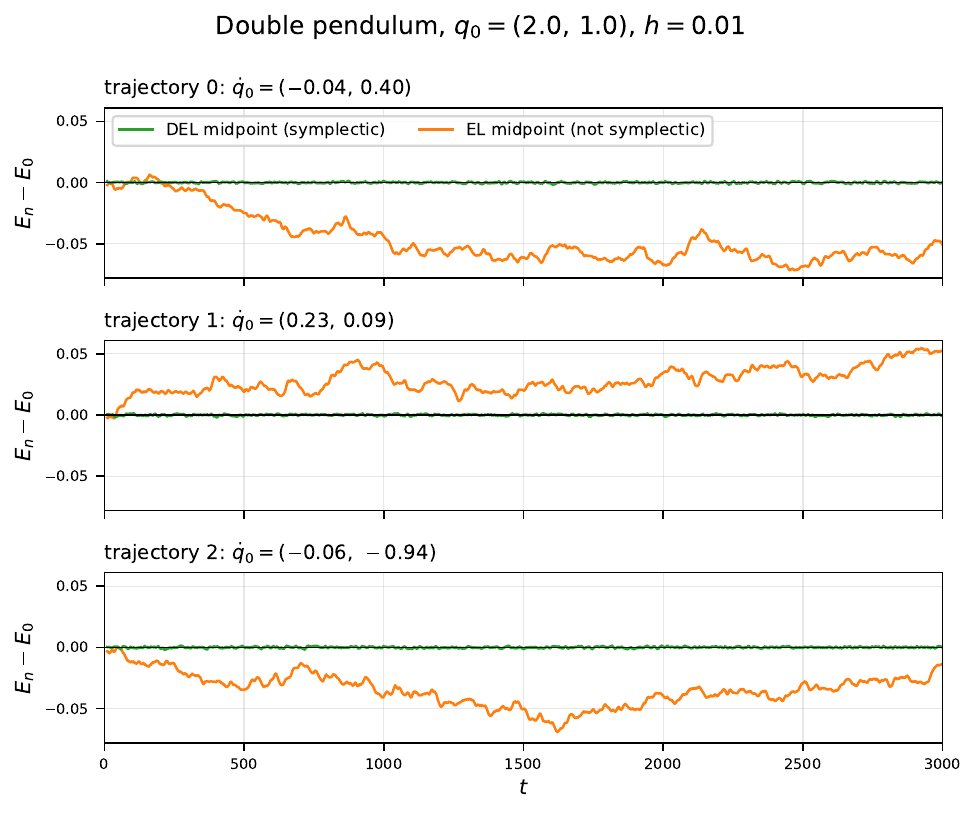}
    \caption{Energy error $E_n - E_0$ for the undamped double pendulum ($F=0$), integrated with the true Lagrangian from initial position $q_0=(2.0, 1.0)$ with step size $h=0.01$, one panel per initial velocity. The plot shows the moving average over $1000$ time steps. Green curves show the midpoint discretization of the Lagrangian (symplectic), while orange curves show the implicit midpoint discretization of the Euler--Lagrange equations in $(q,\dot q)$ (not symplectic).}
    \label{fig:appendix energy-midpoint}
\end{figure}
To show that the two midpoint discretizations behave differently also in practice, we integrate both with the true Lagrangian of the undamped double pendulum ($m_1=m_2=l_1=l_2=1$, $g=9.81$, $F=0$), here no learning is involved. We refer to the midpoint discretization of the Lagrangian, leading to the discrete Euler--Lagrange equations~\eqref{eq:DEL_force}, as the DEL midpoint scheme, and to the implicit midpoint discretization of the Euler--Lagrange equations~\eqref{EL-GLNN} as the EL midpoint scheme. All trajectories start from the same initial position $q_0=(2.0,\,1.0)$. We draw three initial velocities uniformly from $[-1,1]$ and integrate $3\cdot 10^5$ steps with $h=0.01$, that is, until $T=3000$. For these initial conditions both arms swing over the top and the motion is chaotic.
Pointwise errors therefore grow quickly and carry little information, so we compare energies instead. For the DEL midpoint scheme the energy is evaluated from the discrete momentum $p_n = \nabla_2 L_\Delta(q_{n-1}, q_n)$, and for the EL midpoint scheme directly from the state $(q_n, \dot q_n)$. The implicit equations are solved with Newton's method with tolerance $10^{-12}$. Figure~\ref{fig:appendix energy-midpoint} shows the energy error. Both schemes oscillate with a
comparable $\mathcal{O}(h^2)$ amplitude. For the symplectic DEL midpoint scheme the oscillation stays centred at zero over the entire run. For the non-symplectic EL midpoint scheme the centre drifts, and after $3000$ time units it has moved by $0.03$--$0.06$, with the sign of the drift differing between trajectories.

\section{Hyperparameters}\label{sc:Hyperparameters}

\begin{table}[htbp]
    
    \centering
    \begin{tabular}{llccccc}
        \toprule
        Model & Parameter & Task 1 & Task 2 & Task 3 & Task 4a & Task 4b \\
        \midrule
        \multirow{4}{*}{DFLNN}
        & $\omega_\text{physics}$ & 1.0 & 1.0 & 1.0 & 1.0 & 1.0 \\
        & $\omega_\text{reg}$     & 1.0 & 1.0 & 10.0 & 1.0 & 10.0\\
        & $\omega_\text{AE}$      & -   & -   & 1.0 & 1.0 & 1.0 \\
        & $R$                     & 10 & 10 & 20  & 20 & 50 \\
        \bottomrule
    \end{tabular}
    \caption{Loss parameters. '-' indicates that the hyperparameter is not defined for the specific model.}
    \label{tab:loss-params}
\end{table}

To keep $F_{\theta}^{\text{Free}}$ from interfering with the learning of the conservative dynamics specified by the Lagrangian term, we found that using dropout regularization \cite{srivastava2014dropout} was an effective solution. We used a fixed dropout rate at $0.5$. All experiments concentrate on small models and on the robustness of the model architectures in identifying an appropriate set of trainable parameters without requiring extensive hyperparameter tuning. Consequently, all neural network components are implemented as compact feed-forward networks with three hidden layers, a width of 30 neurons, and GELU activation functions.

The hyperparameters related to the loss function are detailed in Table \ref{tab:loss-params}.

\section{Harris-based Relative Image Error}\label{apx:HarrisRelativeError}
To evaluate the quality of predicted images, we introduce a feature-based relative error metric based on Harris corner detection~\cite{HarrisCorner1988}.
Instead of pixel-wise error, structural metrics have been shown to better reflect perceptual and geometrical similarity \cite{wang2004image}

Given a reference image $I^{ref}$ and a predicted image $I^{pred}$, we first compute their Harris corner response maps using the classical Harris detector.
Each response map is thresholded and binarized to obtain corner indicator functions $B^{ref}, B^{pred} \in \{0,1\}^{H \times W}$, where a value of $1$ indicates a detected corner pixel. From these binary maps, we compute the confusion counts:
\begin{equation*}
    \mathrm{TP} = |B^{pred} \cap B^{ref}|, \quad
    \mathrm{FP} = |B^{pred} \setminus B^{ref}|, \quad
    \mathrm{FN} = |B^{ref} \setminus B^{pred}|,
\end{equation*}
with $N = HW$ denoting the total number of pixels.

We then define a counts-based relative error term. In this work we use the balanced error formulation,
\begin{equation*}
    E_{\text{counts}} = \tfrac12\left( \frac{\mathrm{FN}}{|B^{ref}|} + \frac{\mathrm{FP}}{|B^{ref,c}|} \right),
\end{equation*}
which penalizes both missed corner detections (false negatives) and spurious corner detections in the background (false positives).  
Alternative variants such as Dice loss and miss rate were also considered.

To prevent degenerate solutions in which the predicted image becomes nearly constant (i.e.\ visually flat and corner-free), we include a flatness penalty based on image variance:
\begin{equation*}
    E_{\text{flat}} = \max\!\left(0,\,1 - \frac{\mathrm{Var}(I^{pred})}{\mathrm{Var}(I^{ref}) + \varepsilon}\right),
\end{equation*}
which approaches $1$ when the predicted image is much flatter than the reference and vanishes when both images have comparable variance.

The final Harris-based relative error is defined as a convex combination:
\begin{equation*}
    E = \alpha\, E_{\text{counts}} + (1-\alpha)\, E_{\text{flat}},
\end{equation*}
where $\alpha \in [0,1]$ controls the trade-off between structural feature mismatch and flatness regularization.
In all experiments, we set $\alpha = 0.7$.

This metric yields a value close to zero for structurally accurate predictions and increases when salient image features are missing, misplaced, or when predictions collapse to low-variance (nearly uniform) images.

\section{Datasets}

\subsection{Damped Double pendulum}\label{apx:dataset-doublependulum}
Let $\theta_1(t)$ and $\theta_2(t)$ denote the angular positions of the two pendulum links, and $\dot{\theta}_1(t)$, $\dot{\theta}_2(t)$ their angular velocities. 
Defining $\Delta = \theta_1 - \theta_2$, the state vector is
\begin{equation*}
    y(t) = (\theta_1, \dot{\theta}_1, \theta_2, \dot{\theta}_2).
\end{equation*}

The dynamics are governed by the nonlinear system
\begin{align*}
    \dot{\theta}_1 &= \omega_1, \\
    \dot{\theta}_2 &= \omega_2,
\end{align*}
\begin{align*}
    \dot{\omega}_1 &= 
    \frac{
    m_2 l_1 \omega_1^2 \sin(2\Delta)
    + 2 m_2 l_2 \omega_2^2 \sin(\Delta)
    + 2 g m_2 \cos(\theta_2)\sin(\Delta)
    + 2 g m_1 \sin(\theta_1)
    + \gamma_1
    }{
    -2 l_1 \left(m_1 + m_2 \sin^2(\Delta)\right)
    }, \\
    \dot{\omega}_2 &=
    \frac{
    m_2 l_2 \omega_2^2 \sin(2\Delta)
    + 2 (m_1 + m_2) l_1 \omega_1^2 \sin(\Delta)
    + 2 g (m_1 + m_2) \cos(\theta_1)\sin(\Delta)
    + \gamma_2
    }{
    2 l_2 \left(m_1 + m_2 \sin^2(\Delta)\right)
    }.
\end{align*}

Damping is incorporated through generalized forces $\gamma_1$ and $\gamma_2$. 
We use relative damping in which the torques depend on relative angular velocities,
\begin{equation*}
    \tau_1 = - b_1 \dot{\theta}_1 - b_2(\dot{\theta}_1 - \dot{\theta}_2), 
    \qquad
    \tau_2 = - b_2(\dot{\theta}_2 - \dot{\theta}_1),
\end{equation*}
which are inserted into $\gamma_1$ and $\gamma_2$ accordingly.

\subsection{Dissipative Charged particle traveling in a Magnetic Field}\label{apx:dataset-chargedparticle}
Let $\mathbf{r}(t) = (x(t), y(t), z(t)) \in \mathbb{R}^3$ denote the particle position and $\mathbf{v}(t) = \dot{\mathbf{r}}(t)$ its velocity. 
The state vector is
\begin{equation*}
    y(t) = (x, y, z, v_x, v_y, v_z).
\end{equation*}

The dynamics are governed by Newton's second law with Lorentz force and damping,

\begin{align*}
    \dot{\mathbf{r}} &= \mathbf{v}, \\
    m \dot{\mathbf{v}} &= q \, \mathbf{v} \times \mathbf{B}(\mathbf{r}) - b \mathbf{v}, 
\end{align*}
where $m$ is the particle mass, $q$ its electric charge, $b$ the damping coefficient, and $\mathbf{B}(\mathbf{r})$ is the magnetic field.

In the present setup, the magnetic field is taken to be spatially constant,
\begin{equation*}
    \mathbf{B}(\mathbf{r}) = (B_x, B_y, B_z),
\end{equation*}
although the formulation allows for spatial dependence on $r=\sqrt{x^2+y^2}$.

Writing the system in component form yields
\begin{equation*}
    \dot{x} = v_x, \qquad
    \dot{y} = v_y, \qquad
    \dot{z} = v_z,
\end{equation*}
\begin{align*}
    \dot{v}_x &= \frac{q}{m}(v_y B_z - v_z B_y) - \frac{b}{m} v_x, \\
    \dot{v}_y &= \frac{q}{m}(v_z B_x - v_x B_z) - \frac{b}{m} v_y, \\
    \dot{v}_z &= \frac{q}{m}(v_x B_y - v_y B_x) - \frac{b}{m} v_z.
\end{align*}




% \section*{Declaration of generative AI and AI-assisted technologies in the manuscript preparation process}

% During the preparation of this work the author(s) used [NAME OF TOOL/SERVICE] in order to [REASON]. After using this tool/service, the author(s) reviewed and edited the content as needed and take(s) full responsibility for the content of the published article.

% =========================================================================================

% =========================================================================================

\end{document}